\documentclass[iop,apj,twocolappendix,numberedappendix]{emulateapj}
\usepackage{apjfonts}
\usepackage{multirow}
\usepackage{textcomp}
\usepackage{amsmath}
\usepackage{microtype}
\usepackage{subfigure}
\usepackage{xcolor,ulem}
\usepackage[breaklinks=true]{hyperref}

\DeclareMathOperator{\erf}{erf}

\newcommand\edit[1]{{#1}}

\begin{document}

\title{Parameter estimation for binary neutron-star coalescences with realistic noise during the Advanced LIGO era}

\journalinfo{{\textsc{The Astrophysical Journal}}, \href{http://dx.doi.org/10.1088/0004-637X/804/2/114}{804:114 (24pp)}, 2015 May 10}

\author{Christopher P.\ L.\ Berry\altaffilmark{1}}
\email{cplb@star.sr.bham.ac.uk}

\author{Ilya Mandel\altaffilmark{1}}
\author{Hannah Middleton\altaffilmark{1}}
\author{Leo P.\ Singer\altaffilmark{2,3,14}}
\author{Alex L.\ Urban\altaffilmark{4}}
\author{Alberto Vecchio\altaffilmark{1}}
\author{Salvatore Vitale\altaffilmark{5}}

\author{Kipp Cannon\altaffilmark{6}}
\author{Ben Farr\altaffilmark{7,1,8}}
\author{Will M.\ Farr\altaffilmark{1}}
\author{Philip B.\ Graff\altaffilmark{9,10}}
\author{Chad Hanna\altaffilmark{11,12}}
\author{Carl-Johan Haster\altaffilmark{1}}
\author{Satya Mohapatra\altaffilmark{13,5}}
\author{Chris Pankow\altaffilmark{4}}
\author{Larry R.\ Price\altaffilmark{2}}
\author{Trevor Sidery\altaffilmark{1}}
\author{John Veitch\altaffilmark{1}}

\altaffiltext{1}{School of Physics \& Astronomy, University of Birmingham, Birmingham, B15 2TT, UK}
\altaffiltext{2}{LIGO Laboratory, California Institute of Technology, Pasadena, CA 91125, USA}
\altaffiltext{3}{Astrophysics Science Division, NASA Goddard Space Flight Center, Code 661, Greenbelt, MD 20771, USA}
\altaffiltext{4}{Leonard E.\ Parker Center for Gravitation, Cosmology, and Astrophysics, University of Wisconsin\nobreakdashes--Milwaukee, Milwaukee, WI 53201, USA}
\altaffiltext{5}{Massachusetts Institute of Technology, 185 Albany St, Cambridge, MA 02139, USA}
\altaffiltext{6}{Canadian Institute for Theoretical Astrophysics, 60 St.\ George Street, University of Toronto, Toronto, Ontario, M5S 3H8, Canada}
\altaffiltext{7}{Department of Physics and Astronomy \& Center for Interdisciplinary Exploration and Research in Astrophysics (CIERA), Northwestern University, Evanston, IL 60208, USA}
\altaffiltext{8}{Enrico Fermi Institute, University of Chicago, Chicago, IL 60637, USA}
\altaffiltext{9}{Department of Physics, University of Maryland\nobreakdashes--College Park, College Park, MD 20742, USA}
\altaffiltext{10}{Gravitational Astrophysics Lab, NASA Goddard Space Flight Center, Greenbelt, MD 20771, USA}
\altaffiltext{11}{Perimeter Institute for Theoretical Physics, Ontario, N2L 2Y5, Canada}
\altaffiltext{12}{The Pennsylvania State University, University Park, PA 16802, USA}
\altaffiltext{13}{Syracuse University, Syracuse, NY 13244, USA}
\altaffiltext{14}{NASA Postdoctoral Fellow}

\shorttitle{PE with realistic noise during aLIGO}
\shortauthors{Berry et al.}


\begin{abstract}
Advanced ground-based gravitational-wave (GW) detectors begin operation imminently. Their intended goal is not only to make the first direct detection of GWs, but also to make inferences about the source systems. Binary neutron-star mergers are among the most promising sources. We investigate the performance of the parameter-estimation \edit{(PE)} pipeline that will be used during the first observing run of the Advanced Laser Interferometer Gravitational-wave Observatory (aLIGO) in 2015: we concentrate on the ability to reconstruct the source location on the sky, but also consider the ability to measure masses and the distance.  Accurate, rapid sky-localization is necessary to alert electromagnetic (EM) observatories so that they can perform follow-up searches for counterpart transient events. We consider PE accuracy in the presence of \edit{non-stationary}, non-Gaussian noise. We find that the character of the noise makes negligible difference to the PE performance \edit{at a given signal-to-noise ratio}. The source luminosity distance can only be poorly constrained, the median $90\%$ ($50\%$) credible interval scaled with respect to the true distance is $0.85$ ($0.38$). However, the chirp mass is well measured. Our chirp-mass estimates are subject to systematic error because we used gravitational-waveform templates without component spin to carry out inference on signals with moderate spins, but the total error is typically less than $10^{-3} M_\odot$. The median $90\%$ ($50\%$) credible region for sky localization is $\sim600~\mathrm{deg^{2}}$ ($\sim150~\mathrm{deg^{2}}$), with $3\%$ ($30\%$) of detected events localized within $100~\mathrm{deg^{2}}$. Early aLIGO, with only two detectors, will have a sky-localization accuracy for binary neutron stars of hundreds of square degrees; this makes EM follow-up challenging, but not impossible.
\end{abstract}

\section{Introduction}

The goal of gravitational-wave (GW) astronomy is to learn about the Universe through observations of gravitational radiation. This requires not only the ability to detect GWs, but also to infer the properties of their source systems. In this work, we investigate the ability to perform parameter estimation (PE) on signals detected by the upcoming Advanced LIGO (aLIGO) instruments \citep{Harry:2010zz,TheLIGOScientific:2014jea} in the initial phase of their operation \citep{Aasi:2013wya}.

Compact binary coalescences (CBCs), the GW-driven inspiral and merger of stellar-mass compact objects, are a prime source for aLIGO and Advanced Virgo \citep[AdV;][]{AdvVirgo,TheVirgo:2014hva}. Binary neutron-star (BNS) systems may be the most abundant detectable CBCs \citep{Abadie:2010cf}. We focus on BNS mergers in this study.

Following the identification of a detection candidate, we wish to extract the maximum amount of information from the signal. It is possible to make some inferences using selected components of the data. However, full information regarding the source system, including the component objects' masses and spins, is encoded within the gravitational waveform, and can be obtained by comparing the data to theoretical waveform models \citep{Cutler:1994ys,Jaranowski:2007pe}. Doing so can be computationally expensive.

PE is performed within a Bayesian framework. We use algorithms available as part of the \textsc{LALInference} toolkit for the analysis of CBC signals. The most expedient code is \textsc{bayestar} \citep{Singer:2014qca,Singer:2014}, which infers sky location from data returned from the detection pipeline. Exploring the posterior probability densities for the parameters takes longer for models where the parameter space is larger or the likelihood is more complicated. Calculating estimates for parameters beyond sky location is done using the stochastic-sampling algorithms of \textsc{LALInference} \citep{Veitch:2014wba}. There are three interchangeable sampling algorithms: \textsc{LALInference\_nest} \citep{Veitch:2009hd}, \textsc{LALInference\_mcmc} \citep{vanderSluys:2008qx,Raymond:2008im} and \textsc{LALInference\_bambi} \citep{Graff:2011gv}, which we refer to as \textsc{LALInference} for short. These compute waveform templates for use in the likelihood. Using the least computationally expensive waveforms allows for posteriors to be estimated on timescales of hours to days; potentially more accurate estimates can be calculated with more expensive waveforms. In this paper, we discuss what can be achieved using low-latency (\textsc{bayestar}) and medium-latency (\textsc{LALInference} with inexpensive waveforms) PE; a subsequent paper will evaluate what can be achieved on longer timescales using more expensive waveform templates.

With the detection of GWs, it is also possible to perform multi-messenger astronomy, connecting different types of observations of the same event. BNS mergers could be accompanied by an electromagnetic (EM) counterpart \citep{Metzger:2011bv}. To associate an EM event with a GW signal, it is beneficial to have an accurate sky location: timing information can also be used for EM signals that are independently detected, such as gamma-ray bursts \citep{Aasi:2014iia}. To provide triggers for telescopes to follow up a GW detection, it is necessary to provide rapid sky localization.

Several large-scale studies investigated the accuracy with which sky position can be reconstructed from observations with ground-based detector networks. The first only used timing information from a multi-detector network to triangulate the source position on the sky \citep[e.g.,][]{Fairhurst:2009tc,Fairhurst:2010is}. Subsequently, further information about the phase of the gravitational waveform was folded into the timing triangulation (TT) analysis \citep{Grover:2013sha}. The most sophisticated techniques perform a coherent Bayesian analysis to reconstruct probability distributions for the sky location \citep[e.g., ][]{Veitch:2012df,Nissanke:2012dj,Kasliwal:2013yqa,Grover:2013sha,Sidery:2013zua}. \citet{Singer:2014qca} used both \textsc{bayestar} and \textsc{LALInference} to analyse the potential performance of aLIGO and AdV in the first two years of their operation. They assumed the detector noise was stationary and Gaussian. Here, we further their studies (although we use the same analysis pipeline) by using a set of injections into observed noise from initial LIGO detectors recoloured (see section \ref{sec:recolour}) to the expected spectral density of early aLIGO.\footnote{We refer to the noise as recoloured as it is first whitened (removing its colour), to eliminate initial LIGO's frequency dependence, and then passed through a linear response filter (reintroducing colour) so that, on average, it has the aLIGO spectral density.} This provides results closer to those expected in practice, as real interferometer noise includes features such as non-stationary glitches \citep{Aasi:2013wya,Aasi:2014usa}. Our results are just for the first observing run (O1) of aLIGO, expected in the latter half of 2015, assuming that this occurs before the introduction of AdV. As the sensitivity of the detectors will increase with time, and because the introduction of further detectors increases the accuracy of sky localization \citep{Schutz:2011tw}, these set a lower bound for the advanced-detector era. Estimates for sky-localization accuracy in later observing periods can be calibrated using our results.

PE beyond sky localization, considering the source system's mass, spin, distance and orientation, has been subject to similar studies. The initial investigations estimated PE using the Fisher information matrix \citep[e.g.,][]{Cutler:1994ys,Poisson:1995ef,Arun:2004hn}. \edit{This} only gives an approximation to true PE potential \citep{Vallisneri:2007ev}. More reliable (but computationally expensive) results are found by simulating a GW event and analysing it using PE codes, mapping the posterior probability distributions \citep[e.g.,][]{Rover:2006ni,vanderSluys:2007st,Veitch:2009hd,Rodriguez:2013oaa}. This has even been done for a blind injection during the run of initial LIGO \citep{Aasi:2013jjl}. As with sky-localization, general PE can improve with the introduction of more detectors to the network \citep{Veitch:2012df}. 

To be as faithful as possible, our analysis is performed using one of the pipelines intended for use during O1. We make use of \edit{the} LIGO Scientific Collaboration Algorithm Library (LAL).\footnote{\url{http://www.lsc-group.phys.uwm.edu/lal}} In particular, we shall make use of \textsc{GSTLAL},\footnote{\url{https://www.lsc-group.phys.uwm.edu/daswg/projects/gstlal.html}} one of the detection pipelines, to search for signals and \textsc{LALInference} for PE on detection candidates.

We begin by describing the source catalogue and detector sensitivity curve used for this study in section \ref{sec:input}. In section \ref{sec:method} we explain how the data is analysed to produce sky areas and other parameter estimates. \edit{Many details from these two sections are shared with the preceding work of \citet{Singer:2014qca}, which can be consulted for further information}. In section \ref{sec:results} we present the results of our work. We first discuss the set of events that are selected by the detection pipeline in section \ref{sec:detect} (with supplementary information in appendix \ref{ap:mass}); then we examine PE, considering sky-localization accuracy in section \ref{sec:sky-loc}, and mass and distance measurements in section \ref{sec:PE}. We conclude with a discussion of these results in section \ref{sec:end}; this includes in section \ref{sec:three-det} an analysis of estimates for sky localization in later observing periods with reference to our findings. Estimates of the computational costs associated with running \textsc{bayestar} and full \textsc{LALInference} PE are given in appendix \ref{ap:cost}. \edit{A supplementary catalogue of results is described in appendix \ref{ap:online-data}, with data available at \url{http://www.ligo.org/scientists/first2years/}.}

Our main findings are:
\begin{enumerate}
\item The detection pipeline returns a population of sources that is not significantly different from the input astrophysical population, despite a selection bias based upon the chirp mass.
\item Both \textsc{bayestar} and \textsc{LALInference} return comparable sky-localization accuracies (for a two-detector network). The latter takes more computational time (a total CPU time of $\sim 10^6~\mathrm{s}$ per event compared with $\sim 10^3~\mathrm{s}$), but returns estimates for more parameters than just location.
\item At a given signal-to-noise ratio (SNR), the character of the noise does not affect sky localization or other PE.
\item Switching from a detection threshold based upon SNR to one based upon the false alarm rate (FAR) changes the SNR-distribution of detected events. A selection based upon FAR includes more low-SNR events (the distribution at high SNRs is unaffected).
\item TT provides a poor predictor of sky localization for a two-detector network; it does better (on average) for a three-detector network when phase coherence is included, but remains imperfect.
\item Systematic errors from uncertainty in the waveform template are significant for chirp-mass estimation. Neglecting the mass--spin degeneracy by using non-spinning waveforms artificially narrows the posterior distribution.
\end{enumerate} 
For O1, we find that the luminosity distance is not well-measured, the median $50\%$ credible interval (interquartile range) divided by the true distance is $0.38$ and the median $90\%$ credible interval divided by the true distance is $0.85$. Despite being subject to systematic error, the chirp mass is still accurately measured, with the posterior mean being less than $10^{-3}M_\odot$ from the true value \edit{in almost} all ($96\%$) cases. We find that the median area of $50\%$ sky localization credible region is $154~\mathrm{deg^2}$ and the median area of the $90\%$ credible region is $632~\mathrm{deg^2}$; the median searched area (area of the smallest credible region that encompasses the source location) is $132~\mathrm{deg^2}$. EM follow-up to BNS mergers in 2015 will be challenging and require careful planning.

\section{Sources and sensitivities}\label{sec:input}

Our input data consists of two components: simulated detector noise and simulated BNS signals. We describe the details of these in the following subsections, before continuing with the analysis of the data in section \ref{sec:method}.

\subsection{Recoloured 2015 noise}\label{sec:recolour}

\begin{figure}
\centering
\includegraphics[angle=90,width=0.9\columnwidth]{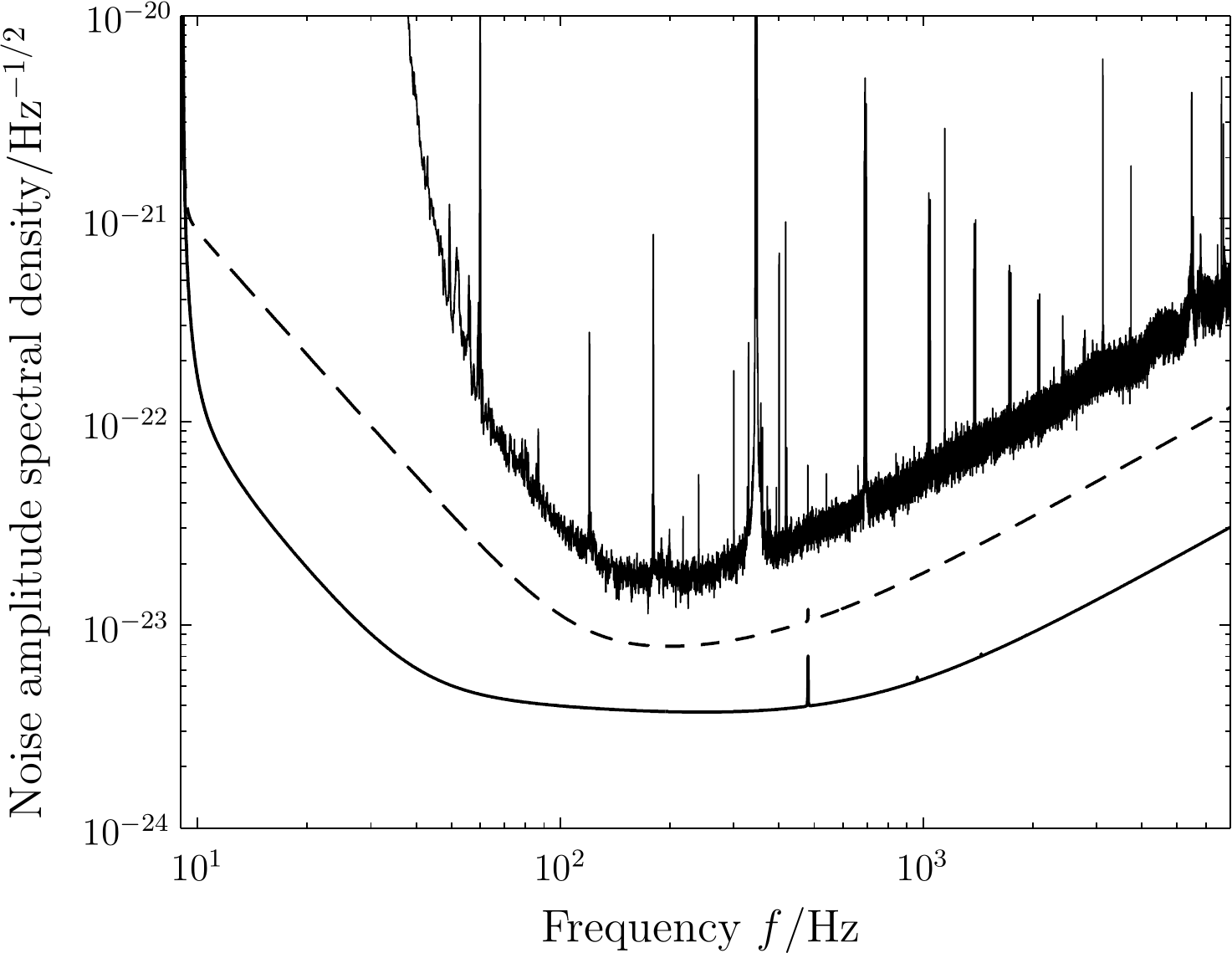}
\caption{Initial and Advanced LIGO noise amplitude spectral densities. The upper line is the measured sensitivity of the initial LIGO Hanford detector during S6 \citep{Aasi:2014usa}. The dashed line shows the early aLIGO sensitivity and the lower solid line the final sensitivity \citep{Barsotti:2012}. The early sensitivity is used as a base here.}
\label{fig:noise}
\end{figure}

We consider the initial operation of the advanced detectors at LIGO Hanford and LIGO Livingston. The sensitivity is assumed to be given by the early curve of \citet{Barsotti:2012}, which has a BNS detection range of $\sim 55~\mathrm{Mpc}$ (assuming Gaussian noise). This configuration corresponds to the 2015 observing scenario in \citet{Aasi:2013wya}. Figure \ref{fig:noise} plots the noise spectral density, the square root of the power spectral density \citep{Moore:2014lga}, as measured during the sixth science (S6) run of initial LIGO,\footnote{\url{http://www.ligo.caltech.edu/~jzweizig/distribution/LSC_Data/}} the early aLIGO sensitivity curve, and final aLIGO curve \citep{Shoemaker:2010}.

The noise is constructed from data from the S6 run of initial LIGO \citep{Christensen:2010zz,Aasi:2014usa}, recoloured to the early aLIGO noise spectral density as was done for \citet{Aasi:2014tra}. \edit{We use real noise, instead of idealised Gaussian noise, to try to capture a realistic detector response including transients; however, the S6 noise can only serve as a proxy for the actual noise in aLIGO since the detectors are different. Two calendar months (21 August 2010--20 October 2010) of S6 data were used. The recoloured data are constructed using \mbox{\textsc{GSTLAL\_fake\_frames}}.\footnote{\url{https://ldas-jobs.ligo.caltech.edu/~gstlalcbc/doc/gstlal-0.7.1/html/gstlal__fake__frames.html}} The recolouring process can be thought of as applying a finite-impulse response filter to whitened noise. The result is a noise stream that, on average, has the same power spectral density as expected for early aLIGO, but contains transients that are similar to those found in S6. Recolouring preserves the non-stationary and non-Gaussian features of the noise, although they are distorted \citep{Aasi:2014tra}. The recoloured noise is the most realistic noise we can construct ahead of having the real noise from aLIGO.}

\subsection{Binary neutron-star events}

BNS systems constitute the most probable and best understood source of signals for advanced ground-based GW detectors. There is a wide range in predicted event rates as a consequence of uncertainty in our knowledge of the astrophysics. \citet{Abadie:2010cf} gives a BNS merger rate for the full-sensitivity aLIGO--AdV network of $0.01$--$10~\mathrm{Mpc^{-3}\,Myr^{-1}}$, with $1~\mathrm{Mpc^{-3}\,Myr^{-1}}$ as the most realistic estimate \citep{Kalogera:2003tn}.

We use the same list of simulated sources as in \citet{Singer:2014qca}. The neutron-star masses are uniformly distributed from $m_\mathrm{min} = 1.2 M_\odot$ to $m_\mathrm{max} = 1.6 M_\odot$, which safely encompasses the observed mass range of BNS systems \citep{Kiziltan:2013oja}. Their (dimensionless) spin magnitudes are uniformly distributed between $a_\mathrm{min} = 0$ and $a_\mathrm{max} = 0.05$. \edit{The most rapidly rotating BNS constituent to be observed in a binary that should merge within a Hubble time} is PSR J0737$-$3039A \citep{Burgay:2003jj,Kramer:2009zza}. This has been estimated to have a spin within this range \citep{Mandel:2009nx,Brown:2012qf}: since we do not know precisely the neutron-star equation of state \citep{Lattimer:2012nd}, it is not possible to exactly convert from a spin period to a spin magnitude. The spin orientations are distributed isotropically. The binaries are uniformly scattered in volume and \edit{isotropically orientated}. This set of parameters is motivated by our understanding of the astrophysical population of BNSs.

The GW signals were constructed using a post-Newtonian (PN) inspiral template, the SpinTaylorT4 approximant \citep{Buonanno:2002fy,Buonanno:2009zt} which is a time-domain approximant accurate to 3.5PN order in phase and 1.5PN order in amplitude. There exist more accurate but more expensive waveforms. This template only contains the inspiral part of the waveform and not the subsequent merger: this should happen outside of the sensitive band of the detector for the masses considered and so should not influence PE \citep{Mandel:2014tca}. \edit{We do not use SpinTalyorT4 templates either for detection or PE}, instead we use a less expensive approximant. In a future study, we shall investigate the effects of using SpinTaylorT4 templates for PE, such that the injection and recovery templates perfectly match.

\section{Analysis pipeline}\label{sec:method}

To accurately forecast sky localization prospects in O1, we run our simulated events through the same data-analysis pipeline as is intended for real data. The results of this pipeline are analysed in the next section (section \ref{sec:results}). A GW search is performed using \textsc{GSTLAL\_inspiral} \citep{Cannon:2010qh,Cannon:2011tb,Cannon:2011vi,Cannon:2012zt}; this is designed to provide GW triggers in real time with $\sim10$--$100~\mathrm{s}$ latency during LIGO--Virgo observing runs. A trigger is followed up for sky localization if its calculated FAR is less than $10^{-2}~\mathrm{yr^{-1}}$, which is roughly equivalent to a network-SNR threshold of $\varrho \gtrsim 12$ \citep{Aasi:2013wya}.

In using the FAR to select triggers, our method differs from that used in \citet{Singer:2014qca}. Since they considered Gaussian noise, which is free of glitches, their FAR would not be representative of those computed using real noise; the FAR calculated with Gaussian noise corresponds to a SNR-threshold that is too low for detection in realistic noise. Therefore, they also imposed a network-SNR cut of $\varrho \geq 12$, in addition to the FAR selection. This joint SNR and FAR threshold was found to differ negligibly from an SNR-only threshold: in effect, they select by SNR alone. While this is a small difference in selection criteria, we shall see in section \ref{sec:sky-loc} that this has an impact on our sky-localization results.

To recover the GW signal, another PN inspiral approximant, TaylorF2 \citep{Damour:2000zb,Damour:2002kr,Buonanno:2009zt}, was used as a template. This is a frequency-domain stationary-phase approximation waveform accurate to 3.5PN order in phase and Newtonian order in amplitude. It does not include the effects of spin, although it can be modified to incorporate these \citep{Mikoczi:2005dn,Arun:2008kb,Bohe:2013cla}. We neglect spin as this should not lead to a significant reduction in detection efficiency for systems with low spins \citep{Brown:2012qf}, which we confirm in section \ref{sec:mass-spin}. TaylorF2 does not incorporate as many physical effects as SpinTaylorT4, notably it does not include precession, but is less computationally expensive, permitting more rapid follow-up.

Rapid sky localization is computed using \textsc{bayestar} \citep{Singer:2014qca}. This reconstructs sky position using a combination of information associated with the triggers: the times, phases and amplitudes of the signals at arrival at each detector. It coherently combines this information to reconstruct posteriors for the sky position. \textsc{bayestar} makes no attempt to infer intrinsic parameters such as the BNS masses and, hence, can avoid computationally expensive waveform calculations. The sky-position distributions can be formulated in under a minute (see appendix \ref{ap:cost}).

Full PE, computing posterior distributions for sky localization parameters as well as the other parameters for the source system like masses, orientation and inclination, is performed using \textsc{LALInference} \citep{Veitch:2014wba}. \textsc{LALInference} maps the posterior probability distribution by stochastically sampling the parameter space \citep[e.g.,][chapter 29]{MacKay:2003}. There are three codes within \textsc{LALInference} to sample these posterior distributions: \textsc{LALInference\_nest} \citep{Veitch:2009hd}, a nested sampling algorithm \citep{Skilling:2006}; \textsc{LALInference\_mcmc} \citep{vanderSluys:2008qx,Raymond:2008im}, a Markov-chain Monte Carlo algorithm \citep[chapter 12]{Gregory:2005}, and \textsc{LALInference\_bambi} \citep{Graff:2011gv}, another nested sampling algorithm \citep{Feroz:2008xx} which incorporates a means of speeding up likelihood evaluation using machine learning \citep{Graff:2013cla}. All three codes use the same likelihood and so should recover the same posteriors; consistency of the codes has been repeatedly checked. While the codes produce the same results, they may not do so in the same times, depending upon the particular problem. All the results here were computed with \textsc{LALInference\_nest}.

TaylorF2 waveforms were used again in constructing the \textsc{LALInference} posterior. Since these do not exactly match the waveforms used for injection, there may be a small bias in the recovered parameters \citep{Buonanno:2009zt}. Using TaylorF2 is much less computationally expensive than using SpinTaylorT4, in this case a \textsc{LALInference} run takes $\sim10^6~\mathrm{s}$ of CPU time (see appendix \ref{ap:cost}).

\section{Results}\label{sec:results}

\subsection{Detection catalogue}\label{sec:detect}

We ran sky-localization codes on a set of $333$ events recovered from the detection pipeline. We shall compare these to the results of \citet{Singer:2014qca} who used Gaussian noise for the same sensitivity curve. They ran \textsc{bayestar} on a sample of $630$ events, but only ran \textsc{LALInference} on a sub-sample of $250$ events. We first consider the set of detected events before moving on to examine sky-localization accuracies in section \ref{sec:sky-loc}, and mass and distance measurement in section \ref{sec:PE}.

\subsubsection{Signal-to-noise ratio distribution}\label{sec:SNRs}

\begin{figure}
  \centering
   \includegraphics[width=0.9\columnwidth]{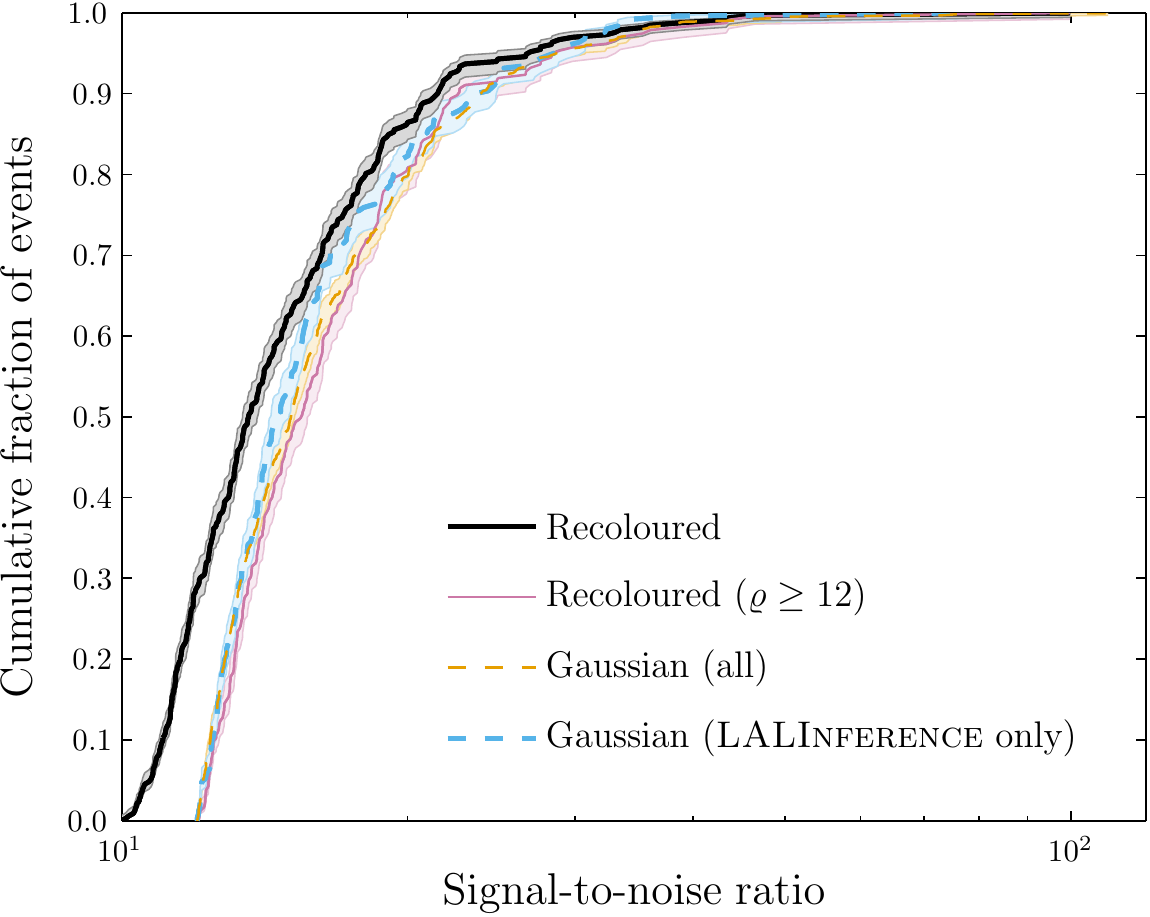}
    \caption{Cumulative fractions of events with network \edit{SNRs} smaller than the abscissa value. The SNR distribution assuming recoloured noise is denoted by the thick solid line; we also show the distribution subject to a lower cutoff of $\varrho \geq 12$, denoted by the thin solid line. The SNR distribution for the complete set of $630$ events with Gaussian noise analysed with \textsc{bayestar} is denoted by the thinner dashed line and the distribution for the subset of $250$ events analysed with both \textsc{bayestar} and \textsc{LALInference} is denoted by the thicker dashed line \citep{Singer:2014qca}. The $68\%$ confidence intervals ($1\sigma$ for a normal distribution) are denoted by the shaded areas, these are estimated from a beta distribution \citep{Cameron:2010bh}.} 
    \label{fig:snr}
\end{figure}

Unsurprisingly, the distribution of SNRs differs between the recoloured and Gaussian data sets. This is shown in figure \ref{fig:snr}.
The recoloured SNR distribution includes a tail at low SNR ($\varrho \simeq 10$--$12$). If we impose a lower threshold $\varrho \geq 12$ for the recoloured data set, as was done for the Gaussian data set, we find that the SNR distributions are similar. With the shared SNR cut, the distributions agree within the expected sampling error; performing a Kolmogorov--Smirnov (KS) test \citep[section 9.5]{DeGroot:1975} comparing the recoloured SNR distribution to the complete (\textsc{LALInference} only) Gaussian SNR distribution returns a $p$-value of $0.311$ ($0.110$).

\begin{figure}
  \centering
   \subfigure[]{\includegraphics[width=0.9\columnwidth]{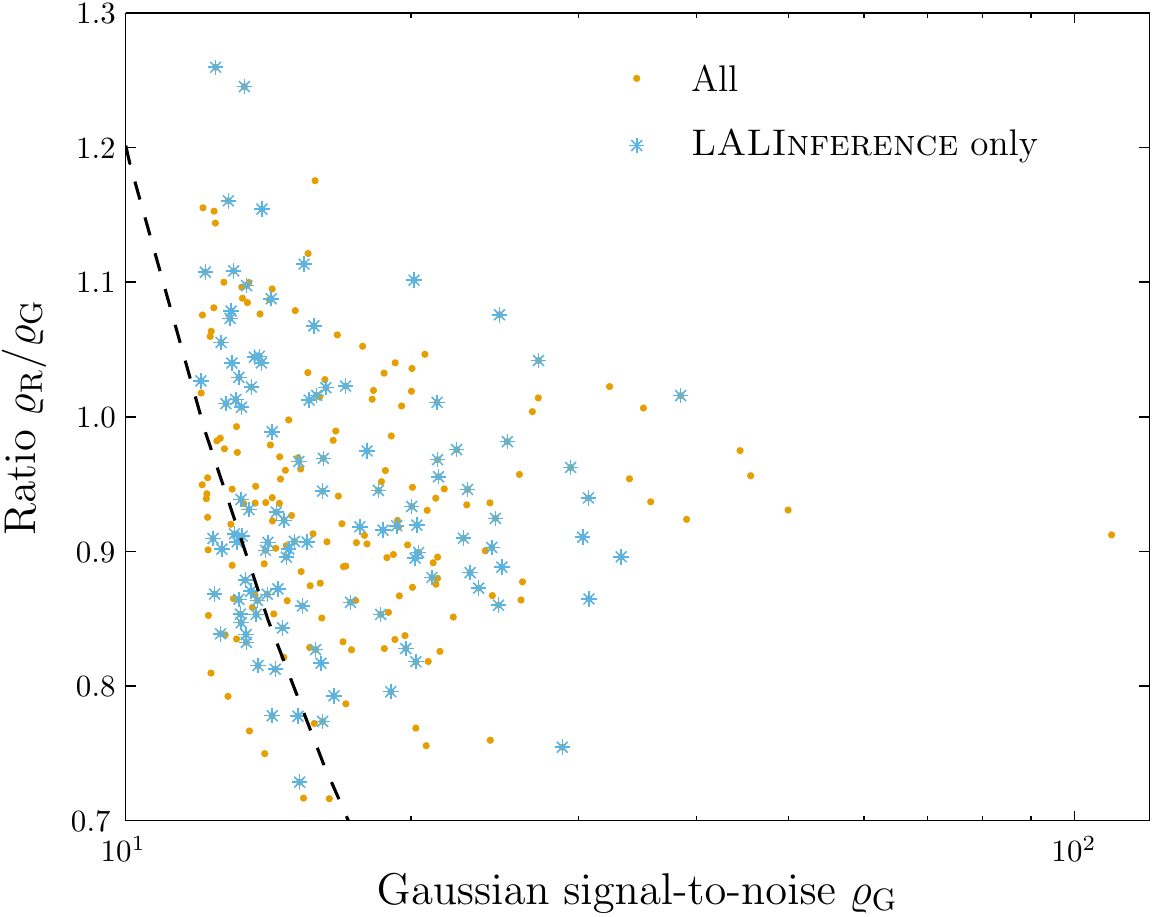}} \\
   \subfigure[]{\includegraphics[width=0.9\columnwidth]{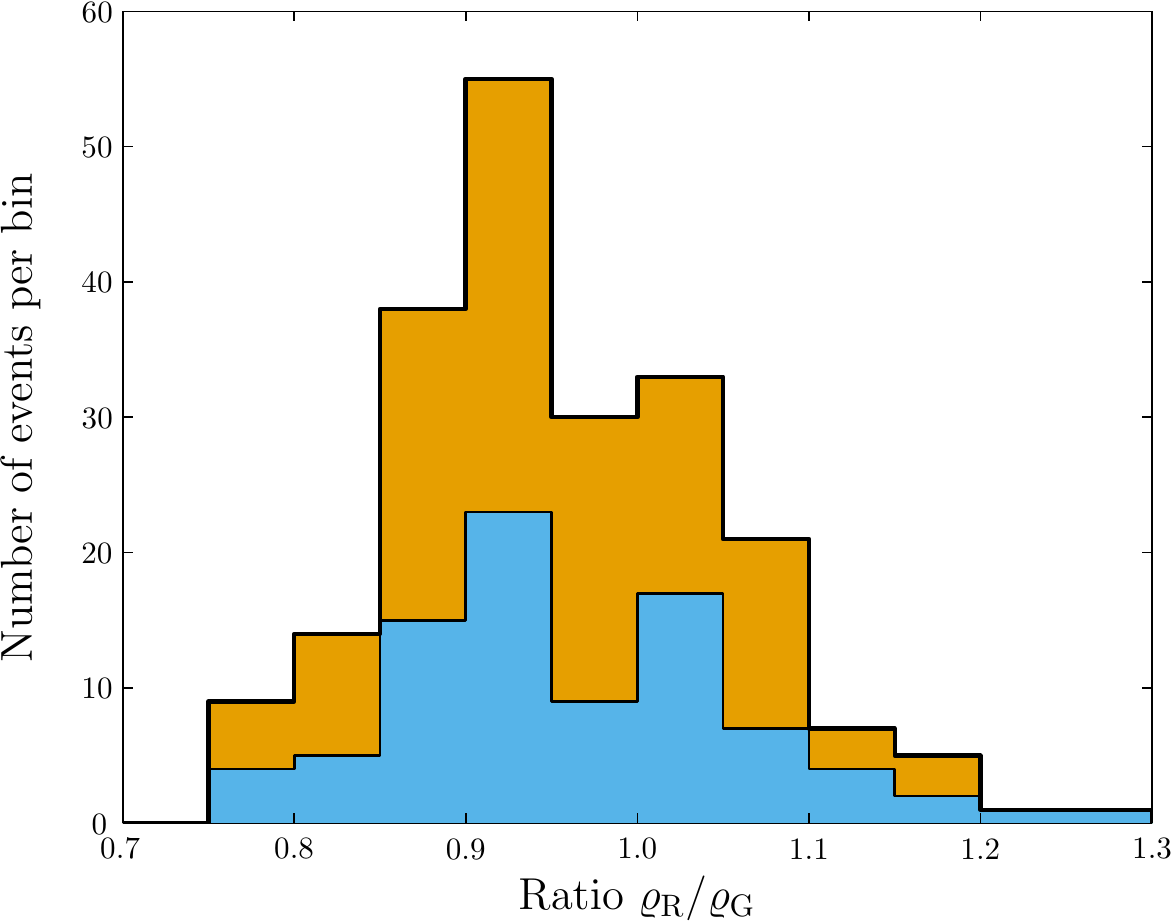}}
    \caption{Comparison of \edit{SNRs} from injections with Gaussian noise $\varrho_\mathrm{G}$ and from injections with recoloured noise $\varrho_\mathrm{R}$. (a) The ratio $\varrho_\mathrm{R}/\varrho_\mathrm{G}$ as a function of $\varrho_\mathrm{G}$. The dashed line shows the locus of $\varrho_\mathrm{R} = 12$. (b) Distribution of $\varrho_\mathrm{R}/\varrho_\mathrm{G}$ with both $\varrho_\mathrm{G} \geq 12$ and $\varrho_\mathrm{R} \geq 12$, using a bin width of $0.5$. Events that fall within the sub-sample of Gaussian events analysed with \textsc{LALInference} are highlighted with blue (star-shaped points) and the complete set of events detected in both the Gaussian and recoloured data sets is indicated by orange (round points).} 
    \label{fig:comparison}
\end{figure}

Comparing injections between the recoloured and Gaussian data sets, there are $255$ events that have been detected in both sets. There are $108$ events shared between the recoloured data set and the sub-sample of the Gaussian data set analysed with \textsc{LALInference}. Considering individual events, we may contrast the SNR for recoloured noise $\varrho_\mathrm{R}$ and Gaussian noise $\varrho_\mathrm{G}$. The ratio of the two SNRs \edit{is shown} in figure \ref{fig:comparison}.
Considering the entire population of shared detections, the mean value of the ratio of SNRs is $\varrho_\mathrm{R}/\varrho_\mathrm{G} = 0.938 \pm 0.006$, showing a small downwards bias as an effect of the differing cutoffs used for the two samples. To limit selection effects that could skew the distribution of the ratio of SNRs, we can impose an SNR cut of $\varrho_\mathrm{R} \geq 12$. This reduces the number of events detected in both noise sets to $214$ using the full Gaussian set and $88$ for the \textsc{LALInference} Gaussian sub-sample. There is a small difference between the SNR as calculated with Gaussian noise and with recoloured noise. This does not appear to be a strong function of the SNR. However, the scatter in the ratio decreases as SNR increases, approximately decreasing as $\varrho^{-1}$. This is as expected as the inclusion of random noise realisations in the signal should produce fluctuations in the SNR of order $\pm 1$; these fluctuations become less significant for louder events. After imposing the cut $\varrho \geq 12$ on both sets, the mean value of the ratio of SNRs is $\varrho_\mathrm{R}/\varrho_\mathrm{G} = 0.955 \pm 0.006$. Although there is a small difference in SNRs, we shall see that this does not impact our PE results.

\subsubsection{Selection effects}\label{sec:mass-spin}

The population of detected events should not match exactly the injected distribution; depending upon their parameters, some systems are louder and hence easier to detect. Here, we look at the selection effects of the most astrophysically interesting parameters: mass and spin. We expect there to be a selection based upon mass, as the component masses set the amplitude of the waveform. We do not expect there to be a dependence upon the spin because the spin magnitude is small, but since we injected with a spinning waveform and recovered with a non-spinning waveform, there could potentially be a selection effect due to waveform mismatch. Checking these distributions confirms the effectiveness of the detection pipeline for this study.

To leading order, the GW amplitude is determined by the ($5/6$ power of the) chirp mass \citep{Sathyaprakash:2009xs}
\begin{equation}
\mathcal{M}_\mathrm{c} = \frac{(m_1 m_2)^{3/5}}{(m_1 + m_2)^{1/5}},
\end{equation}
where $m_1$ and $m_2$ are the individual component masses. We therefore expect to preferentially select systems with larger chirp masses.

\begin{figure}
  \centering
   \includegraphics[width=0.9\columnwidth]{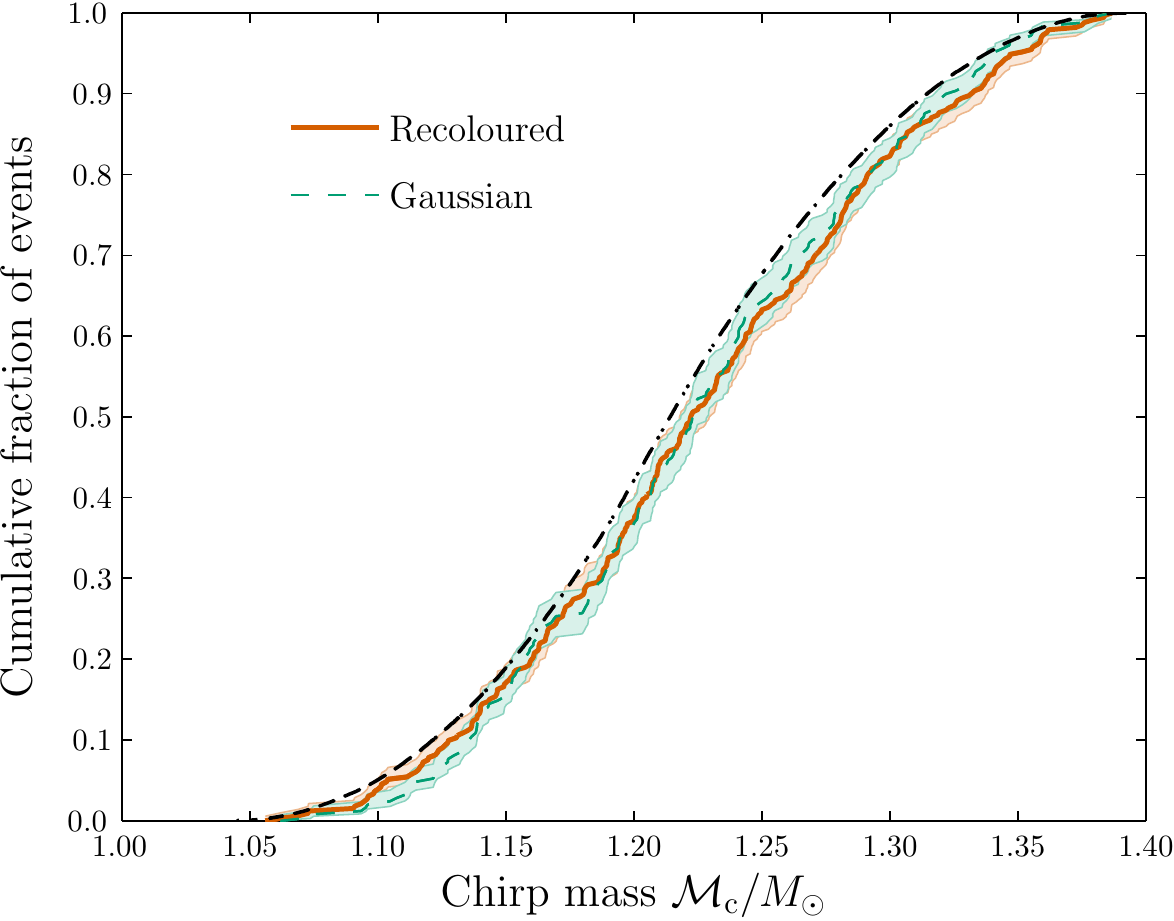}
    \caption{Cumulative fractions of detected events with chirp masses smaller than the abscissa value. Results using recoloured noise are denoted by the solid line, and results from the subset of $250$ events with Gaussian noise analysed with \textsc{LALInference} are denoted by the dashed line \citep{Singer:2014qca}. The $68\%$ confidence intervals are denoted by the shaded areas. The injection distribution, based upon a uniform distribution of component masses, is indicated by the dot--dashed line.} 
    \label{fig:M-chirp}
\end{figure}

Figure \ref{fig:M-chirp} shows the recovered distribution of (injected) chirp masses and the injection distribution (which is calculated numerically).
We do detect fewer systems with smaller chirp masses (and more with larger chirp masses), as indicated by the curve for the recovered distribution lying below the curve for the injection distribution. However, this selection effect does not alter the overall character of population. The difference is only marginally statistically significant with this number of events (a KS test with the injection distribution yields $p$-values of $0.315$ and $0.068$ for the Gaussian and recoloured noise respectively). This is consistent with expectations for this narrow chirp-mass distribution; in appendix \ref{ap:mass} we use a simple theoretical model to predict that we would need $\sim 10^3$ detections (or a broader distribution of chirp masses in the injection set) to see a significant difference between the injected and recovered populations. The character of the noise does not influence the chirp-mass distribution (a KS test gives a $p$-value of $0.999$).

For completeness, in appendix \ref{ap:mass} we present the distributions for the individual component masses, the asymmetric mass ratio and the total mass. The selection effects on these depend upon their correlation with the chirp mass; the total mass, which is most strongly correlated with the chirp mass, shows the most noticeable difference between injection and detected distributions.

\begin{figure}
  \centering
   \includegraphics[width=0.9\columnwidth]{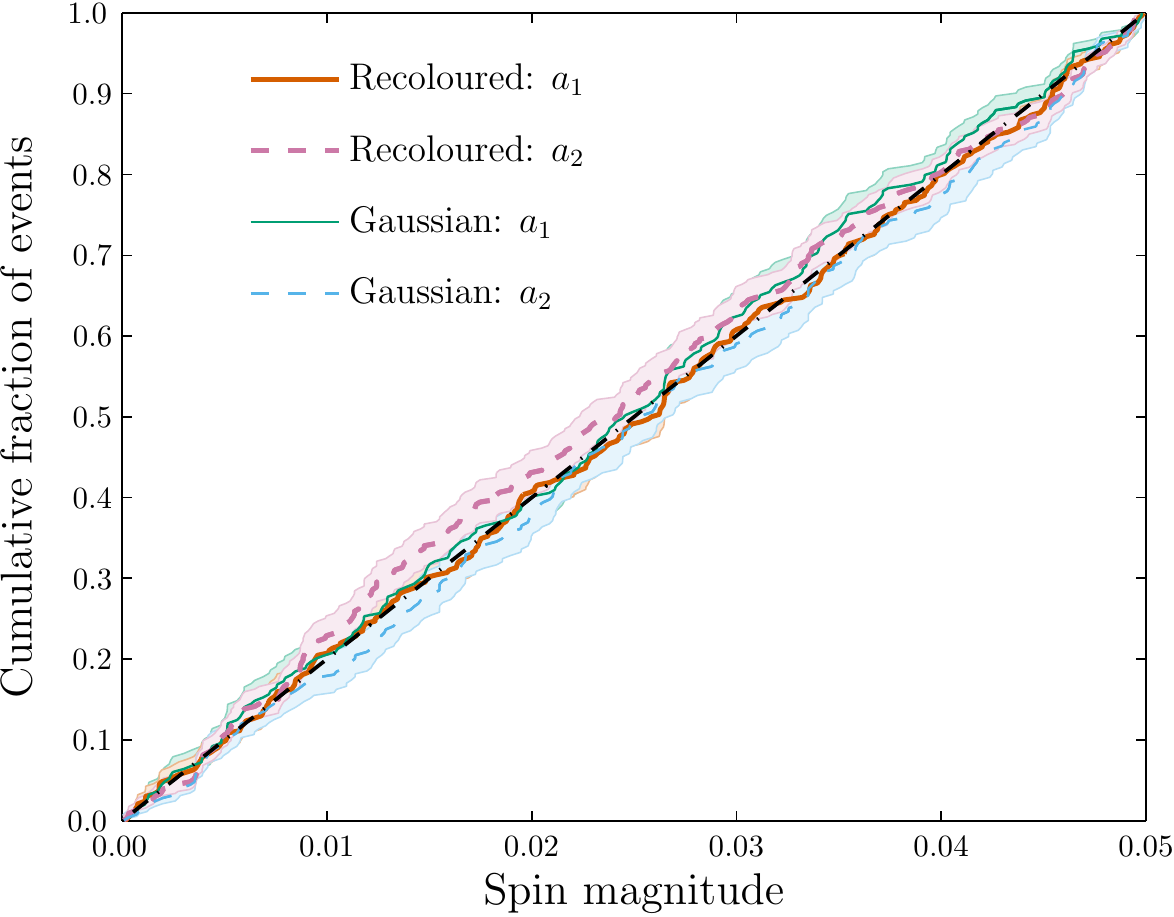}
    \caption{Cumulative fractions of detected events with spin magnitudes smaller than the abscissa value. The spin distribution for the first neutron star $a_1$ is denoted by the solid line, and the distribution for the second neutron star $a_2$ is denoted by the dashed line. Results using recoloured noise are denoted by the thicker red--purple lines, and results from the subset of $250$ events with Gaussian noise analysed with \textsc{LALInference} are denoted by the thinner blue--green lines \citep{Singer:2014qca}. The $68\%$ confidence intervals are denoted by the shaded areas. The expected distribution for spins uniform from $a_\mathrm{min} = 0$ to $a_\mathrm{max} = 0.05$ is indicated by the black dot--dashed line.}
    \label{fig:spin}
\end{figure}

Since we injected with a spinning waveform and recovered with a non-spinning waveform, there could also be a selection bias depending upon the spin magnitude. Figure \ref{fig:spin} shows the recovered distribution of (injected) spins.
The detected events are consistent with having the uniform distribution of spins used for the injections. \edit{We conclude that the presence of spins with magnitudes $a \le 0.05$ does not affect the detection efficiency for BNS systems}, in agreement with \citet{Brown:2012qf}.

\subsection{Sky-localization accuracy}\label{sec:sky-loc}

\begin{figure*}
  \centering
   \subfigure[\label{fig:bayestar-map}]{\includegraphics[width=0.9\columnwidth]{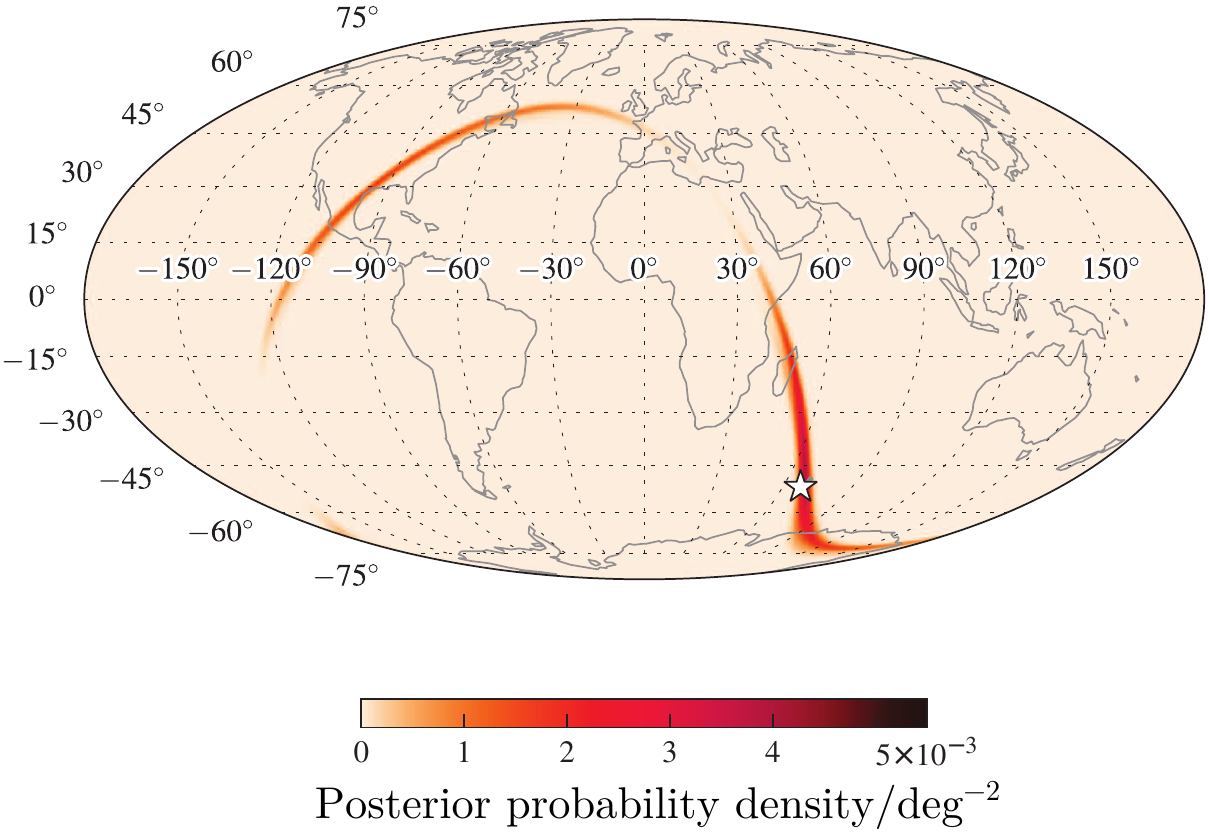}} \qquad
   \subfigure[\label{fig:lalinference-map}]{\includegraphics[width=0.9\columnwidth]{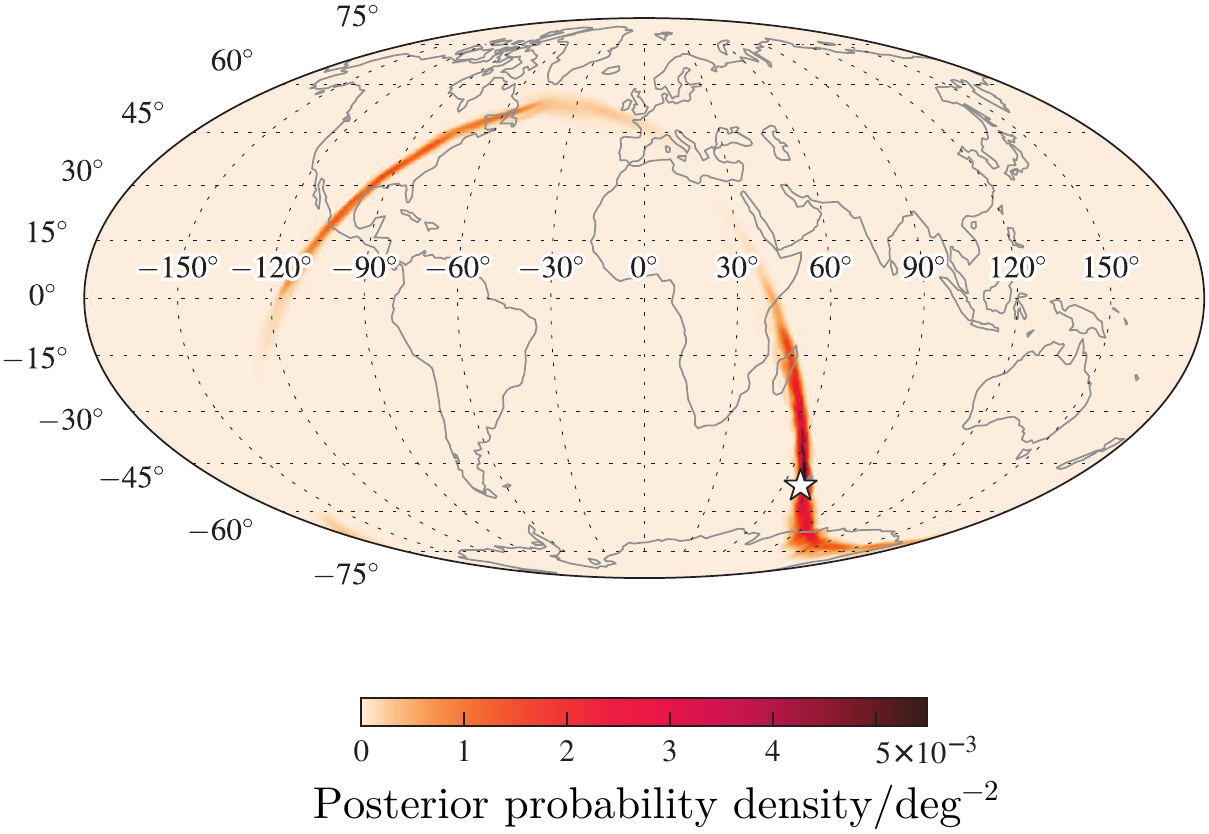}}
    \caption{Posterior probability density for sky location, plotted in a Mollweide projection in geographic coordinates. The star indicates the true source location. (a) Computed by \textsc{bayestar}. (b) Computed by \textsc{LALInference}. The event has simulation ID 1243 and a network SNR of $\varrho = 13.2$. \edit{Versions of these plots, and all the other events using in this study, can be found online at \url{http://www.ligo.org/scientists/first2years/}.}
}  
    \label{fig:sky-map}
\end{figure*}

The recovered sky positions from \textsc{bayestar} and \textsc{LALInference} appear in good agreement. A typical example of the recovered posterior probability density is shown in figure \ref{fig:sky-map}.
This is a bimodal distribution, reflecting the symmetry in the sensitivity of the detectors, which is common \citep{Singer:2014qca}. We use geographic coordinates to emphasise the connection to the position of the detectors. \edit{A catalogue of results can be viewed online at \url{http://www.ligo.org/scientists/first2years/} (see appendix \ref{ap:online-data}).}

To quantify the accuracy of sky localization, we use credible regions: areas of the sky that include a given total posterior probability. We denote the credible region for a total posterior probability $p$ as $\mathrm{CR}_p$: it is defined as
\begin{equation}
\mathrm{CR}_p \equiv \min{A}
\label{eq:CR}
\end{equation}
such that the sky area $A$ satisfies
\begin{equation}
p = \int_A \mathrm{d}\boldsymbol{\Omega} P_{\Omega}(\boldsymbol{\Omega}),
\end{equation}
where $P_{\Omega}(\boldsymbol{\Omega})$ is the posterior probability density over sky position $\boldsymbol{\Omega}$ \citep{Sidery:2013zua}. A smaller $\mathrm{CR}_p$ at a given $p$ indicates more precise sky localization.

We also consider the searched area: the area of the smallest credible region that includes the true location, and, hence, the area of the sky that we expect would have to be observed before the true source was found.

\begin{figure}
  \centering
   \includegraphics[width=0.7\columnwidth]{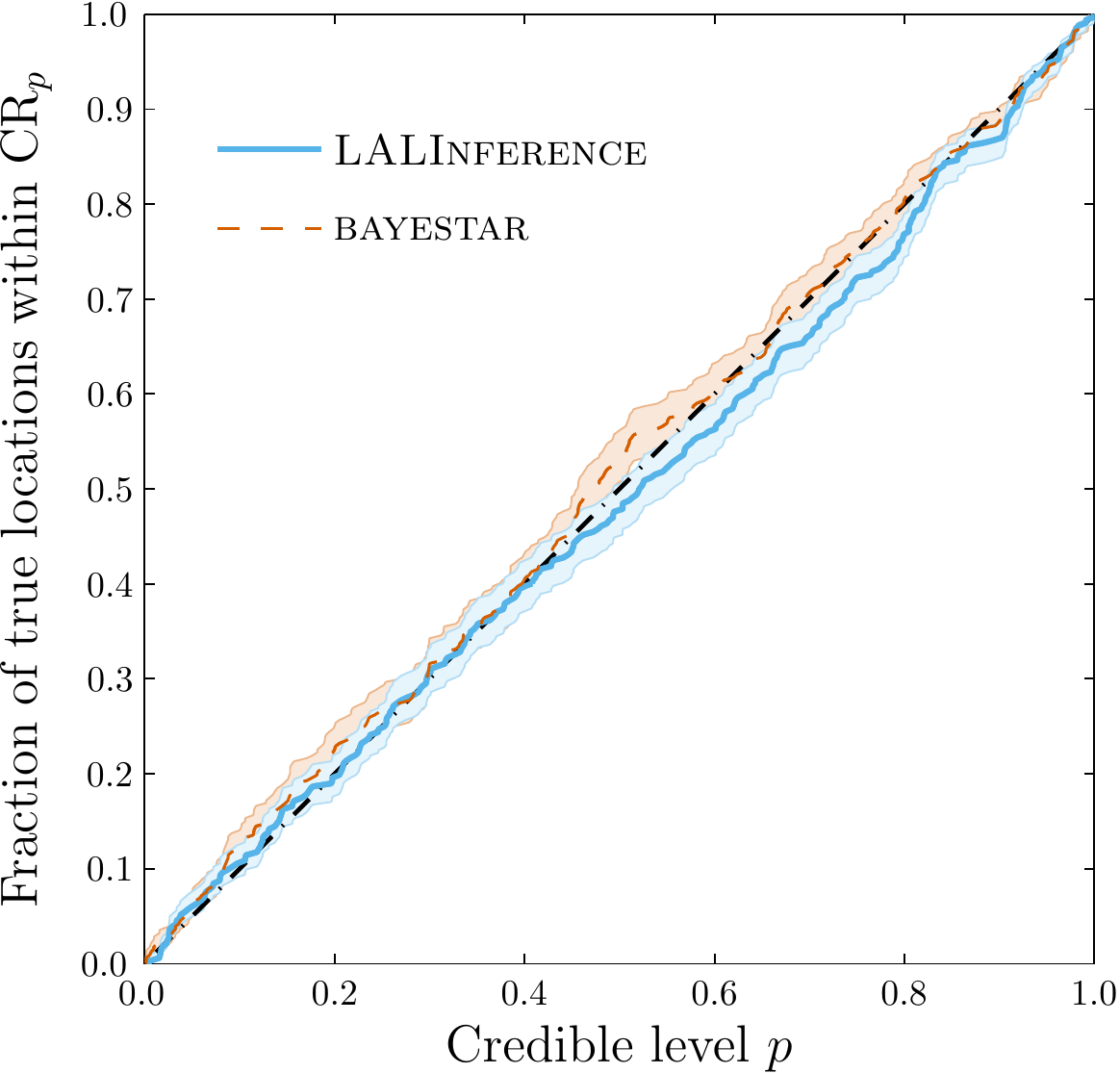}
    \caption{Fraction of true locations found within a credible region as a function of encompassed posterior probability. Results from \textsc{LALInference} are indicated by the solid line, results from \textsc{bayestar} are indicated by the dashed line and the expected distribution is indicated by the dot--dashed diagonal line. The $68\%$ confidencee interval is enclosed by the shaded regions, this accounts for sampling errors and is estimated from a beta distribution \citep{Cameron:2010bh}.} 
    \label{fig:pp}
\end{figure}

The self-consistency of our sky areas can be checked by calculating the fraction of events that fall within the credible region at the given probability. We expect that a fraction $p$ of true sky positions are found within $\mathrm{CR}_p$; that is the frequentist confidence region agrees with our Bayesian credible region \citep{Sidery:2013zua}. Figure \ref{fig:pp} shows the fraction of events found within a given $\mathrm{CR}_p$ as a function of $p$.
The distributions are consistent with expectations: performing a KS test with the predicted distribution yields $p$-values of $0.455$ and $0.546$ for \textsc{LALInference} and \textsc{bayestar} respectively. Both \textsc{LALInference} and \textsc{bayestar} produce self-consistent and unbiased sky areas in the presence of recoloured noise.

\begin{figure}
  \centering
   \subfigure[]{\includegraphics[width=0.9\columnwidth]{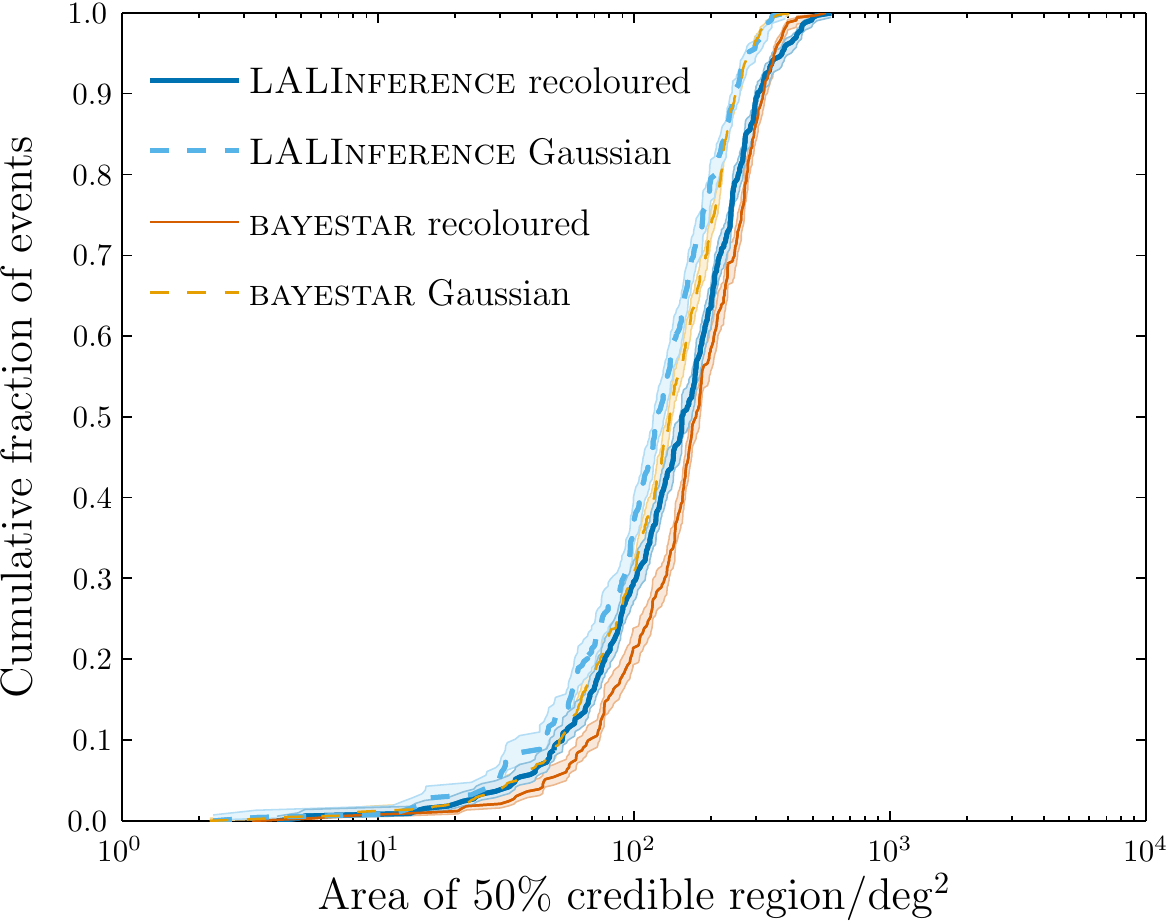}} \\
   \subfigure[]{\includegraphics[width=0.9\columnwidth]{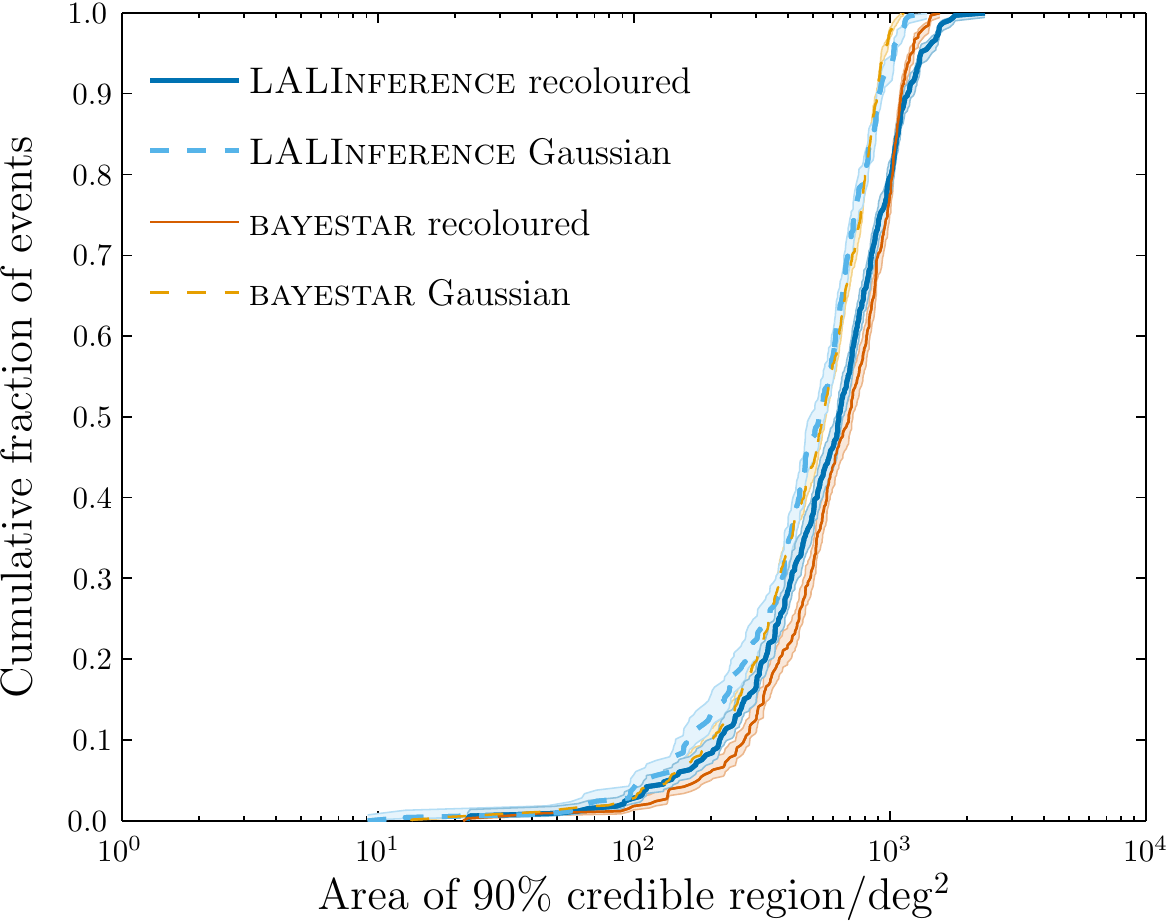}} \\
   \subfigure[]{\includegraphics[width=0.9\columnwidth]{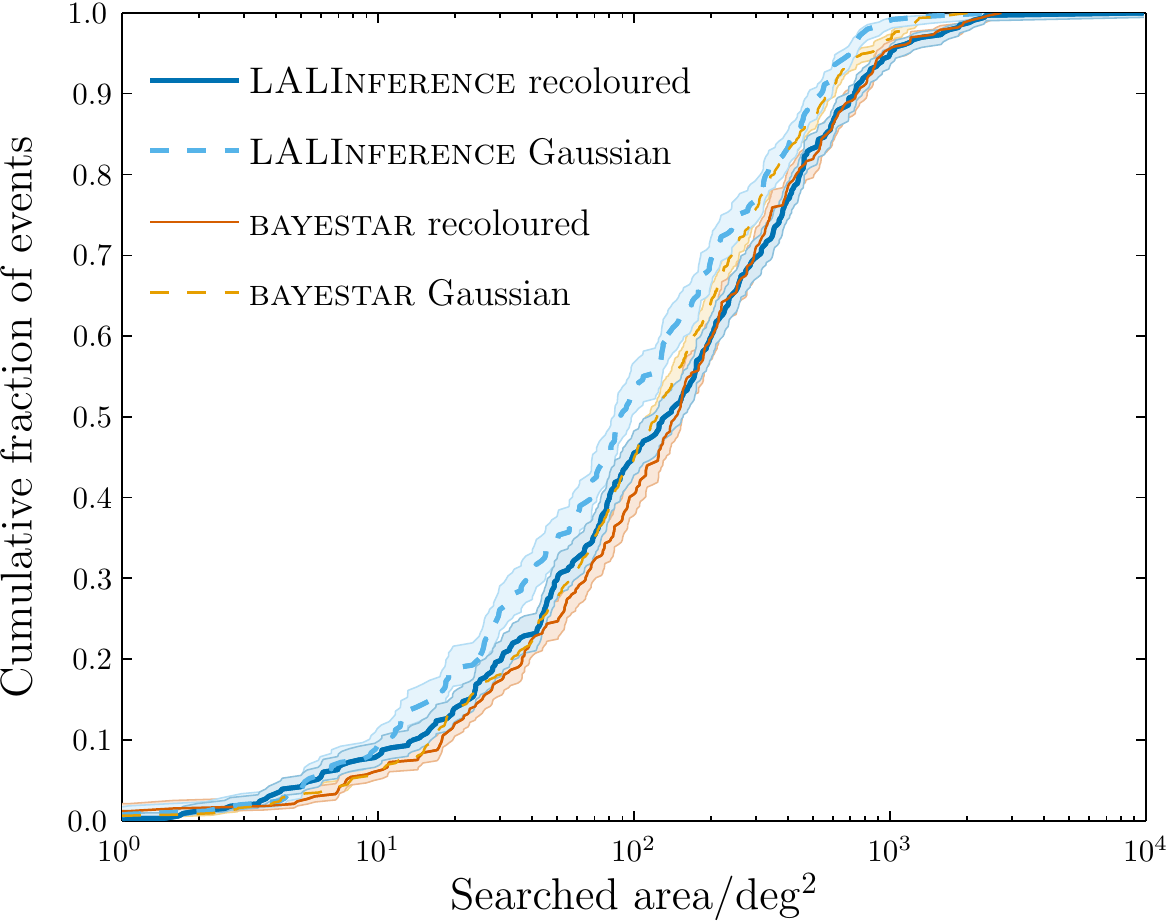}}
    \caption{Cumulative fractions of events with sky-localization areas smaller than the abscissa value. (a) Sky area of $50\%$ credible region $\mathrm{CR}_{0.5}$, the (smallest) area enclosing $50\%$ of the total posterior probability. (b) Sky area of $\mathrm{CR}_{0.9}$. (c) Searched area $A_\ast$, the area of the smallest credible region containing the true position. \textsc{LALInference} and \textsc{bayestar} results are denoted by thicker blue and thinner red--orange lines respectively. The results of this study are indicated by a solid line, while the results of \citet{Singer:2014qca}, which uses Gaussian noise, are indicated by a dashed line. The $68\%$ confidence intervals are denoted by the shaded areas.} 
    \label{fig:areas}
\end{figure}

The recovered sky areas are plotted in figure \ref{fig:areas}. This shows the cumulative distribution of areas for $\mathrm{CR}_{0.5}$, $\mathrm{CR}_{0.9}$ and searched areas $A_\ast$ as recovered from \textsc{LALInference} and \textsc{bayestar}. \edit{We plot} both the results using recoloured noise and the results using Gaussian noise from \citet{Singer:2014qca}.
All the results are similar. \textsc{LALInference} produces (marginally) more accurate sky localizations than \textsc{bayestar}, but the rapid code does a successful job of reconstructing the sky position in a much shorter time (see appendix \ref{ap:cost} for estimates of computation time). The recovered areas are (generally) marginally smaller for \textsc{LALInference} as this makes use of more information and so is expected to perform better (a KS test returns $p$-values of $0.740$ when comparing $\mathrm{CR}_{0.9}$ for Gaussian noise and $0.181$ for recoloured noise).

The difference between the Gaussian and recoloured results can be understood as a consequence of the SNR distribution (see figure \ref{fig:snr}). The SNR is the dominant factor affecting sky localization. For example, there is no strong correlation between the time delay between detection at the two LIGO sites and the sky-localization accuracy. The inclusion of more low-SNR events means that, on average, the results using recoloured noise are worse.

\begin{figure*}
  \centering
   \subfigure[]{\includegraphics[width=0.9\columnwidth]{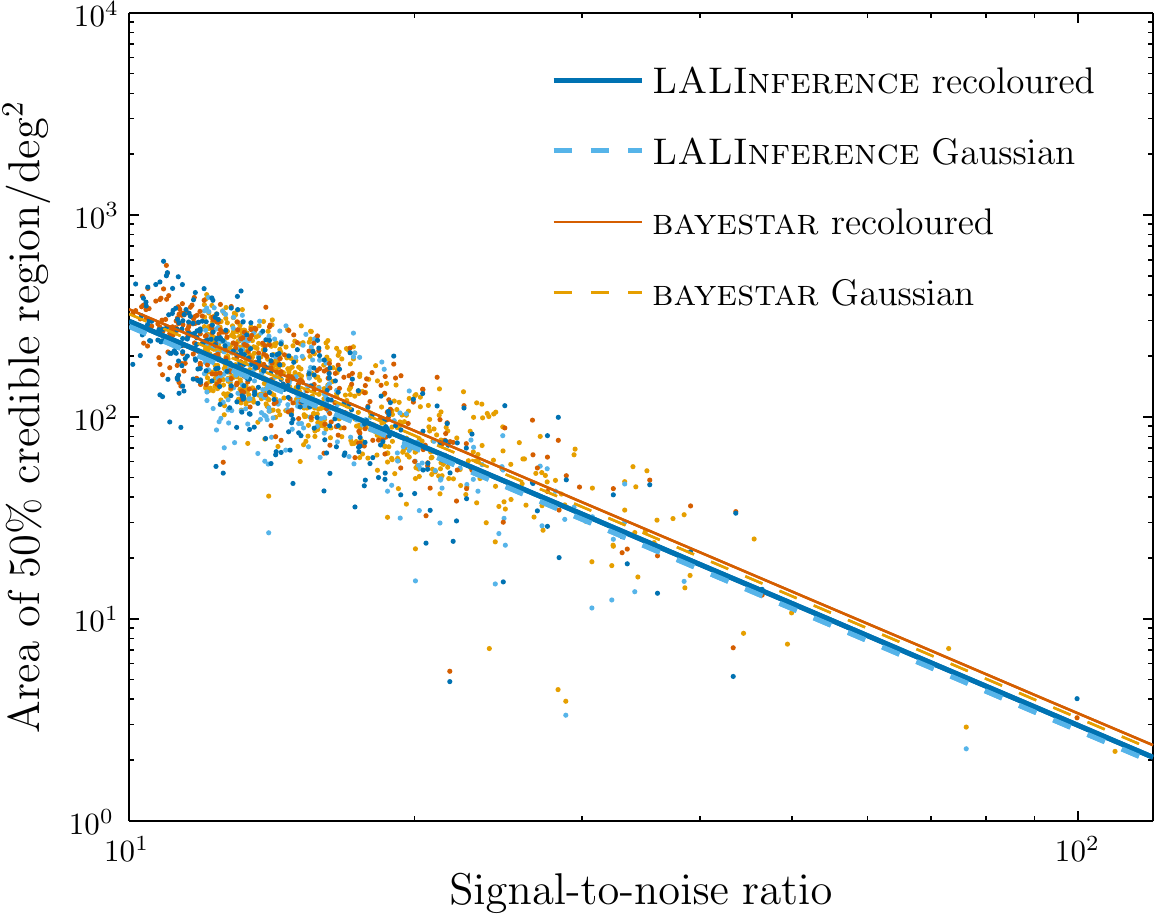}} \quad
   \subfigure[]{\includegraphics[width=0.9\columnwidth]{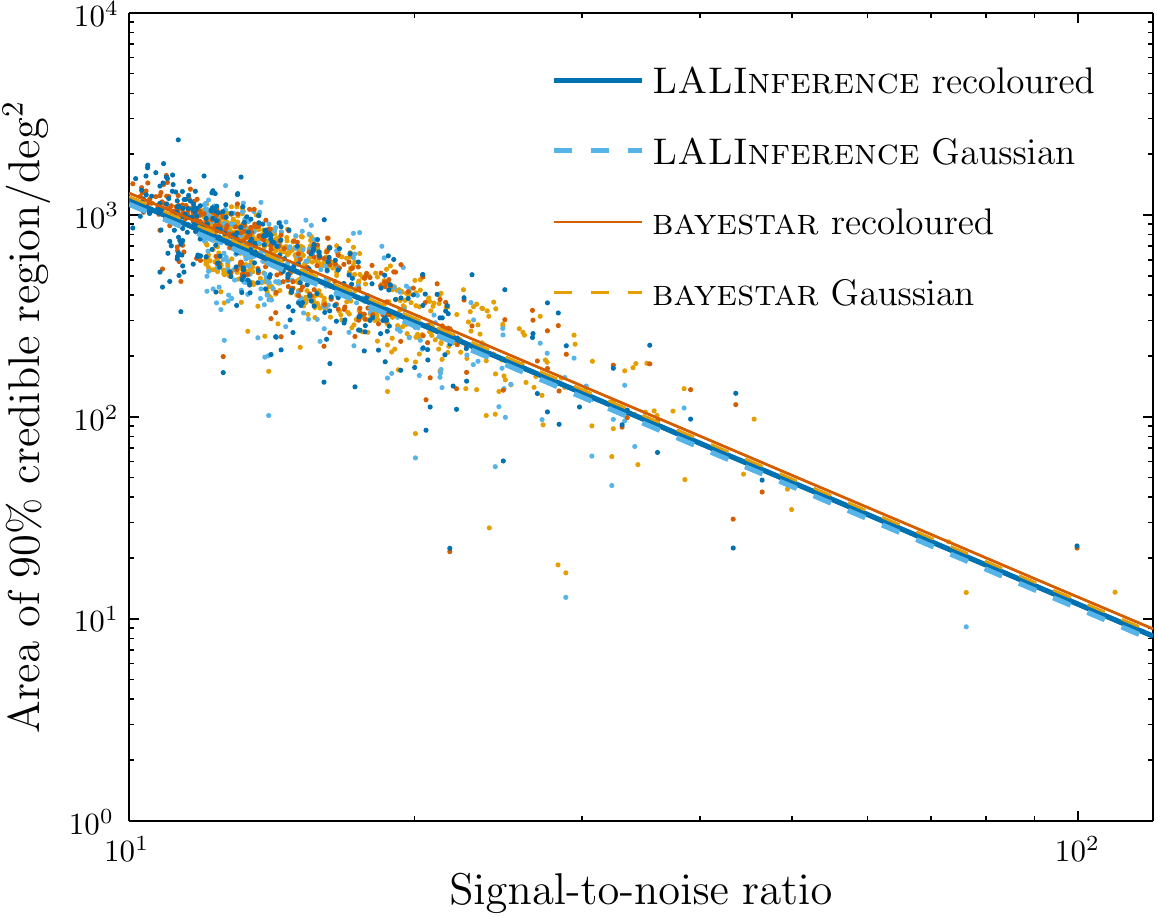}}
    \caption{Sky-localization areas as a function of \edit{SNR} $\varrho$. (a) Sky area of $50\%$ credible region $\mathrm{CR}_{0.5}$. (b) Sky area of $\mathrm{CR}_{0.9}$. Individual results are indicated by points. We include simple best-fit lines assuming that the area $A \propto \varrho^{-2}$. \textsc{LALInference} and \textsc{bayestar} results are denoted by thicker blue and thinner red--orange lines respectively. The results of this study are indicated by a solid line, while the results of \citet{Singer:2014qca}, which uses Gaussian noise, are indicated by a dashed line.} 
    \label{fig:snr-area}
\end{figure*}

The sky-localization accuracy is expected to scale \edit{as $\varrho^{-2}$}. The uncertainty in each direction on the sky scales inversely with the SNR, hence the area scales inversely with the square of the SNR \citep[cf.][]{Fairhurst:2009tc,Fairhurst:2010is}. This SNR scaling can be verified by plotting recovered sky areas as a function of $\varrho$ as shown in figure \ref{fig:snr-area}.
The recovered areas do show the expected correlation, although there is considerable scatter resulting from the variation in intrinsic parameters.

We have plotted fiducial best-fit lines with the expected scaling. The fitting was done simply using a naive least-squares method, fitting a straight line to $\log\varrho$ and $\log A$ for each sky area $A$. Allowing the slope of the line to vary from $-2$ yields negligible change to the fit. There is little difference between the trends for the recoloured and Gaussian results, indicating that the variation in the sky-localization accuracies is primarily an effect of the different distribution of SNRs. There is a small discrepancy between \textsc{LALInference} and \textsc{bayestar} in both cases, but the difference is not significant and is within the uncertainty expected from the scatter of results. The general trend for the sky-localization areas can be approximated as
\begin{subequations}\label{eq:SNR-fit}
\begin{align}
\log_{10}\left(\frac{\mathrm{CR}_{0.5}}{\mathrm{deg^2}}\right) \approx {} & -2 \log_{10} \varrho + 4.46, \\
\log_{10}\left(\frac{\mathrm{CR}_{0.9}}{\mathrm{deg^2}}\right) \approx {} & -2 \log_{10} \varrho + 5.06.
\end{align}
\end{subequations}
Sky-localization accuracy (at a given SNR) does not appear to be sensitive to the Gaussianity of the noise.

From our fits (\ref{eq:SNR-fit}), we can immediately see that the ratio $\mathrm{CR}_{0.9}/\mathrm{CR}_{0.5}$ is about $10^{0.6} \simeq 4$. Considering this ratio for each posterior, the mean value of $\log_{10}(\mathrm{CR}_{0.9}/\mathrm{CR}_{0.5})$ is $0.60$ and the standard deviation is $0.07$. For comparison, if the posterior were a $1$-d Gaussian, we would expect the ratio to be $\erf^{-1}(0.9)/\erf^{-1}(0.5) \simeq 2.4 \simeq 10^{0.39}$, and if it were a $2$-d Gaussian, the ratio would be $\ln(1-0.9)/\ln(1-0.5) \simeq 3.3 \simeq 10^{0.52}$ \citep{Fairhurst:2009tc,Fairhurst:2010is}. Neither of these agree well. The sky-location posteriors can have complicated shapes, and cannot be accurately modelled by a simple Gaussian description.

\begin{figure}
  \centering
   \subfigure[]{\includegraphics[width=0.9\columnwidth]{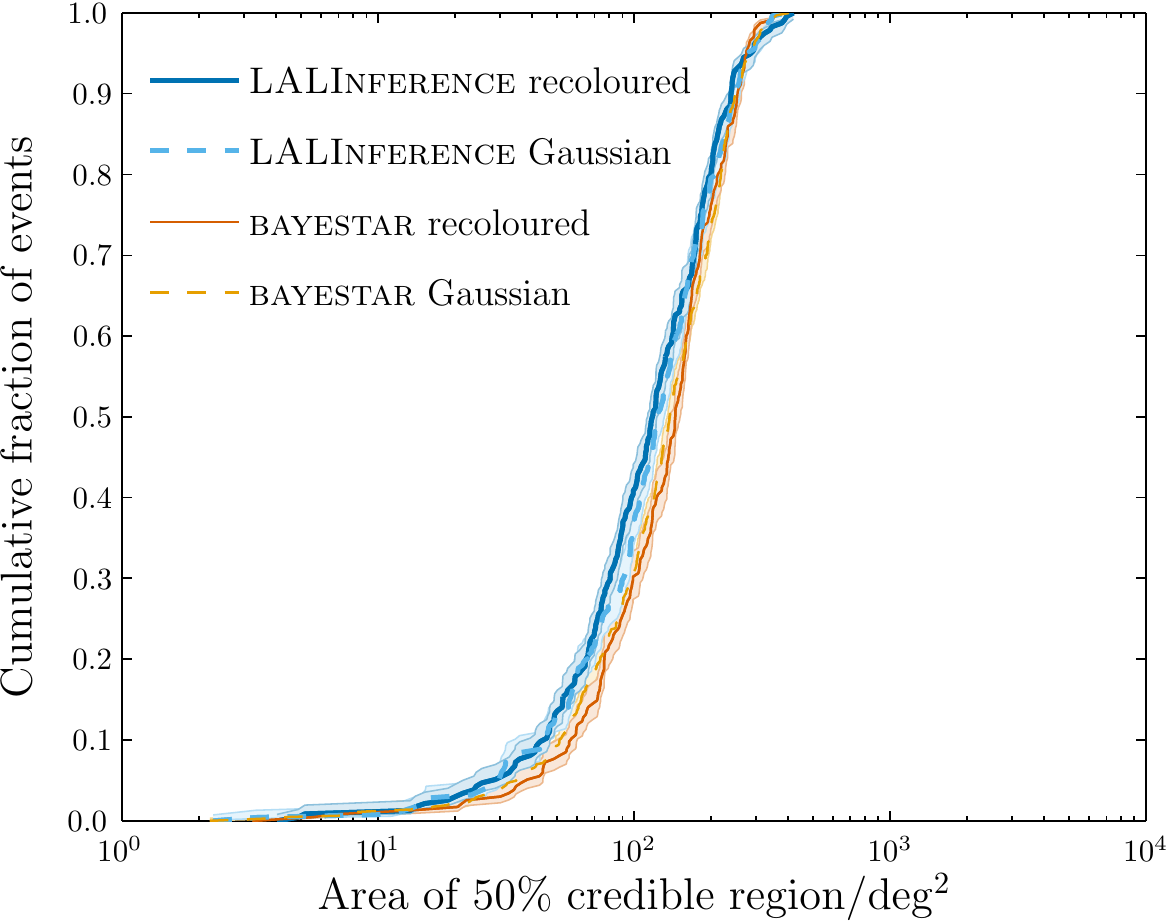}} \\
   \subfigure[]{\includegraphics[width=0.9\columnwidth]{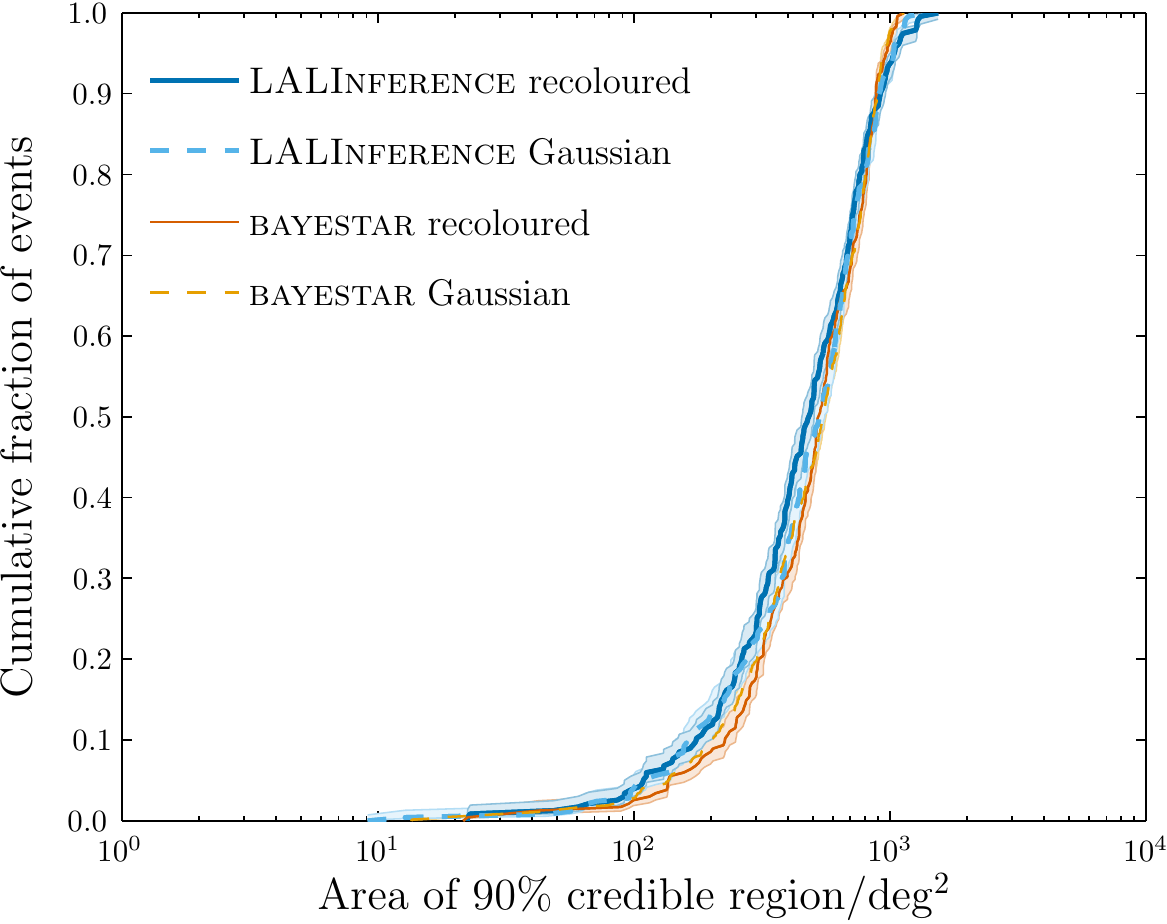}} \\
   \subfigure[]{\includegraphics[width=0.9\columnwidth]{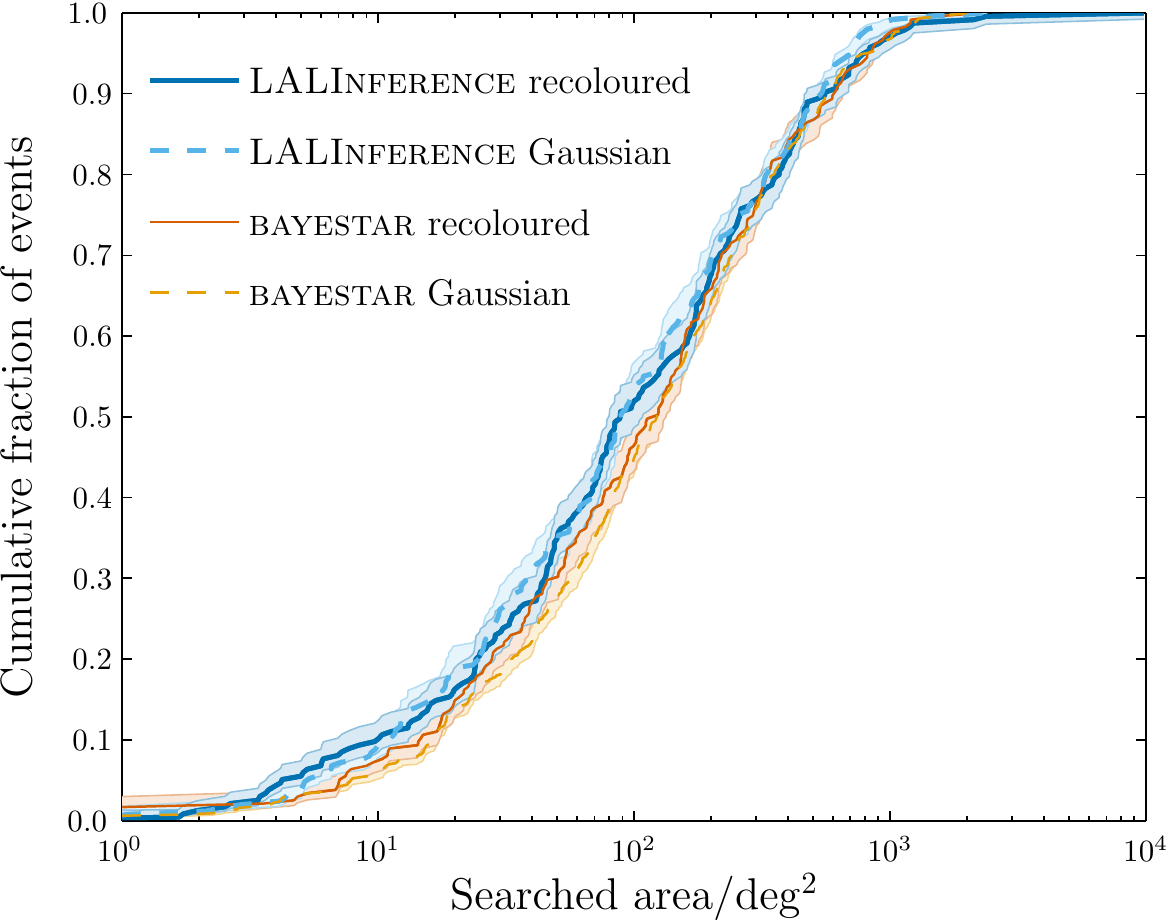}}
    \caption{Cumulative fractions of events with sky-localization areas smaller than the abscissa value as in figure \ref{fig:areas} but imposing an SNR cut of $\varrho_\mathrm{R} \geq 12$. (a) Sky area of $\mathrm{CR}_{0.5}$. (b) Sky area of $\mathrm{CR}_{0.9}$. (c) Searched area $A_\ast$. \textsc{LALInference} and \textsc{bayestar} results are denoted by thicker blue and thinner red--orange lines respectively. The results of this study are indicated by a solid line, while the results of \citet{Singer:2014qca}, which uses Gaussian noise, are indicated by a dashed line. The $68\%$ confidence intervals are denoted by the shaded areas.} 
    \label{fig:areas-cut}
\end{figure}

To verify that SNR distribution is the dominant cause of difference between the Gaussian and recoloured results, we impose a cut on the recoloured data set of $\varrho \geq 12$ to match the Gaussian set. This reduces the number of events from $333$ to $236$. The cumulative distribution of sky-localization areas for results with $\varrho \geq 12$ are shown in figure \ref{fig:areas-cut}.
The distributions do overlap as expected: the Gaussian and recoloured results are in agreement (a KS test on $\mathrm{CR}_{0.9}$ gives a $p$-value of $0.550$ when comparing \textsc{LALInference} results between noise realizations and $0.673$ for \textsc{bayestar}).

The key numbers describing the distributions are given in tables \ref{tab:fraction} and \ref{tab:median}; the former gives the fraction of events with sky-localization areas smaller than fiducial values, and the latter gives median sky-localization areas. Our results are discussed further in section \ref{sec:ob-scenario}.
\begin{deluxetable*}{crcccccc}
\tabletypesize{\footnotesize}
\tablecaption{Fractions of events with sky-localization areas smaller than a given size from this study using recoloured noise and \citet{Singer:2014qca}, which uses Gaussian noise. Results are quoted for the full catalogue of results with recoloured noise and imposing a SNR cut of $\varrho \geq 12$ to match the Gaussian catalogue. Figures for the $50\%$ credible region $\mathrm{CR}_{0.5}$, the $90\%$ credible region $\mathrm{CR}_{0.9}$ and the searched area $A_\ast$ are included. A dash (---) is used for fractions less than $0.01$.\label{tab:fraction}}
\tablewidth{0pt}
\tablehead{
\colhead{} & \colhead{} & \multicolumn{2}{c}{Gaussian noise} & \multicolumn{2}{c}{Recoloured noise} & \multicolumn{2}{c}{Recoloured noise $\varrho \geq 12$} \\
\colhead{} & \colhead{} & \multicolumn{1}{c}{\textsc{bayestar}} & \multicolumn{1}{c}{\textsc{LALInference}} & \multicolumn{1}{c}{\textsc{bayestar}} & \multicolumn{1}{c}{\textsc{LALInference}} & \multicolumn{1}{c}{\textsc{bayestar}} & \multicolumn{1}{c}{\textsc{LALInference}}
}
\startdata
\multirow{5}{*}{$\mathrm{CR}_{0.5} \leq$} & $5~\mathrm{deg^2}$ & \multicolumn{1}{c}{---} & \multicolumn{1}{c}{---} & \multicolumn{1}{c}{---} & \multicolumn{1}{c}{---} & \multicolumn{1}{c}{---} & \multicolumn{1}{c}{---} \\
 & $20~\mathrm{deg^2}$ & $0.02$ & $0.03$ & $0.01$ & $0.02$ & $0.02$ & $0.03$ \\
 & $100~\mathrm{deg^2}$ & $0.30$ & $0.37$ & $0.21$ & $0.30$ & $0.30$ & $0.41$ \\
 & $200~\mathrm{deg^2}$ & $0.74$ & $0.80$ & $0.58$ & $0.64$ & $0.76$ & $0.80$ \\
 & $500~\mathrm{deg^2}$ & $1.00$ & $1.00$ & $1.00$ & $0.99$ & $1.00$ & $1.00$ \\
 & $1000~\mathrm{deg^2}$ & $1.00$ & $1.00$ & $1.00$ & $1.00$ & $1.00$ & $1.00$ \\ \tableline
\multirow{5}{*}{$\mathrm{CR}_{0.9} \leq$ } & $5~\mathrm{deg^2}$ & \multicolumn{1}{c}{---} & \multicolumn{1}{c}{---} & \multicolumn{1}{c}{---} & \multicolumn{1}{c}{---} & \multicolumn{1}{c}{---} & \multicolumn{1}{c}{---} \\
 & $20~\mathrm{deg^2}$ & \multicolumn{1}{c}{---} & \multicolumn{1}{c}{---} & \multicolumn{1}{c}{---} & \multicolumn{1}{c}{---} & \multicolumn{1}{c}{---} & \multicolumn{1}{c}{---} \\
 & $100~\mathrm{deg^2}$ & $0.03$ & $0.04$ & $0.02$ & $0.03$ & $0.03$ & $0.04$ \\
 & $200~\mathrm{deg^2}$ & $0.10$ & $0.13$ & $0.06$ & $0.08$ & $0.09$ & $0.12$ \\
 & $500~\mathrm{deg^2}$ & $0.44$ & $0.48$ & $0.31$ & $0.38$ & $0.44$ & $0.52$ \\
 & $1000~\mathrm{deg^2}$ & $0.98$ & $0.93$ & $0.78$ & $0.80$ & $0.96$ & $0.94$ \\ \tableline
\multirow{5}{*}{$A_\ast \leq$} & $5~\mathrm{deg^2}$ & $0.03$ & $0.04$ & $0.03$ & $0.04$ & $0.03$ & $0.06$ \\
 & $20~\mathrm{deg^2}$ & $0.14$ & $0.19$ & $0.12$ & $0.14$ & $0.15$ & $0.16$ \\
 & $100~\mathrm{deg^2}$ & $0.45$ & $0.54$ & $0.40$ & $0.45$ & $0.47$ & $0.52$ \\
 & $200~\mathrm{deg^2}$ & $0.64$ & $0.70$ & $0.60$ & $0.60$ & $0.66$ & $0.68$ \\
 & $500~\mathrm{deg^2}$ & $0.87$ & $0.89$ & $0.82$ & $0.83$ & $0.87$ & $0.89$ \\
 & $1000~\mathrm{deg^2}$ & $0.97$ & $0.99$ & $0.96$ & $0.95$ & $0.98$ & $0.97$
\enddata
\end{deluxetable*}
\begin{deluxetable*}{cccccccc}
\tabletypesize{\footnotesize}
\tablecaption{Median sky-localization areas from this study using recoloured noise, and \citet{Singer:2014qca}, which uses Gaussian noise. Results are quoted for the full catalogue of results with recoloured noise and imposing a SNR cut of $\varrho \geq 12$ to match the Gaussian catalogue. Figures for the $50\%$ credible region $\mathrm{CR}_{0.5}$, the $90\%$ credible region $\mathrm{CR}_{0.9}$ and the searched area $A_\ast$ are included.\label{tab:median}}
\tablewidth{0pt}
\tablehead{
\colhead{} & \colhead{} & \multicolumn{2}{c}{Gaussian noise} & \multicolumn{2}{c}{Recoloured noise} & \multicolumn{2}{c}{Recoloured noise $\varrho \geq 12$} \\
 &  & \multicolumn{1}{c}{\textsc{bayestar}} & \multicolumn{1}{c}{\textsc{LALInference}} & \multicolumn{1}{c}{\textsc{bayestar}} & \multicolumn{1}{c}{\textsc{LALInference}} & \multicolumn{1}{c}{\textsc{bayestar}} & \multicolumn{1}{c}{\textsc{LALInference}}
}
\startdata
\multirow{3}{*}{Median} & \multicolumn{1}{c}{$\mathrm{CR}_{0.5}$} & $138~\mathrm{deg^2}$ & $124~\mathrm{deg^2}$ & $175~\mathrm{deg^2}$ & $154~\mathrm{deg^2}$ & $145~\mathrm{deg^2}$ & $118~\mathrm{deg^2}$ \\
 & \multicolumn{1}{c}{$\mathrm{CR}_{0.9}$} & $545~\mathrm{deg^2}$ & $529~\mathrm{deg^2}$ & $692~\mathrm{deg^2}$ & $632~\mathrm{deg^2}$ & $524~\mathrm{deg^2}$ & $481~\mathrm{deg^2}$ \\
 & \multicolumn{1}{c}{$A_\ast$} & $123~\mathrm{deg^2}$ & $\hphantom{0}88~\mathrm{deg^2}$ & $145~\mathrm{deg^2}$ & $132~\mathrm{deg^2}$ & $118~\mathrm{deg^2}$ & $\hphantom{0}88~\mathrm{deg^2}$ \enddata
\end{deluxetable*}

\subsection{Mass and distance estimation}\label{sec:PE}

Independent of any EM counterpart, GW astronomy is still informative. GW observations allow for measurement of various properties of the source system. Here, we examine the ability to measure luminosity distance and mass (principally the chirp mass of the system).

Accurate mass and distance measurements have many physical applications. Measurement of the chirp-mass distribution can constrain binary evolution models \citep{Bulik:2003ap}. Determining the maximum mass of a neutron star would shed light on its equation of state \citep[e.g.,][]{Read:2009yp}, and, potentially, on the existence of a mass gap between neutron stars and black holes \citep{Ozel:2010,Farr:2010,Kreidberg:2012ud}. Combining mass and distance measurement, it may be possible to construct a new (independent) measure of the Hubble constant \citep{Taylor:2011fs}. GW observations shall give us unique insight into the properties of BNS systems.

In addition to component masses and the distance to the source, the component spins are of astrophysical importance \citep[e.g.,][]{Mandel:2009nx}. Unfortunately, we cannot estimate the component spins as we are using non-spinning waveform templates. Measurement of the spins will be examined in a future study investigating PE using SpinTaylorT4 waveforms.

\subsubsection{Luminosity distance}

Quantifying the precision of distance estimation is simpler than for sky localization as we are now working in a single dimension. The equivalent of a credible region is a credible interval. We denote the distance credible interval for a total posterior probability $p$ as $\mathrm{CI}^{D}_p$. It is defined to exclude equal posterior probabilities in each of the tails; it is given by
\begin{equation}
\mathrm{CI}^{D}_p = C^{-1}_{D}\left(\frac{1+p}{2}\right) - C^{-1}_{D}\left(\frac{1-p}{2}\right),
\label{eq:CI}
\end{equation}
where $C^{-1}_{D}(p)$ is the inverse of the cumulative distribution function
\begin{equation}
C_{D}(D) = \int_0^{D} \mathrm{d}D' P_D(D')
\end{equation}
for distance posterior $P_D(D)$. The same symmetric definition for the credible interval was used by \citet{Aasi:2013jjl}. A smaller $\mathrm{CI}^{D}_p$ for a given $p$ indicates more precise distance estimation.

\begin{figure}
  \centering
   \includegraphics[width=0.7\columnwidth]{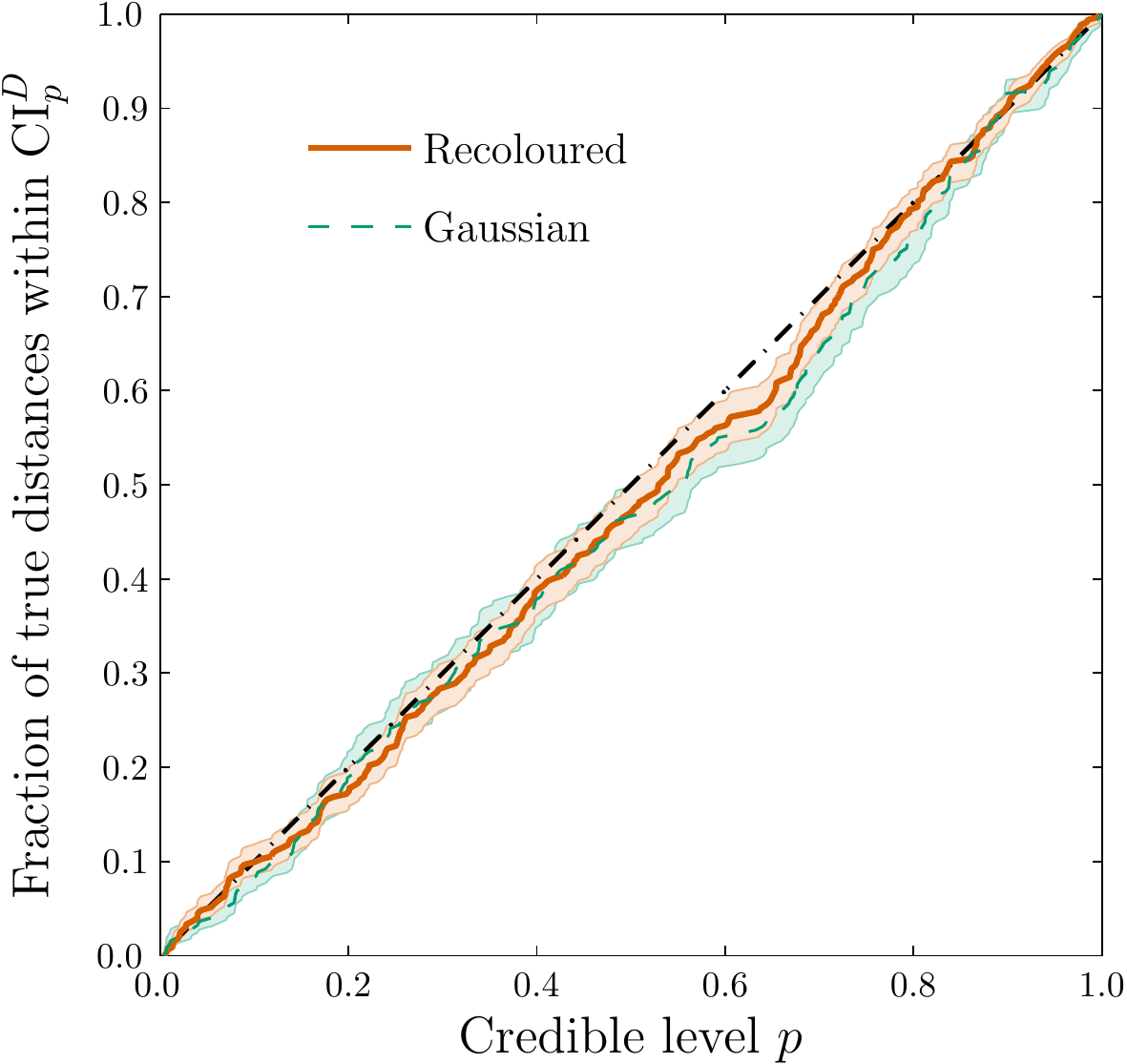}
    \caption{Fraction of true luminosity distances found within a credible interval as a function of encompassed posterior probability. Results using recoloured noise are indicated by a solid line, while the results using Gaussian noise \citep{Singer:2014qca} are indicated by a dashed line. The expected distribution is indicated by the dot--dashed diagonal line. The shaded regions enclose the $68\%$ confidence intervals accounting for sampling errors.} 
    \label{fig:pp-D}
\end{figure}

The self-consistency of our distance estimates can be verified by calculating the fraction of true values that fall within the credible interval at a given $p$. This is shown in figure \ref{fig:pp-D} for results from both the Gaussian and recoloured noise results.
Both distributions are consistent with expectations (performing a KS test with the predicted distribution yields $p$-values of $0.168$ and $0.057$ for the recoloured and Gaussian noise respectively). \textsc{LALInference} does return self-consistent distance estimates.

\begin{figure*}
  \centering
   \subfigure[]{\includegraphics[width=0.9\columnwidth]{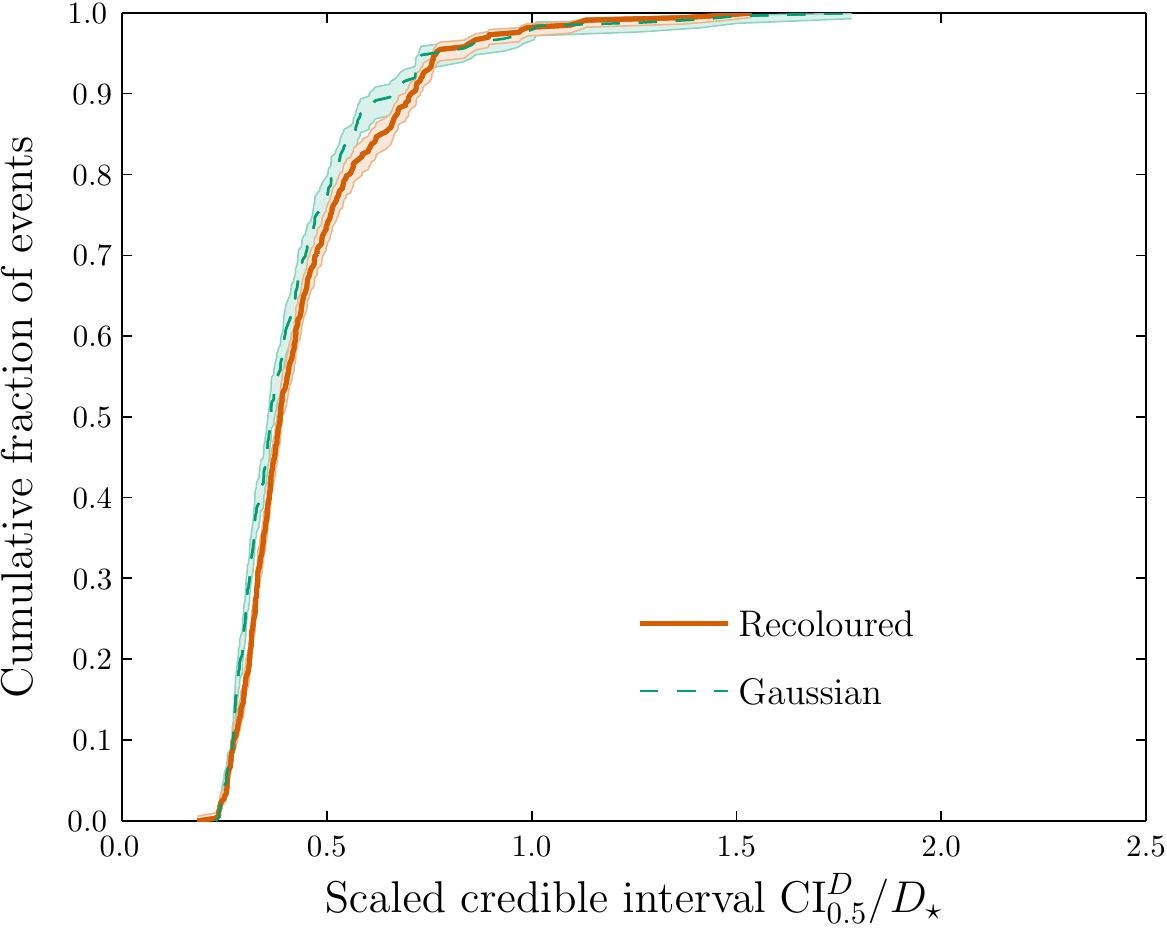}} \quad
   \subfigure[]{\includegraphics[width=0.9\columnwidth]{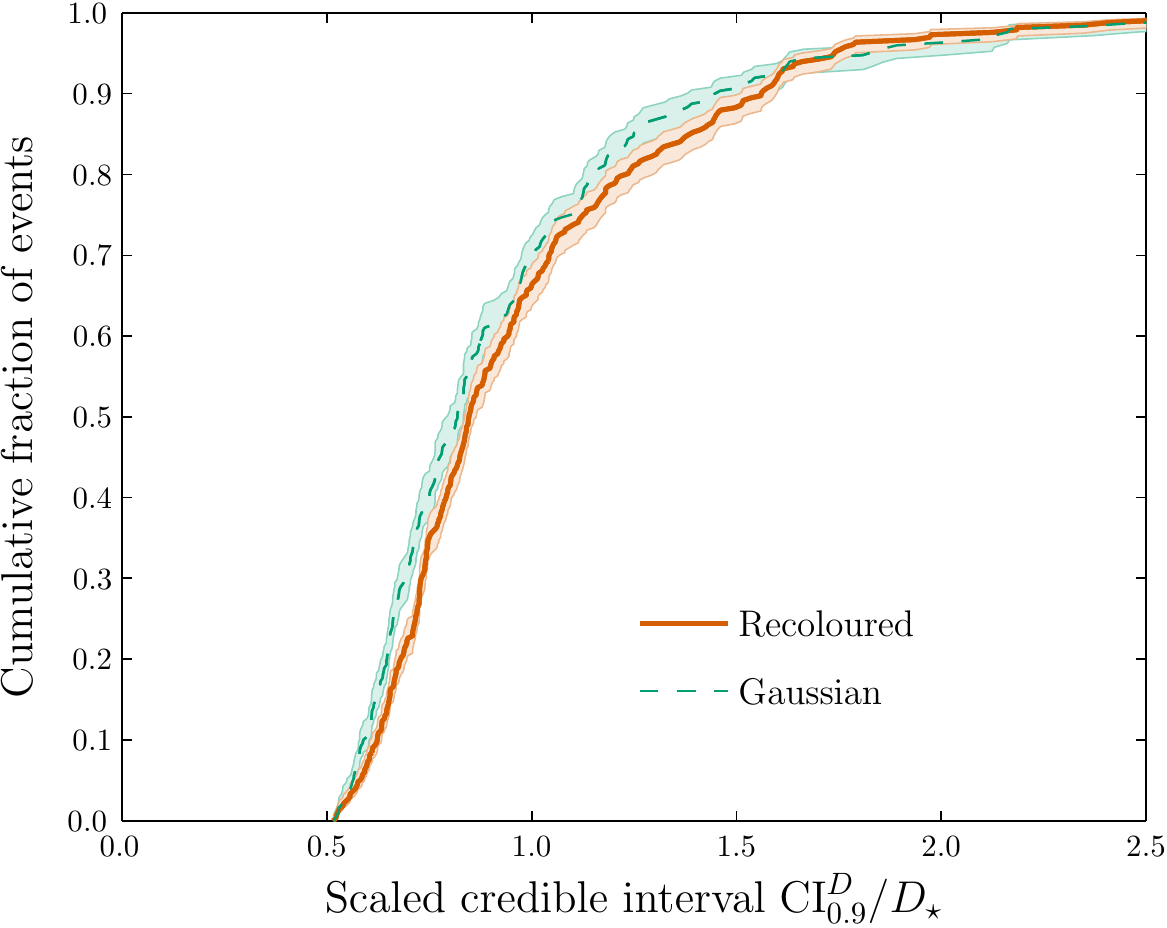}}
    \caption{Cumulative fractions of events with luminosity-distance credible intervals (divided by the true distance) smaller than the abscissa value. (a) Scaled $50\%$ credible interval $\mathrm{CI}^{D}_{0.5}/D_\star$. (b) Scaled $90\%$ interval $\mathrm{CI}^{D}_{0.9}/D_\star$. Results using recoloured noise are indicated by a solid line and the results using Gaussian noise \citep{Singer:2014qca} are indicated by a dashed line. The $68\%$ confidence intervals are denoted by the shaded areas.} 
    \label{fig:dist-CI}
\end{figure*}

The cumulative distributions of credible intervals are plotted in figure \ref{fig:dist-CI}. We divide the credible interval by the true (injected) distance $D_\star$; this gives an approximate analogue of twice the fractional uncertainty.
The \edit{quantity} $\mathrm{CI}^{D}_p/D_\star$ appears insensitive to the detection cut-off (a KS test between $\mathrm{CI}^{D}_{0.9}/D_\star$ for the recoloured and Gaussian results gives a $p$-value of $0.077$). This appears in contrast to the case for sky areas, but the differing SNR distributions are accounted for by scaling with respect to the distance (which is inversely proportional to the SNR). The estimation of the distance, like that for sky areas, does not depend upon the character of the noise.

Distance estimation is imprecise: the posterior widths are frequently comparable the the magnitude of the distance itself. This is a consequence of a degeneracy between the distance and the inclination \citep{Cutler:1994ys,Aasi:2013jjl}. The key numbers summarising distance estimation are given in tables \ref{tab:fraction-dist} and \ref{tab:median-dist}; the former gives the fraction of events with $\mathrm{CI}^{D}_p/D_\star$ smaller than fiducial values, and the latter gives median values.

\begin{deluxetable}{cccc}
\tabletypesize{\footnotesize}
\tablecaption{Fractions of events with fractional distance estimate uncertainties smaller than a given size. Results using recoloured noise and Gaussian noise are included \citep{Singer:2014qca}. Figures for the $50\%$ credible interval $\mathrm{CI}^{D}_{0.5}$ and the $90\%$ credible interval $\mathrm{CI}^{D}_{0.9}$ are included, both are scaled with respect to the true distance $D_\star$. A dash (---) is used for fractions less than $0.01$.\label{tab:fraction-dist}}
\tablewidth{0pt}
\tablehead{
\colhead{} & \colhead{} & \colhead{Gaussian noise} & \colhead{Recoloured noise}
}
\startdata
\multirow{5}{*}{$\displaystyle \frac{\mathrm{CI}^{D}_{0.5}}{D_\star} \leq$} & $0.25$ & $0.04$ & $0.03$ \\
 & $0.50$ & $0.77$ & $0.74$ \\
 & $0.75$ & $0.95$ & $0.93$ \\
 & $1.00$ & $0.98$ & $0.98$ \\
 & $2.00$ & $1.00$ & $1.00$ \\ \tableline
\multirow{5}{*}{$\displaystyle \frac{\mathrm{CI}^{D}_{0.9}}{D_\star} \leq$ } & $0.25$ & \multicolumn{1}{c}{---} & \multicolumn{1}{c}{---} \\
 & $0.50$ & \multicolumn{1}{c}{---} & \multicolumn{1}{c}{---} \\
 & $0.75$ & $0.40$ & $0.35$ \\
 & $1.00$ & $0.70$ & $0.66$ \\
 & $2.00$ & $0.96$ & $0.97$
\enddata
\end{deluxetable}
\begin{deluxetable}{cccc}
\tabletypesize{\footnotesize}
\tablecaption{Median distance credible intervals (divided by the true distance) using recoloured noise and Gaussian noise \citep{Singer:2014qca}. Figures for the $50\%$ credible interval $\mathrm{CI}^{D}_{0.5}$ and the $90\%$ credible interval $\mathrm{CI}^{D}_{0.9}$ are included.\label{tab:median-dist}}
\tablewidth{0pt}
\tablehead{
\colhead{} & \colhead{} & \colhead{Gaussian noise} & \colhead{Recoloured noise}
}
\startdata
\multirow{2}{*}{Median} & \multicolumn{1}{c}{$\mathrm{CI}^{D}_{0.5}/D_\star$} & $0.36$ & $0.38$ \\
 & \multicolumn{1}{c}{$\mathrm{CI}^D_{0.9}/D_\star$} & $0.82$ & $0.85$ \enddata
\end{deluxetable}

\subsubsection{Chirp mass}

The chirp mass should be precisely measured as it determines the GW phase evolution. We again use the credible interval to quantify measurement precision; the chirp-mass credible interval $\mathrm{CI}^{\mathcal{M}_c}_p$ is defined equivalently to its distance counterpart in (\ref{eq:CI}).

\begin{figure}
  \centering
   \includegraphics[width=0.7\columnwidth]{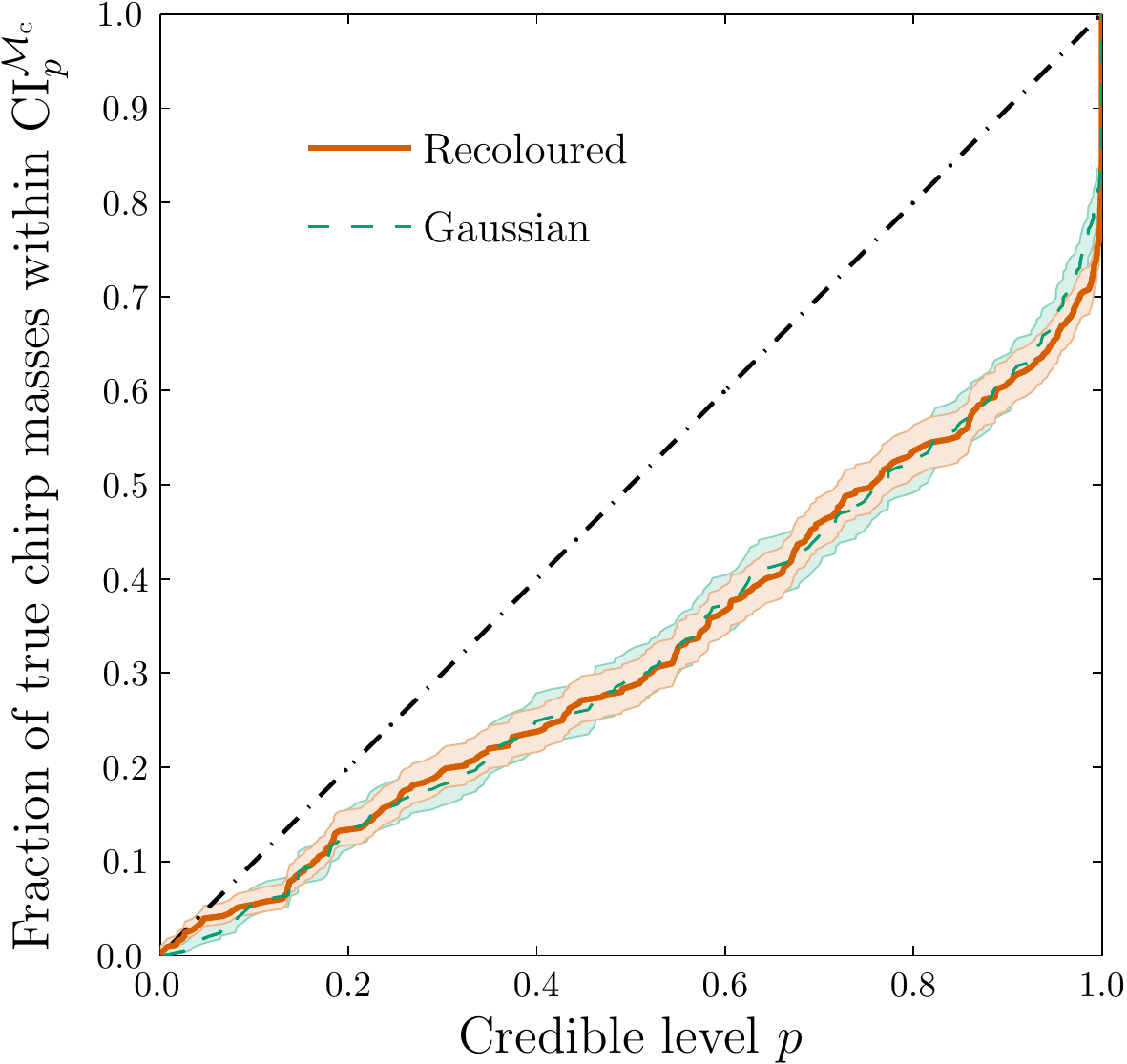}
    \caption{Fraction of true source chirp masses found within a credible interval as a function of encompassed posterior probability. Results using recoloured noise are indicated by a solid line, while the results using Gaussian noise \citep{Singer:2014qca} are indicated by a dashed line. The expected distribution is indicated by the dot--dashed diagonal line. The shaded regions enclose the $68\%$ confidence intervals accounting for sampling errors.} 
    \label{fig:pp-Mc}
\end{figure}

The fraction of true chirp masses that fall within $\mathrm{CI}^{\mathcal{M}_c}_p$ at a given $p$ is plotted in figure \ref{fig:pp-Mc}.
Neither the results calculated using Gaussian noise nor those using recoloured noise fit our expectations: the posteriors are not well calibrated. However, the two sets of results are entirely consistent with each other (a KS test between the two gives a $p$-value of $0.524$), indicating that the PE is not affected by the noise. There appears to be a systematic error in our posterior distributions of the chirp mass.

The discrepancies between our posterior estimates for the chirp masses and their true values are a consequence of our use of non-spinning TaylorF2 waveform templates. This has two effects. First, by using a non-spinning waveform, we do not explore the degeneracy between mass and spin \citep{Cutler:1994ys,vanderSluys:2007st,Baird:2012cu}. This results in an artificially narrow marginalized posterior for mass parameters such as the chirp mass. In effect, we are pinning the spin to be zero, which is information we should not have a priori. Second, we have used a template that does not exactly match the injected waveform (SpinTaylorT4). The small difference in approximants results in a mismatch in estimated parameters \citep{Buonanno:2009zt,Aasi:2013jjl}. Since the posterior on the chirp mass is narrow, because it is intrinsically well-measured and because we have not included degeneracy with spin, even a small difference in templates is sufficient to offset the posterior from the true chirp mass by a statistically significant amount. 

\begin{figure}
  \centering
   \includegraphics[width=0.9\columnwidth]{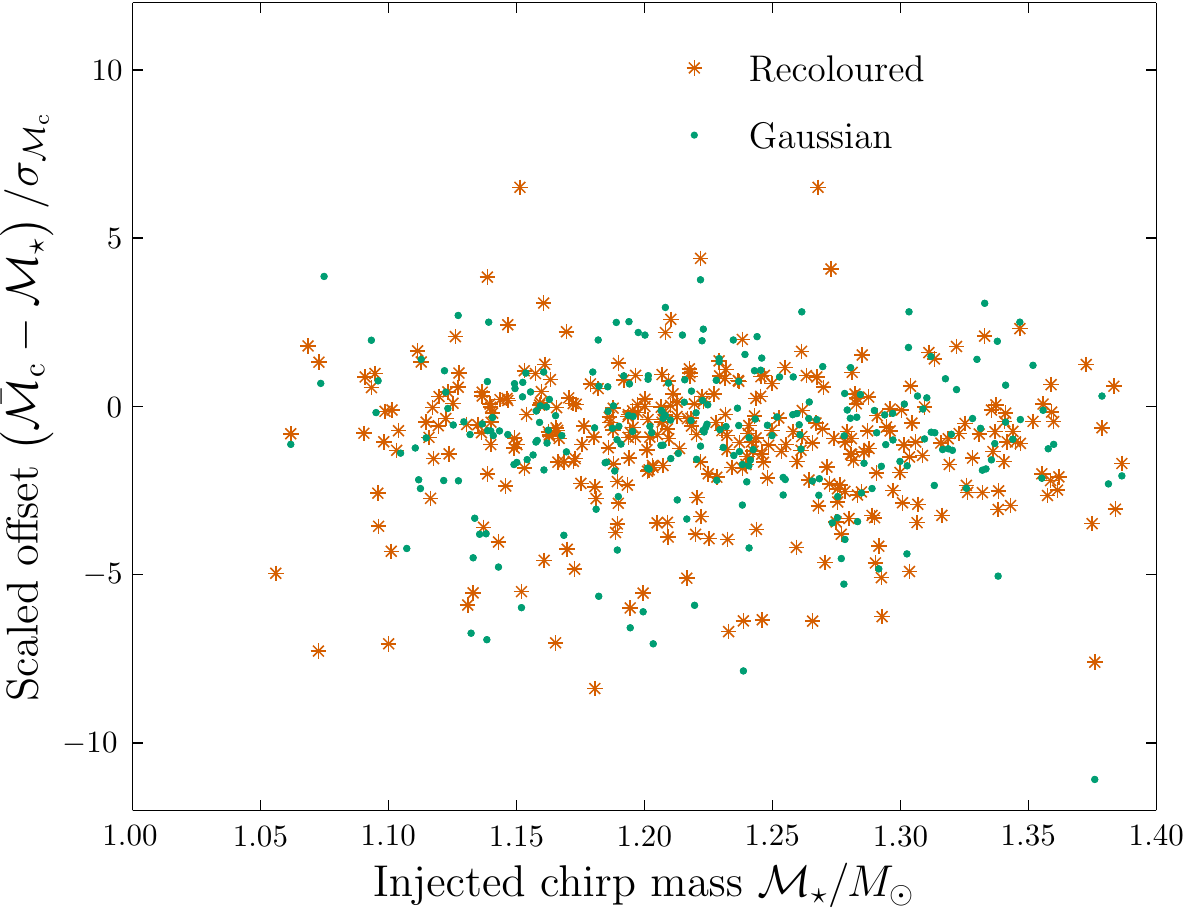}
    \caption{Offset between the posterior mean estimate for the chirp mass $\bar{\mathcal{M}_\mathrm{c}}$ and the true (injected) value $\mathcal{M}_\star$ divided by the standard deviation of the posterior distribution $\sigma_{\mathcal{M}_\mathrm{c}}$. The round (green) points are for the results using Gaussian noise \citep{Singer:2014qca} and the star-shaped (red) points are for results using recoloured noise.} 
    \label{fig:Mc-scatter}
\end{figure}

To examine the offset between the estimated and true chirp masses, we plot in figure \ref{fig:Mc-scatter} the difference between the posterior mean $\bar{\mathcal{M}_\mathrm{c}}$ and the true value $\mathcal{M}_\star$ divided by the standard deviation of the posterior $\sigma_{\mathcal{M}_\mathrm{c}}$.
Using the median in place of the mean, or $\mathrm{CI}^{\mathcal{M}_\mathrm{c}}_{0.68}/2$ in place of $\sigma_{\mathcal{M}_\mathrm{c}}$, gives only a small quantitative difference. Over this narrow mass range, the offset is not a strong function of the chirp mass. The offset is a combination of both error introduced by the presence of noise and theoretical error from the mismatch between the injected waveform and template waveforms \citep{Cutler:2007mi}. If only the former were significant, we would expect the mean offset to be zero, and the typical scatter of offsets to be of order of the posterior's standard deviation. Neither of these is the case. The average scaled offset $(\bar{\mathcal{M}_\mathrm{c}} - \mathcal{M}_\star)/\sigma_{\mathcal{M}_\mathrm{c}}$ across the recoloured (Gaussian) data set is $-1.3 \pm 0.1$ ($-0.9 \pm 0.1$). This shows that there is a systematic error. However, it is not as simple as just systematically underestimating the chirp mass; there is a large scatter in the offsets, the standard deviation of the scaled offset for the recoloured (Gaussian) data set is $2.07 \pm 0.08$ ($2.09 \pm 0.09$). This is consistent with our expectation that the mass--spin degeneracy should broaden the posterior; these results imply that the posterior should be a factor of $\sim 2$ wider \citep[cf.][]{Poisson:1995ef}.

\begin{figure}
  \centering
   \subfigure[\label{fig:Mc-50}]{\includegraphics[width=0.9\columnwidth]{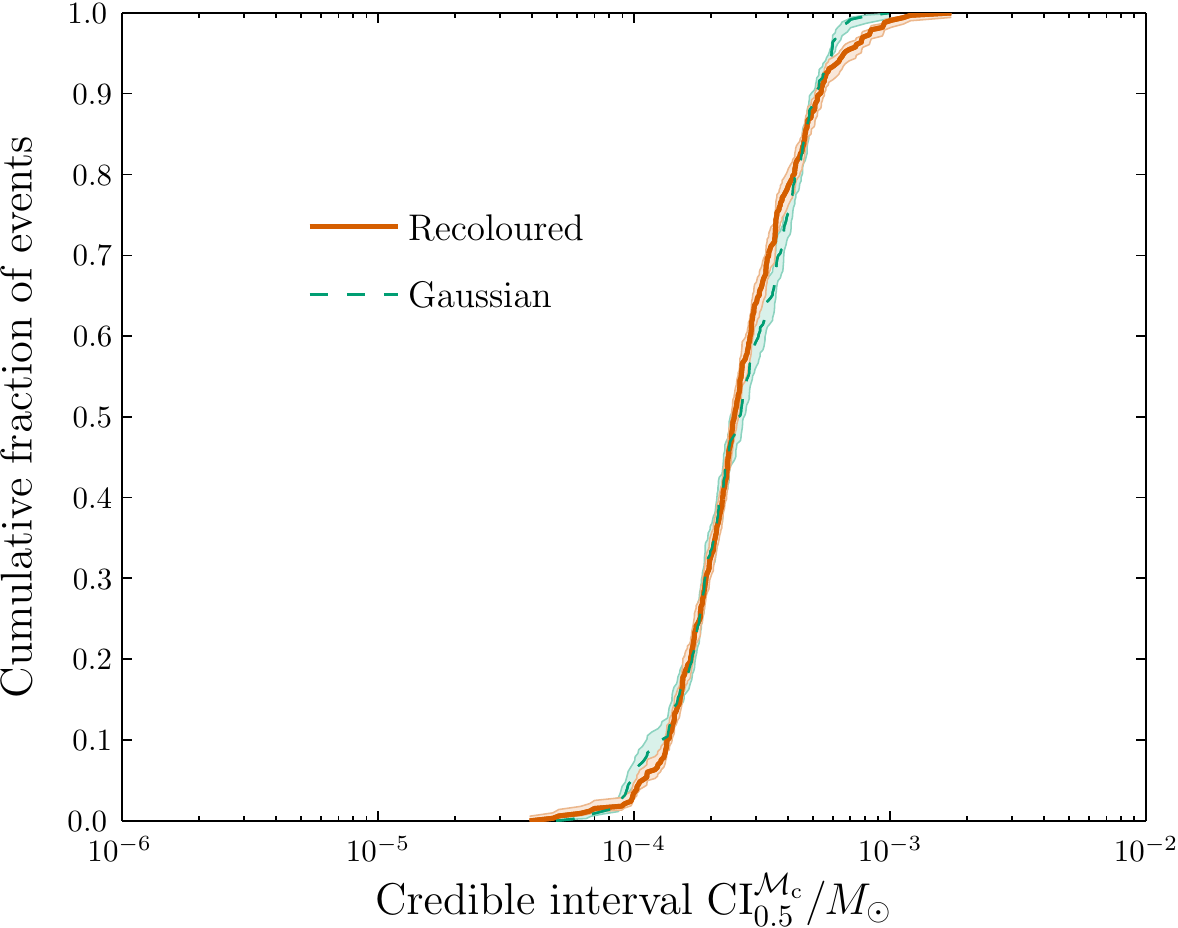}} \\
   \subfigure[\label{fig:Mc-90}]{\includegraphics[width=0.9\columnwidth]{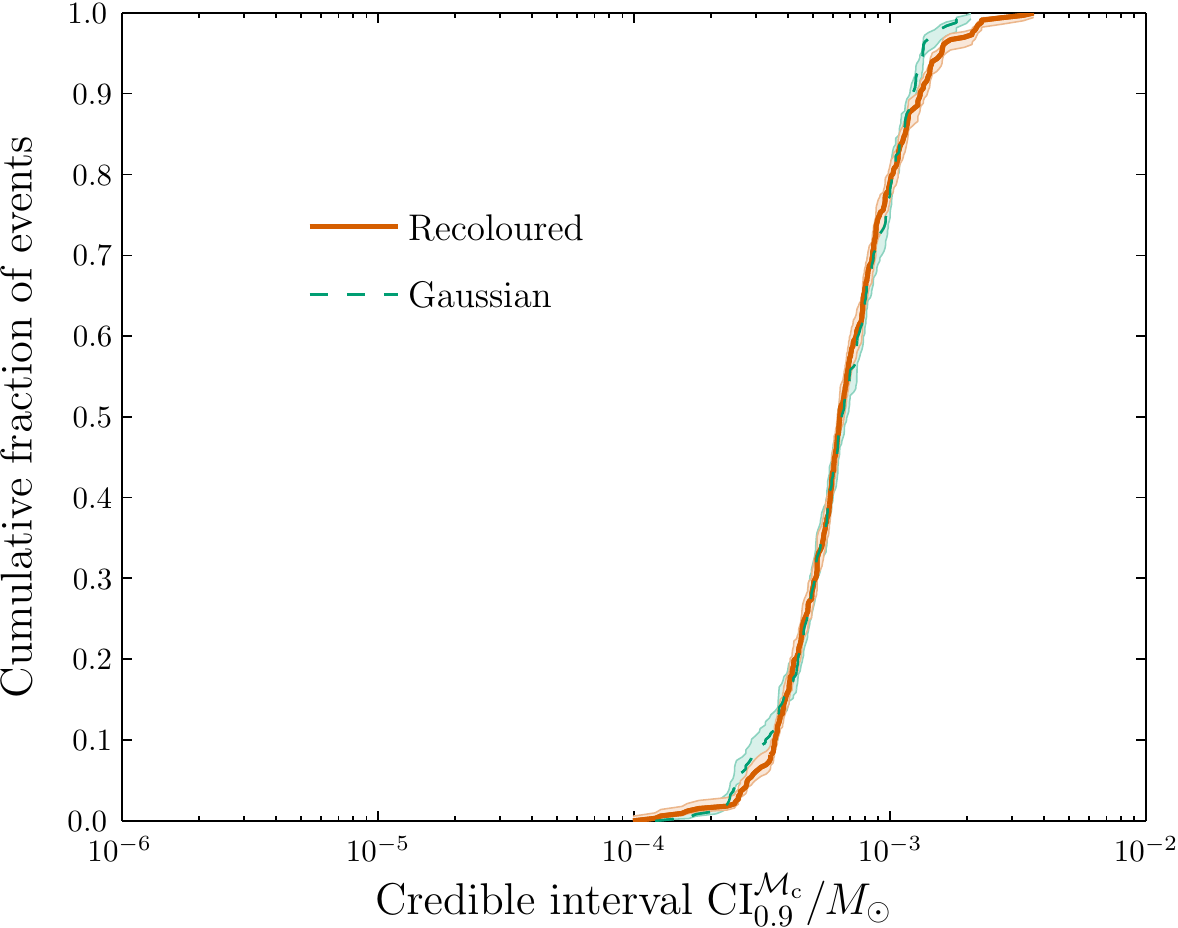}} \\
   \subfigure[\label{fig:Mc-abs}]{\includegraphics[width=0.9\columnwidth]{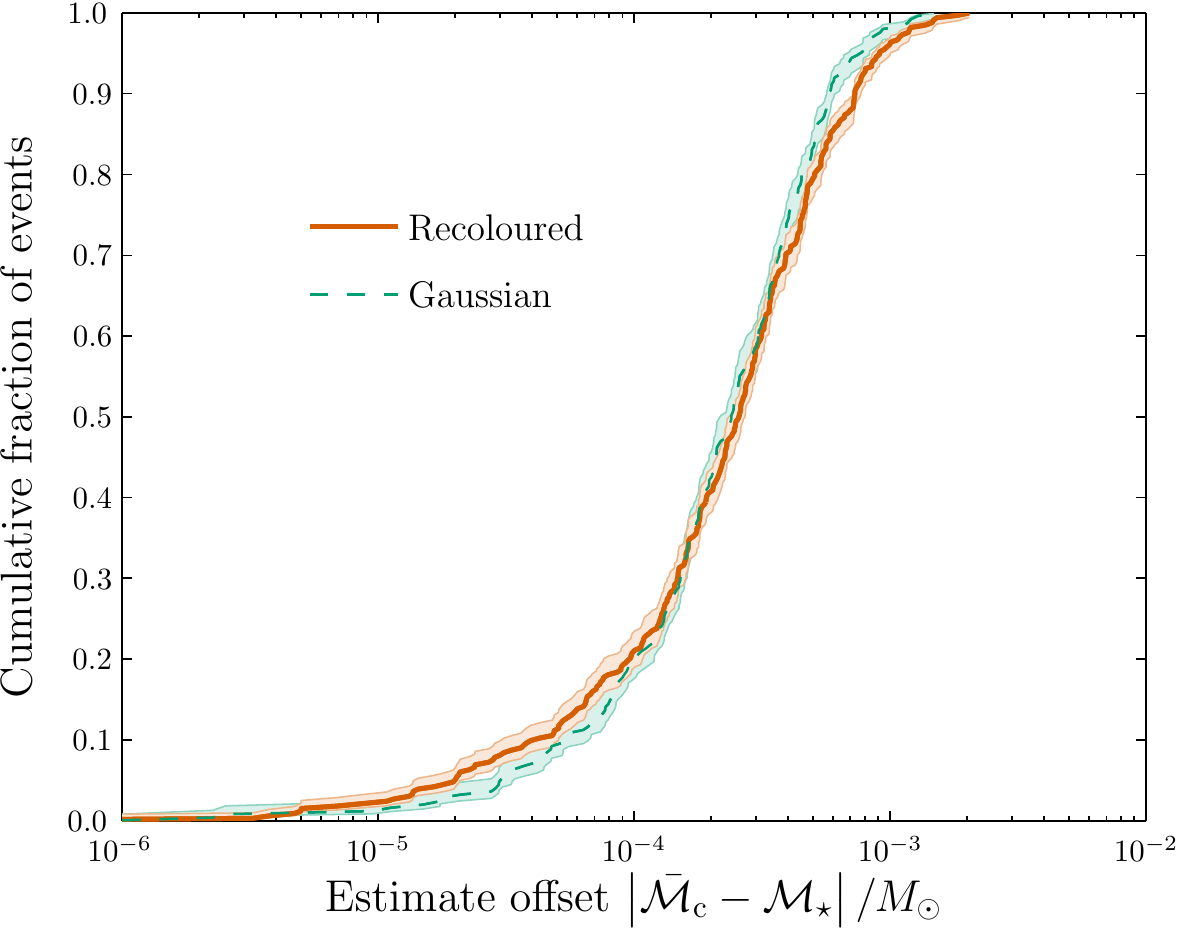}}
    \caption{Cumulative fractions of events with (a) $50\%$ chirp-mass credible interval, (b) $90\%$ credible interval, and (c) offsets between the posterior mean and true chirp mass smaller than the abscissa value. Results using recoloured noise are indicated by a solid line and the results using Gaussian noise \citep{Singer:2014qca} are indicated by a dashed line. The $68\%$ confidence intervals are denoted by the shaded areas.} 
    \label{fig:Mc-errors}
\end{figure}

While the theoretical error is important in determining the accuracy to which we can infer the chirp mass, it does not completely dominate the noise error. To illustrate the scale of the errors, we plot distribution of the $50\%$ and $90\%$ credible intervals in figures \ref{fig:Mc-50} and \ref{fig:Mc-90}, and the absolute magnitudes of the offsets in figure \ref{fig:Mc-abs}.
For a well calibrated posterior, we would expect the offset to be smaller than $\mathrm{CI}^{\mathcal{M}_\mathrm{c}}_{0.9}/2$ ($\mathrm{CI}^{\mathcal{M}_\mathrm{c}}_{0.5}/2$) in approximately $90\%$ ($50\%$) of events. Figure \ref{fig:pp-Mc} shows that this is not the case, that we do have systematic error. Figure \ref{fig:Mc-scatter} confirms this and shows that the theoretical error is of a comparable size to the noise error. In figure \ref{fig:Mc-errors}, we see that the presence of theoretical error does not radically affect the distribution of offsets. The median value of the offsets are $(2.6 \times 10^{-4})M_\odot$ and $(2.4 \times 10^{-4})M_\odot$, and the median values of $\mathrm{CI}^{\mathcal{M}_\mathrm{c}}_{0.5}/2$ are $(1.2 \times 10^{-4})M_\odot$ and $(1.3 \times 10^{-4})M_\odot$ for the recoloured and Gaussian data sets respectively; the theoretical error approximately doubles the total uncertainty on the chirp mass. The key numbers summarising the distributions are given in tables \ref{tab:fraction-Mc} and \ref{tab:median-Mc}, which give the fraction of events with uncertainties smaller than fiducial values and the median uncertainties respectively.

Furthermore, figure \ref{fig:Mc-errors} shows that the (in)ability to measure the chirp mass is not significantly influenced by the character of the noise or the detection threshold used (a KS test comparing the $\mathrm{CI}^{\mathcal{M}_\mathrm{c}}_{0.9}$ and $|\bar{\mathcal{M}_\mathrm{c}} - \mathcal{M}_\star|$ distributions between the Gaussian and recoloured data sets gives $p$-values of $0.805$ and $0.507$ respectively). The latter is a consequence of both thresholds recovering equivalent chirp-mass distributions (figure \ref{fig:M-chirp}).    

It should be possible to incorporate knowledge of theoretical waveform error into PE by marginalizing out the uncertainty. This can be done using parametric models for the uncertainty if a specific form of the waveform error is suspected, or non-parametrically if we wish to be agnostic. The effect of folding in this additional uncertainty is to broaden the posteriors and possibly shift their means; doing so should make posterior estimates consistent with the true values. 

While we cannot correctly reconstruct the posterior distribution for the chirp mass, the error in the estimate is still small. We can measure the chirp mass accurately, even though we are affected by systematic error.

\begin{deluxetable}{crcc}
\tabletypesize{\footnotesize}
\tablecaption{Fractions of events with chirp-mass estimate errors smaller than a given value. Results using recoloured noise and Gaussian noise are included \citep{Singer:2014qca}. Included are figures for the $50\%$ credible interval $\mathrm{CI}^{\mathcal{M}_\mathrm{c}}_{0.5}$ and the $90\%$ credible interval $\mathrm{CI}^{\mathcal{M}_\mathrm{c}}_{0.9}$, which only include statistical error from the noise, and for the posterior mean offset relative to the true chirp mass $|\bar{\mathcal{M}_\mathrm{c}} - \mathcal{M}_\star|$, which includes both noise error and theoretical error. A dash (---) is used for fractions less than $0.01$.\label{tab:fraction-Mc}}
\tablewidth{0pt}
\tablehead{
\colhead{} & \colhead{} & \colhead{Gaussian noise} & \colhead{Recoloured noise}
}
\startdata
\multirow{6}{*}{$\mathrm{CI}^{\mathcal{M}_\mathrm{c}}_{0.5} \leq$} & $(5 \times 10^{-5})M_\odot$ & \multicolumn{1}{c}{---} & \multicolumn{1}{c}{---} \\
 & $(1 \times 10^{-4})M_\odot$ & $0.05$ & $0.03$ \\
 & $(2 \times 10^{-4})M_\odot$ & $0.34$ & $0.33$ \\
 & $(5 \times 10^{-4})M_\odot$ & $0.89$ & $0.88$ \\
 & $(1 \times 10^{-3})M_\odot$ & $1.00$ & $0.99$ \\
 & $(2 \times 10^{-3})M_\odot$ & $1.00$ & $1.00$ \\ \tableline
\multirow{6}{*}{$\mathrm{CI}^{\mathcal{M}_\mathrm{c}}_{0.9} \leq$} & $(5 \times 10^{-5})M_\odot$ & \multicolumn{1}{c}{---} & \multicolumn{1}{c}{---} \\
 & $(1 \times 10^{-4})M_\odot$ & \multicolumn{1}{c}{---} & \multicolumn{1}{c}{---} \\
 & $(2 \times 10^{-4})M_\odot$ & $0.01$ & $0.02$ \\
 & $(5 \times 10^{-4})M_\odot$ & $0.29$ & $0.29$ \\
 & $(1 \times 10^{-3})M_\odot$ & $0.77$ & $0.79$ \\
 & $(2 \times 10^{-3})M_\odot$ & $1.00$ & $0.97$ \\ \tableline
\multirow{6}{*}{$\left|\bar{\mathcal{M}_\mathrm{c}} - \mathcal{M}_\star\right| \leq$} & $(5 \times 10^{-5})M_\odot$ & $0.09$ & $0.11$ \\
 & $(1 \times 10^{-4})M_\odot$ & $0.20$ & $0.21$ \\
 & $(2 \times 10^{-4})M_\odot$ & $0.42$ & $0.41$ \\
 & $(5 \times 10^{-4})M_\odot$ & $0.83$ & $0.79$ \\
 & $(1 \times 10^{-3})M_\odot$ & $0.98$ & $0.96$ \\
 & $(2 \times 10^{-3})M_\odot$ & $1.00$ & $1.00$
\enddata
\end{deluxetable}
\begin{deluxetable}{cccc}
\tabletypesize{\footnotesize}
\tablecaption{Median chirp mass credible intervals and posterior estimate offset using recoloured noise and Gaussian noise \citep{Singer:2014qca}. Included are figures for the $50\%$ credible interval $\mathrm{CI}^{\mathcal{M}_\mathrm{c}}_{0.5}$ and the $90\%$ credible interval $\mathrm{CI}^{\mathcal{M}_\mathrm{c}}_{0.9}$, and the posterior mean offset relative to the true value $|\bar{\mathcal{M}_\mathrm{c}} - \mathcal{M}_\star|$.\label{tab:median-Mc}}
\tablewidth{0pt}
\tablehead{
\colhead{} & \colhead{} & \colhead{Gaussian noise} & \colhead{Recoloured noise}
}
\startdata
\multirow{3}{*}{Median} & \multicolumn{1}{c}{$\mathrm{CI}^{\mathcal{M}_\mathrm{c}}_{0.5}$} & $(2.6 \times 10^{-4})M_\odot$ & $(2.5 \times 10^{-4})M_\odot$ \\
 & \multicolumn{1}{c}{$\mathrm{CI}^{\mathcal{M}_\mathrm{c}}_{0.9}$} & $(6.4 \times 10^{-4})M_\odot$ & $(6.4 \times 10^{-4})M_\odot$ \\
 & \multicolumn{1}{c}{$\left|\bar{\mathcal{M}_\mathrm{c}} - \mathcal{M}_\star\right|$} & $(2.4 \times 10^{-4})M_\odot$ & $(2.6 \times 10^{-4})M_\odot$ \enddata
\end{deluxetable}

\subsubsection{Component masses}

The chirp mass is a combination of the component masses; in some cases it can be used to infer whether the source is a BNS or a binary black-hole system \citep{Hannam:2013uu,Vitale:2013bma}, but the component masses are of greater interest. The mass--spin degeneracy affects our ability to construct accurate estimates for the individual masses. Since we have already seen a systematic error in the chirp mass, we expect an analogous (larger) phenomenon here.

\begin{figure}
  \centering
   \includegraphics[width=0.7\columnwidth]{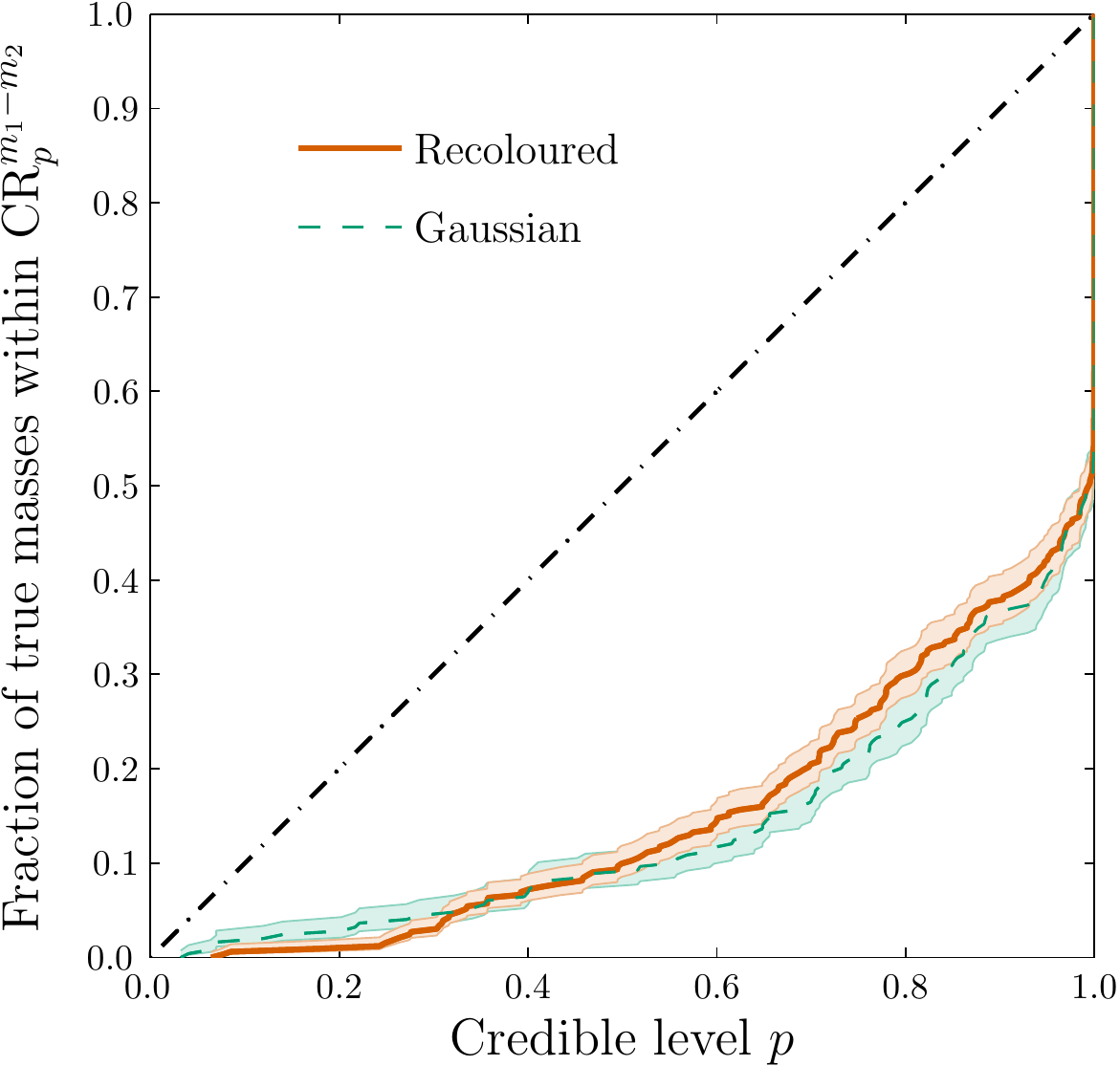}
    \caption{Fraction of true source component masses $(m_1,\,m_2)$ found within a credible region as a function of encompassed posterior probability. Results using recoloured noise are indicated by a solid line, while the results using Gaussian noise \citep{Singer:2014qca} are indicated by a dashed line. The expected distribution is indicated by the dot--dashed diagonal line. The shaded regions enclose the $68\%$ confidence intervals accounting for sampling errors.} 
    \label{fig:pp-m1m2}
\end{figure}

We are again working in two dimensions, so we use credible regions to quantify PE precision. The mass-space credible region $\mathrm{CR}^{m_1\mathrm{-}m_2}_p$ is defined analogously to its sky-area counterpart in (\ref{eq:CR}); it is easier to compute as we do not have to contend with the spherical geometry of the sky or with as intricate posterior distributions. We plot in figure \ref{fig:pp-m1m2} the fraction of injected masses that fall within $\mathrm{CR}^{m_1\mathrm{-}m_2}_p$ at a given $p$.
As for the chirp mass, the posterior is not well calibrated, approximately $40\%$ ($38\%$ for results with recoloured noise and $42\%$ for Gaussian) of the true component masses lie altogether outside the range of the estimated posterior, but the two sets of results are consistent with each other (performing a KS test gives a $p$-value of $0.969$). We cannot accurately reconstruct the component masses using our non-spinning waveforms.

\begin{figure}
  \centering
   \includegraphics[width=0.9\columnwidth]{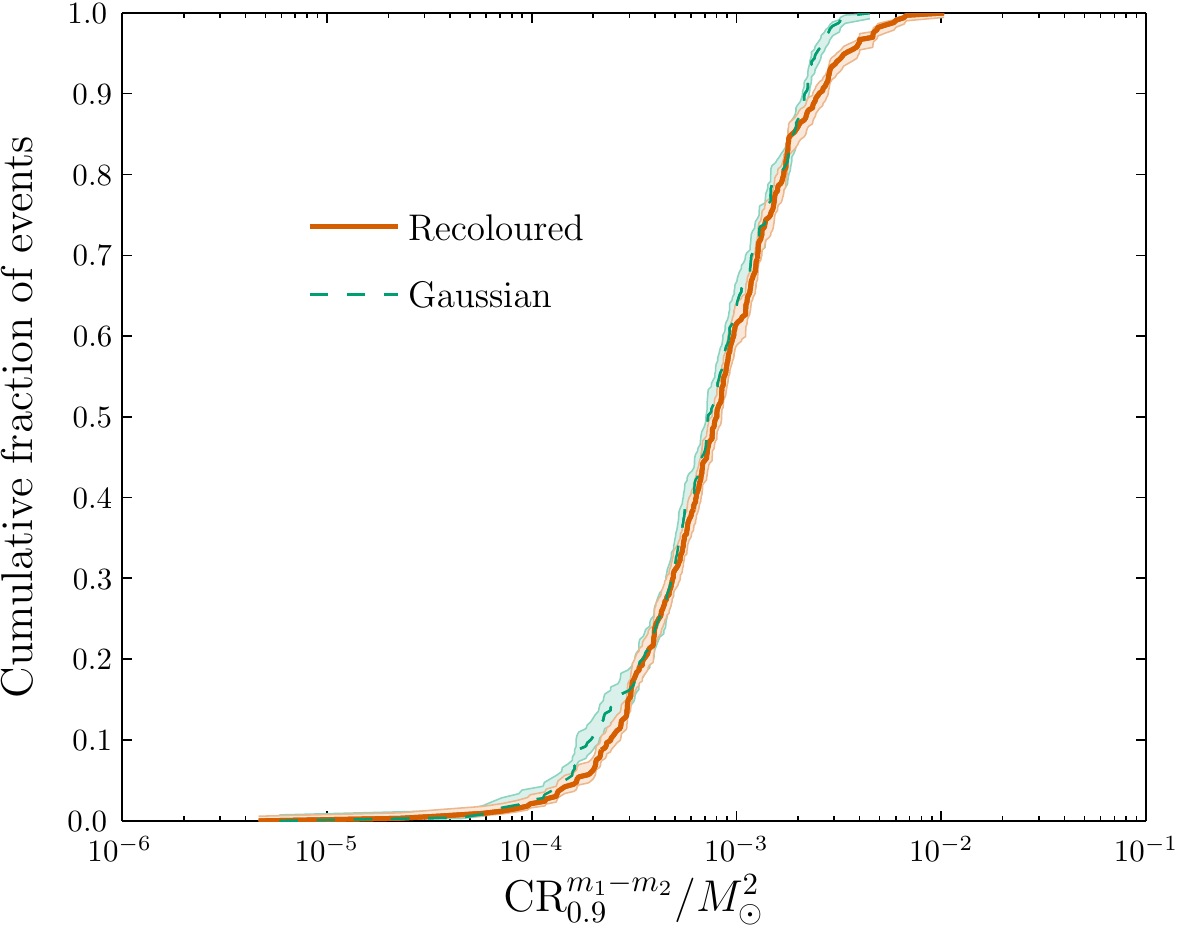}
    \caption{Cumulative fractions of events with $m_1$--$m_2$ $90\%$ credible regions smaller than the abscissa value. Results using recoloured noise are indicated by a solid line and the results using Gaussian noise \citep{Singer:2014qca} are indicated by a dashed line. The $68\%$ confidence intervals are denoted by the shaded areas. These results show the typical posterior width using non-spinning waveforms, the failure to include the mass--spin degeneracy means that these posteriors are too narrow.} 
    \label{fig:m1m2-90}
\end{figure}

To give an indication of the scale of the uncertainty in $m_1$--$m_2$ space, we plot the $90\%$ credible region in figure \ref{fig:m1m2-90}.
Since our estimates for the component masses are inaccurate, with many true values lying outside the posterior, $\mathrm{CR}^{m1\mathrm{-}m2}_p$ is a lower bound on the typical scale for measurement accuracy. This does not reflect how well we can actually measure the component masses, to produce accurate estimates, we must include the mass--spin degeneracy which broadens the posterior. 

It is apparent that a statement regarding measurement of component masses must wait until an analysis is done using waveforms that include spin. We will return this question in a future publication.

\section{Discussion and conclusions}\label{sec:end}

\subsection{Observing scenarios}\label{sec:ob-scenario}

Having determined the sky-localization accuracy expected for O1, we now use our results to compare with current predictions for observing scenarios in the advanced-detector era. In section \ref{sec:two-det} we consider the two-detector network of O1. In section \ref{sec:three-det} we extend our discussion to consider predictions for sky-localization in subsequent observing runs using a three-detector network.

\subsubsection{Two-detector sky-localization accuracy}\label{sec:two-det}

Prospects for sky localization in the advanced-detector era are specified by \citet{Aasi:2013wya}. This states that any events detected in 2015 would not be well localized. This has been shown to not be the case \citep[e.g.,][]{Nissanke:2011ax,Kasliwal:2013yqa,Singer:2014qca}. We see that while only a small fraction of events have well-localized sources, this fraction is non-zero. The $90\%$ credible region is almost always smaller than $10^3~\mathrm{deg^2}$. The 2015 observing scenario of \citet{Aasi:2013wya} does not give any figures for potential sky-localization accuracy, but we can now be specific using the results of this work.

The sky-localization figures currently included in \citet{Aasi:2013wya} are calculated using TT \citep{Fairhurst:2009tc,Fairhurst:2010is}. This is a convenient means of predicting sky-localization accuracy; it is not a method used to reconstruct the sky-position posterior of detected signals. For a two-detector network, triangulation predicts an unbroken annulus on the sky. The area of this ring linearly scales with the uncertainty on the timing measurement, which is inversely proportional to the SNR. \edit{Our results show that}, when using a coherent Bayesian approach, the recovered sky area is not (always) a ring, see figure \ref{fig:sky-map}, and the area scales inversely with the square of the SNR \citep{Raymond:2008im}. Hence, TT is a poor fit in this case.

\begin{figure*}
  \centering
   \subfigure[]{\includegraphics[width=0.9\columnwidth]{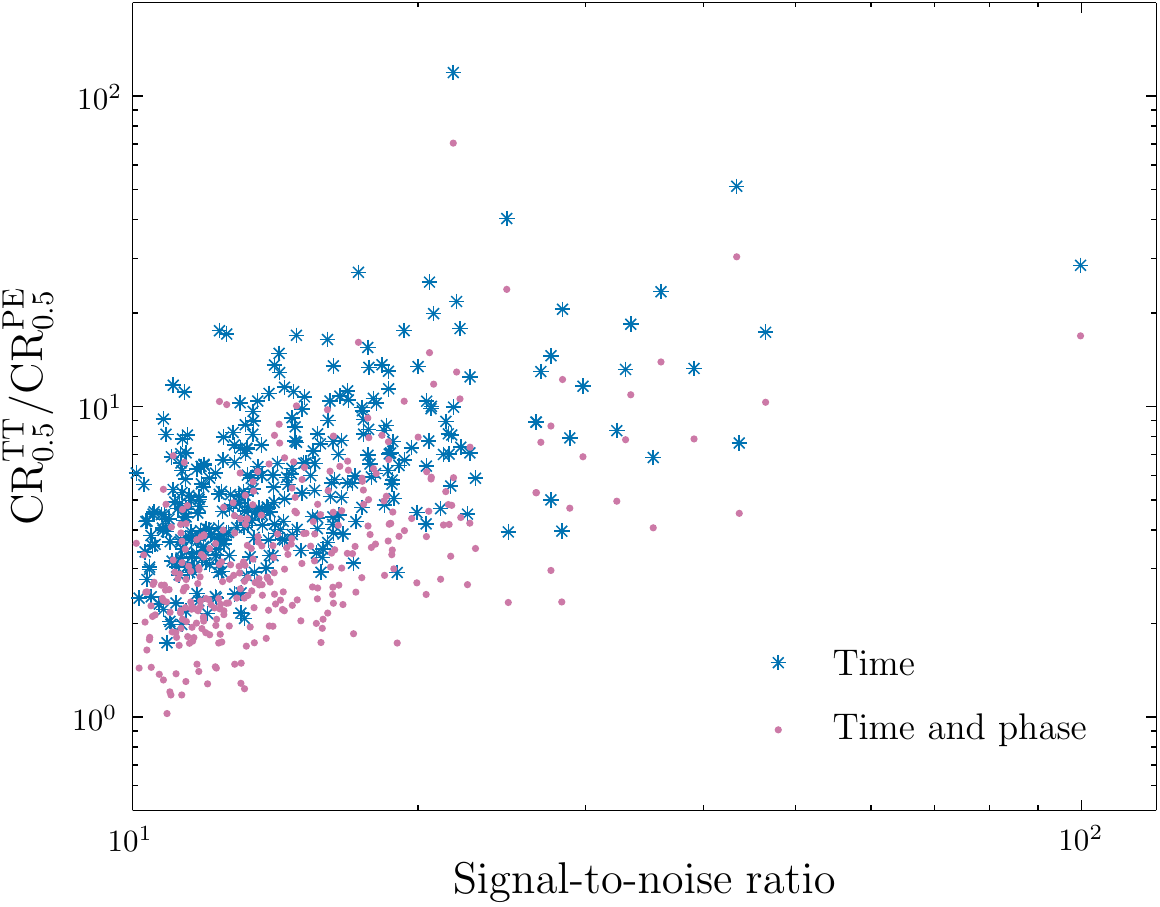}} \quad
   \subfigure[]{\includegraphics[width=0.9\columnwidth]{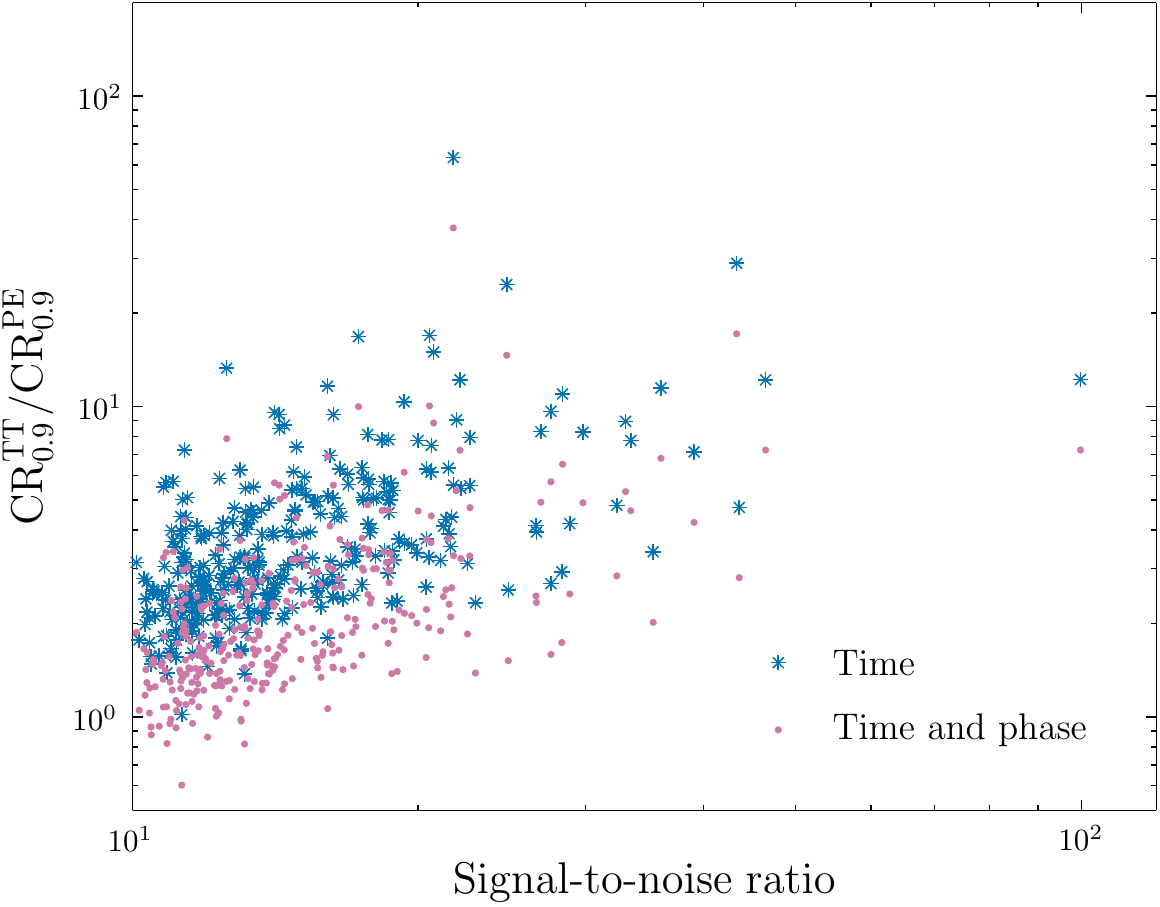}}
    \caption{Ratio of the area of credible regions calculated using TT and PE as a function of the SNR. (a) Ratio of $50\%$ credible regions $\mathrm{CR}_{0.5}$. (b) Ratio of $\mathrm{CR}_{0.9}$. TT results are calculated using just time of arrivals \citep{Fairhurst:2009tc,Fairhurst:2010is}, indicated by the star-shaped (blue) points, and by also including phase coherence \citep{Grover:2013sha}, indicated by the round (purple) points. PE results are calculated from the posteriors returned by \textsc{LALInference}.} 
    \label{fig:TT2}
\end{figure*}

In figure \ref{fig:TT2} we plot the ratio of the predicted credible region calculated using TT, to the actual credible region calculated using \textsc{LALInference} PE.
We include predictions from both standard TT and also TT including phase coherence \citep{Grover:2013sha}. The former method estimates timing accuracy (and hence the width of the sky annulus) as a function of the SNR and detector bandwidth.\footnote{In calculating these values we have corrected typos in both equation (28) of \citet{Fairhurst:2009tc}, where the prefactor should be $\sqrt{2}\erf^{-1}(0.9) \approx 1.65$ rather than $3.3$, and equation (15) of \citet{Fairhurst:2010is}, which has an unnecessary factor of $D$.} The latter method introduces the requirement of phase consistency between detectors, which can significantly aid source localization. These effects are modelled via a correction factor, whose value depends on how marginalization over polarization is taken into account. Here, we use the larger of the two correction factors proposed in \citet{Grover:2013sha}, their equation (16), although the degeneracy between phase and polarization means that the correction factor is probably too large for the two-detector network. The time and phase method does better, but neither technique does a good job at matching the true localization: both are too pessimistic. Agreement worsens at higher SNR as a consequence of the different SNR scalings. We cannot naively use TT to predict sky-localization accuracy for a two-detector network. 

We have found that sky areas recovered during O1 are likely to be hundreds of square degrees. Covering such a large area to sufficient depth to detect the most plausible EM counterparts \citep[$r \gtrsim 22$--$26~\mathrm{mag}$;][]{Metzger:2011bv,Barnes:2013wka,Metzger:2014yda} is challenging for current EM observatories \citep{Kasliwal:2013yqa}; furthermore, posterior distributions for the sky location are commonly multimodal or feature long, narrow arcs making them awkward to cover. It will be necessary to carefully consider how to most efficiently point telescopes to maximise the probability of observing a counterpart; using galaxy catalogues could be one means of increasing \edit{this chance} \citep{Nuttall:2010nk,Nissanke:2012dj,Hanna:2013yda,Bartos:2014spa,Fan:2014kka}.

\subsubsection{Three-detector sky-localization accuracy}\label{sec:three-det}

For 2016 onwards, we expect that AdV would also be in operation. The addition of a third detector should significantly improve sky-localization accuracy \citep{Singer:2014qca}.

\citet{Aasi:2013wya} give figures for sky-localization accuracies in the three-detector era. In 2016, \citet{Aasi:2013wya} predicts that $2\%$ ($5$--$12\%$) of BNS detections shall be localized within $5~\mathrm{deg^2}$ ($20~\mathrm{deg^2}$) at $90\%$ confidence. These values are calculated from TT. Ideally, we would like to compare these to results using Bayesian PE using recoloured noise, but performing three-detector PE runs for later observing periods is outside the range of this study. However, we have demonstrated that the properties of the noise do not impact sky-localization accuracies, provided that the chosen detection threshold yields similar SNR distributions in all cases. Consequently, we can use the three-detector, Gaussian-noise \textsc{LALInference} results of \citet{Singer:2014qca} as a reference. For comparison, they find that $2\%$ ($14\%$) of events have $\mathrm{CR}_{0.9}$ smaller than $5~\mathrm{deg^2}$ ($20~\mathrm{deg^2}$). PE with \textsc{LALInference} provides more optimistic sky-localization accuracies than TT. 

\begin{figure*}
  \centering
   \subfigure[]{\includegraphics[width=0.9\columnwidth]{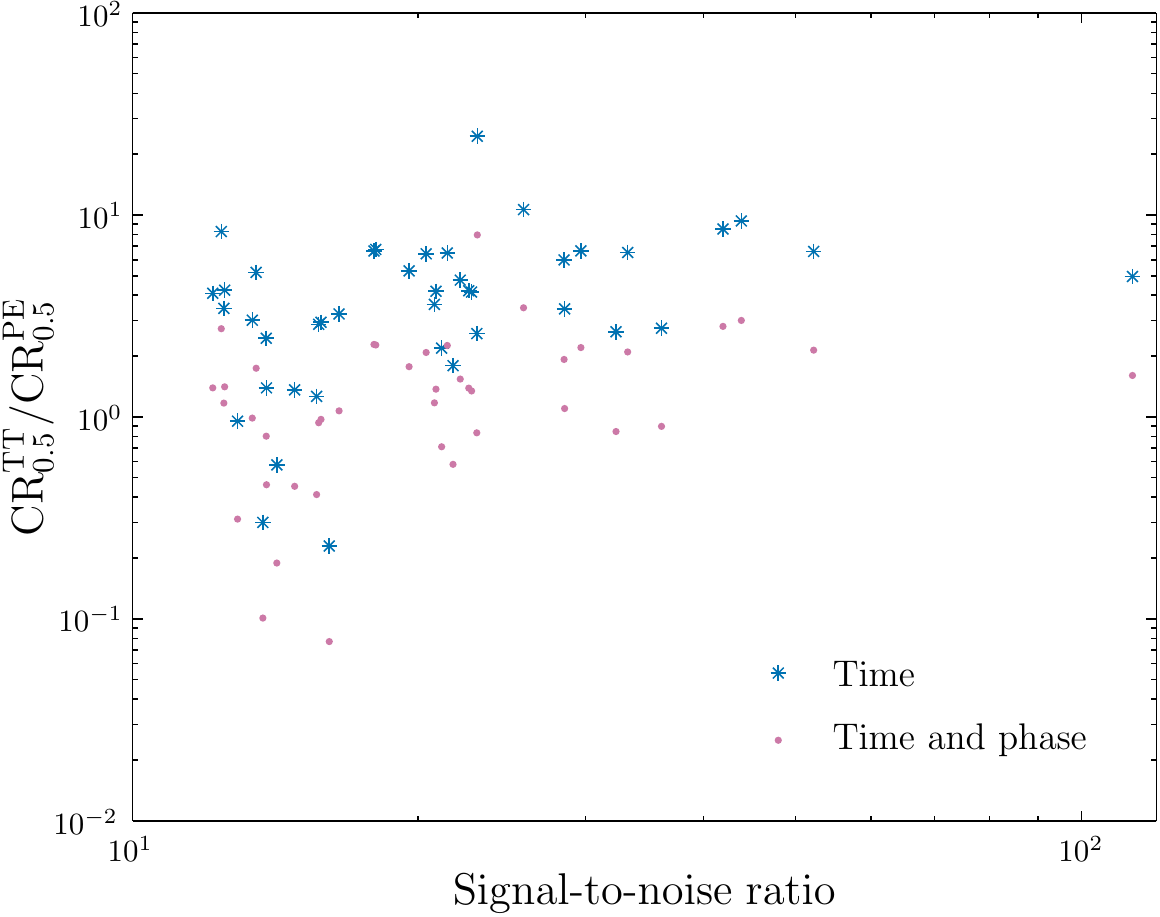}} \quad
   \subfigure[]{\includegraphics[width=0.9\columnwidth]{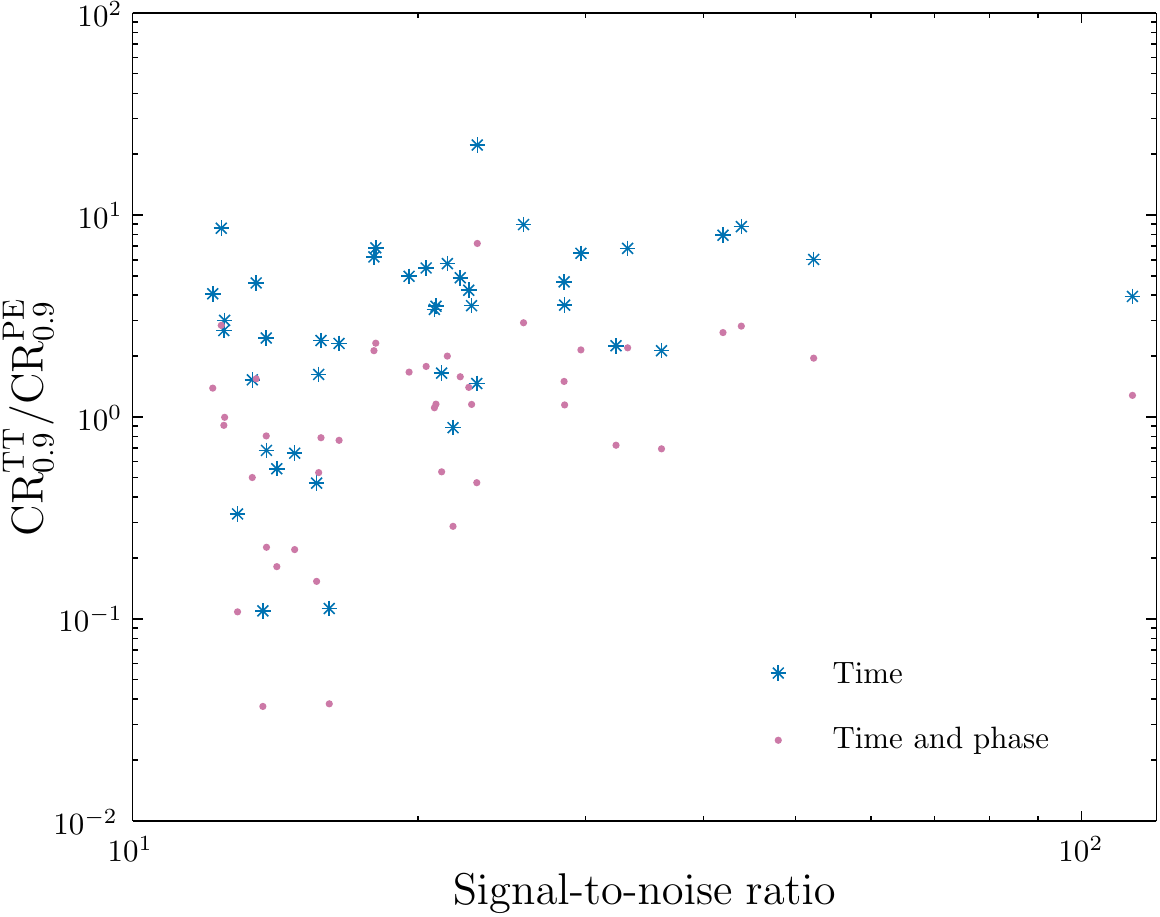}}
    \caption{Ratio of the area of credible regions calculated as a function of the SNR as in figure \ref{fig:TT2}, but for a three-detector network as expected in 2016. (a) Ratio of $\mathrm{CR}_{0.5}$. (b) Ratio of $\mathrm{CR}_{0.9}$. TT results are calculated using just time of arrivals \citep{Fairhurst:2009tc,Fairhurst:2010is}, indicated by the star-shaped (blue) points, and by also including phase coherence \citep{Grover:2013sha}, indicated by the round (purple) points. PE results with Gaussian noise are calculated from the posteriors returned by \textsc{LALInference} \citep{Singer:2014qca}.} 
    \label{fig:TT3}
\end{figure*}

In figure \ref{fig:TT3} we compare the three-detector results of \citet{Singer:2014qca} to the equivalent results calculated using TT.
These results are for 2016, assuming the mid noise curve of \citet{Barsotti:2012} for the aLIGO detectors, and the geometric mean of the high and low bounds of the early curve of \citet{Aasi:2013wya} for the Virgo interferometer. Both triangulation and PE produce sky areas that scale with $\varrho^{-2}$, such that their ratio shows no significant trend with SNR, although the scatter seems to decrease as SNR increases.

Comparing the entire population of points, we can calculate average values, which are given in table \ref{tab:TT-calibration}. We consider the logarithm of the ratio, which should be $\log_{10}(1) = 0$ for perfect agreement. The median $\log_{10}(\mathrm{CR}^\mathrm{TT}_{0.5}/\mathrm{CR}^\mathrm{PE}_{0.5})$ using only time of arrival is $0.61$, in complete agreement with the findings of \citet{Grover:2013sha}; using time and phase, the median value is $0.13$. The TT and PE results have different ratios $\mathrm{CR}_{0.9}/\mathrm{CR}_{0.5}$. The mean value of $\log_{10}(\mathrm{CR}^\mathrm{PE}_{0.9}/\mathrm{CR}^\mathrm{PE}_{0.5})$ is approximately $0.64$ and the standard deviation is $0.13$; again (see section \ref{sec:sky-loc}), this does not fit well with a Gaussian model. The $90\%$ credible regions for triangulation and PE are in better agreement with each other, with the time-and-phase triangulation average areas consistent with those from \textsc{LALInference}. The time-and-phase method produces a reasonable estimate when averaged over the entire population. However, for individual events there is large scatter because TT models are purely predictive and do not take into account the actual data realization.

Despite the good average agreement, there is a large tail of events at low SNRs where credible regions are too small, and the results suggest that at high SNRs the credible regions may be too large; this may introduce errors when considering the sub-populations of the best localized or worst localized events (or if the distribution of events is significantly different from that considered here). Given all these findings, we can be confident that the TT results of \citet{Aasi:2013wya} are overly pessimistic.

There remains one further caveat before we can state that the sky-localization accuracies of \citet{Aasi:2013wya} should be revised to give better results. We have seen that using a realistic FAR cut allows us to detect signals with $\varrho < 12$. These low-SNR results shift the distribution of sky-localization accuracies, such that the performance appears worse. Thus, while we can be confident that the events currently included should have a better accuracy than assumed for \citet{Aasi:2013wya}, the total population of detectable events is potentially larger than previously estimated, and may include some low-SNR events with poorer localization.

\begin{deluxetable}{ccccc}
\tabletypesize{\footnotesize}
\tablecaption{Average values of the logarithm of the ratio of credible regions calculated using TT to those calculated from PE $\log_{10}(\mathrm{CR}^\mathrm{TT}_{p}/\mathrm{CR}^\mathrm{Full}_{p})$. TT results are calculated using just time of arrivals \citep{Fairhurst:2009tc,Fairhurst:2010is} and by also including phase coherence \citep{Grover:2013sha}. PE results with Gaussian noise are calculated from the posteriors returned by \textsc{LALInference} \citep{Singer:2014qca}.\label{tab:TT-calibration}}
\tablewidth{0pt}
\tablehead{
\colhead{Triangulation method} & \colhead{$p$} & \colhead{Mean} & \colhead{Median} & \colhead{Standard deviation}
}
\startdata
\multirow{2}{*}{Time only} & \multicolumn{1}{c}{$0.5$} & $\hphantom{-}0.53$ & $0.61$ & $0.39$ \\
 & \multicolumn{1}{c}{$0.9$} & $\hphantom{-}0.42$ & $0.55$ & $0.49$ \\
\multirow{2}{*}{Time and phase} & \multicolumn{1}{c}{$0.5$} & $\hphantom{-}0.05$ & $0.13$ & $0.39$ \\
 & \multicolumn{1}{c}{$0.9$} & $-0.07$ & $0.07$ & $0.49$
\enddata
\end{deluxetable}

\subsection{Summary}

We provide realistic prospects for sky localization and EM follow-up of CBC sources in the O1 era by simulating a search for BNS sources with a two-detector aLIGO network at anticipated 2015 sensitivity. Our analysis is designed to be as similar as possible to recent work investigating sky-localization capability in the first two years of the advanced-detector era \citep{Singer:2014qca}. That study assumed Gaussian noise whereas our analysis incorporates more realistic noise, using real data from the S6 observing period recoloured to the anticipated 2015 noise spectrum.

We use the same list of simulated BNS sources as previously used in \citet{Singer:2014qca}. The simulated events are passed through the \textsc{GSTLAL\_inspiral} data-analysis pipeline which will be used online in O1. Detection triggers from this search with a FAR of $\leq 10^{-2}~\mathrm{yr^{-1}}$ are then followed up with sky-localization and PE codes.

The pipeline should not significantly distort the population of signals detected compared with the astrophysical population. There appears to be no selection based upon BNS spin. There is a selection effect determined by the chirp mass (systems with smaller chirp masses are harder to detect), but this translates to only a small difference for a small number ($\lesssim 10^2$) of detections.

Comparison of sky-localization areas from \textsc{bayestar} and \textsc{LALInference} demonstrates that while the former only uses a selection of the information available and employs a number of approximations, it does successfully reconstruct sky position. Furthermore, \textsc{bayestar} does this with sufficiently low latency to be of use for rapid EM follow-up.

Rapid sky-localization with \textsc{bayestar} takes on average $900~\mathrm{s}$ of CPU time per event (appendix \ref{ap:cost}). If it is parallelized in a $32$-way configuration (the baseline for online analysis), this correspond to a wall time of $30~\mathrm{s}$. None of our runs would take longer than $60~\mathrm{s}$ to complete.

PE using \textsc{LALInference\_nest} with (non-spinning) TaylorF2 waveforms requires a total CPU time of $\sim2 \times 10^6~\mathrm{s}$ per event (appendix \ref{ap:cost}). Five CPUs were used for each \textsc{LALInference\_nest} run, hence the wall time, as a first approximation, can be estimated as $\sim100~\mathrm{hr}$. These PE results can be produced within a few days, although with more expensive waveforms, the time taken is longer. Ongoing technical improvements should reduce the computational cost in the near future \citep{Veitch:2014wba}.

Considering sky-localization, the median area of $\mathrm{CR}_{0.9}$ ($\mathrm{CR}_{0.5}$) as estimated by \textsc{LALInference} is $632~\mathrm{deg^2}$ ($154~\mathrm{deg^2}$), and the median searched area is $132~\mathrm{deg^2}$. \textsc{LALInference} finds that $2\%$ of events have $\mathrm{CR}_{0.5}$ smaller than $20~\mathrm{deg^2}$; fewer than $1\%$ of events have $\mathrm{CR}_{0.5}$ smaller than $5~\mathrm{deg^2}$ or $\mathrm{CR}_{0.9}$ smaller than $20~\mathrm{deg^2}$, but $14\%$ of events have searched areas smaller than $20~\mathrm{deg^2}$ and $4\%$ have searched areas smaller than $5~\mathrm{deg^2}$. These are worse than predicted using Gaussian noise because of the inclusion of more low-SNR events, but if these additional events are excluded, the results calculated using both types of noise are in agreement. \edit{The non-stationarity and non-Gaussianity of the recoloured noise does not noticeably affect sky-localization accuracy, and sky areas are consistent if the same SNR threshold is applied to the recoloured and Gaussian data sets}.

The 2015 observing scenario of \citet{Aasi:2013wya} currently states that any events detected would not be well localized. This is not the case, although recovered areas are still large.

While \citet{Aasi:2013wya} does not have sky-localization figures for 2015, it does have them for later years. These are calculated using a \edit{TT} method \citep{Fairhurst:2009tc,Fairhurst:2010is}. The Gaussian results of \citet{Singer:2014qca} show that we can achieve better sky localization than expected from TT alone; this improvement can principally be explained by the incorporation of phase consistency \citep{Grover:2013sha}. Hence, the figures in \citep{Aasi:2013wya} may be pessimistic. However, from this study we also know that results using Gaussian noise are liable to be optimistic because they exclude events by using of a detection threshold of $\varrho \geq 12$; in practise, when using a FAR threshold, there is a tail of lower SNR events that skew the distribution. This must be accounted for when quoting the fraction of events located to within a given area. Therefore, updating the numbers in the observing scenarios for later years is not straightforward.

The \textsc{LALInference} runs also return posteriors for other parameters. We looked at the source luminosity distance, the chirp mass and the component masses. The distance is not well measured; the median $\mathrm{CI}^D_{0.9}/D_\star$ ($\mathrm{CI}^D_{0.5}/D_\star$) is $0.85$ ($0.38$). As a consequence of our use of non-spinning waveform templates that do not exactly match the injected waveforms, the chirp-mass estimates are subject to theoretical error of a size roughly equal to the uncertainty introduced by the noise. This means our posteriors are not well calibrated: they are both (on average) offset from the true position and too narrow (by a factor of $\sim 1/2$). Using spinning waveforms, such that the mass--spin degeneracy can be explored, will broaden the posteriors and resolve this problem, but we will always face a potential systematic bias unless we exactly know the true waveforms of Nature. Despite the systematic effects, the posterior mean of the chirp-mass distribution is within $10^{-3}M_\odot$ of the true chirp mass in $96\%$ of events, and the median absolute difference between the two is $(2.6\times 10^{-4})M_\odot$. A larger difference could occur if there is a larger discrepancy between the waveform template and the true waveform, but we expect it to be of a similar order of magnitude. While we can still accurately measure the chirp mass using non-spinning waveforms, the same does not apply for component masses. Estimates for these must be performed using spinning waveforms; we shall examine this in a future study.

\edit{Aggregate PE accuracy is the same for the population of signals with Gaussian noise and the population with recoloured noise. The inclusion of non-stationary and non-Gaussian noise features does not degrade our average PE ability at a given SNR. The recoloured S6 noise is used as a surrogate for real aLIGO noise; while it is more realistic than pure Gaussian noise, it does not necessary reflect the true form of the noise that will be recorded in O1. However, since we do not observe any difference in PE performance using recoloured noise, we can be confident that the non-Gaussianity of real noise should not significantly affect our PE ability (unless the noise characteristics are qualitatively different than anticipated). We expect that the non-stationary and non-Gaussian noise of the advanced detectors will not be a detriment to PE for BNSs.}

\acknowledgements

The authors are grateful for useful suggestions from the CBC group of the LIGO--Virgo Science Collaboration and in particular Yiming Hu.

This work was supported by the Science and Technology Facilities Council. PBG acknowledges NASA grant NNX12AN10G. SV acknowledges the support of the National Science Foundation and the LIGO Laboratory. JV was supported by STFC grant ST/K005014/1. LIGO was constructed by the California Institute of Technology and Massachusetts Institute of Technology with funding from the National Science Foundation and operates under cooperative agreement PHY-0757058.

Results were produced using the computing facilities of the LIGO DataGrid including: the Nemo computing cluster at the Center for Gravitation and Cosmology at the University of Wisconsin--Milwauke under NSF Grants PHY-0923409 and PHY-0600953; the Atlas computing cluster at the Albert Einstein Institute, Hannover; the LIGO computing clusters at Caltech, and the facilities of the Advanced Research Computing @ Cardiff (ARCCA) Cluster at Cardiff University. We are especially grateful to Paul Hopkins of ARCCA for assistance.

Some results were produced using the post-processing tools of the \texttt{plotutils} library at \url{http://github.com/farr/plotutils}, and some were derived using HEALPix \citep{Gorski:2004by}.

This paper is has been assigned LIGO document reference LIGO-P1400232. It contains some results originally included in LIGO technical report LIGO-T1400480.

\appendix

\section{Detection and component masses}\label{ap:mass}

In section \ref{sec:mass-spin}, we examined selection effects of the detection pipeline. In particular, we looked at the detected distribution of chirp masses as this sets the GW amplitude. The magnitude of the selection effect depends on the details of the chirp-mass distribution, but can be estimated using a simple model. For low-mass signals whose inspiral spans the sensitive band of the detector, the amplitude of the waveform is proportional to $\mathcal{M}_\mathrm{c}^{5/6}$ \citep{Sathyaprakash:2009xs}. The sensitive volume is proportional to the cube of this, or $\mathcal{M}_\mathrm{c}^{5/2}$. Suppose that half of the injections are made at a chirp mass of $\bar{\mathcal{M}_\mathrm{c}} - \delta{\mathcal{M}_\mathrm{c}}$ and the other half at a chirp mass value of $\bar{\mathcal{M}_\mathrm{c}} + \delta{\mathcal{M}_\mathrm{c}}$, with $\delta{\mathcal{M}_\mathrm{c}} \ll \bar{\mathcal{M}_\mathrm{c}}$. Then the expected fraction of higher-mass systems among all detected systems is
\begin{align}
\mathcal{F}_\mathrm{high} = & {} \frac{\left(\bar{\mathcal{M}_\mathrm{c}} + \delta{\mathcal{M}_\mathrm{c}}\right)^{5/2}}{\left(\bar{\mathcal{M}_\mathrm{c}} + \delta{\mathcal{M}_\mathrm{c}}\right)^{5/2} + \left(\bar{\mathcal{M}_\mathrm{c}} - \delta{\mathcal{M}_\mathrm{c}}\right)^{5/2}}\nonumber \\
 \simeq & {} \frac{1}{2} + \frac{5}{4}\frac{\delta{\mathcal{M}_\mathrm{c}}}{\bar{\mathcal{M}_\mathrm{c}}}.
\end{align}
If $N$ detections are made in total and the selection effects played no role, the expected number of detections from the higher-mass set would be $N/2$ with a standard deviation of $\sqrt{N}/2$.  However, in our model, there is a predicted excess of $5 N \delta{\mathcal{M}_\mathrm{c}} / (4 \bar{\mathcal{M}_\mathrm{c}})$ high-mass detections because of selection effects. Consequently, we expect to have $x$-$\sigma$ confidence in observing a selection effect on chirp mass, where
\begin{equation}
x = \frac{5\sqrt{N}}{2} \frac{\delta\mathcal{M}_\mathrm{c}}{\bar{\mathcal{M}_\mathrm{c}}}.
\end{equation}
We can estimate $\bar{\mathcal{M}_\mathrm{c}}$ from the mean of the chirp-mass distribution, and $\delta{\mathcal{M}_\mathrm{c}}$ from the standard deviation; for our injections set, $\delta{\mathcal{M}_\mathrm{c}} / \bar{\mathcal{M}_\mathrm{c}} \approx 0.06$. For the Gaussian data set $N = 250$, and so we expect to observe selection effects at only the $\sim 2$-$\sigma$ confidence level; the actual measurements are roughly consistent with this. For such a narrow chirp-mass distribution, $\gtrsim 10^3$ detections are needed to confidently observe the selection effects.

While the chirp mass is of prime importance to GW astronomers (it is their most precisely determined mass parameter), other combinations of mass are of interest in other contexts. Parameters which are correlated with the chirp mass, are also subject to selection effects. However, their significance is proportional to the level of correlation of the parameters with chirp mass; given that selection effects on chirp mass are small, we do not expect statistically significant effects for other mass parameters. Here, we present the distributions of the individual component masses, the asymmetric mass ratio and the total mass.

\begin{figure}
  \centering
   \includegraphics[width=0.9\columnwidth]{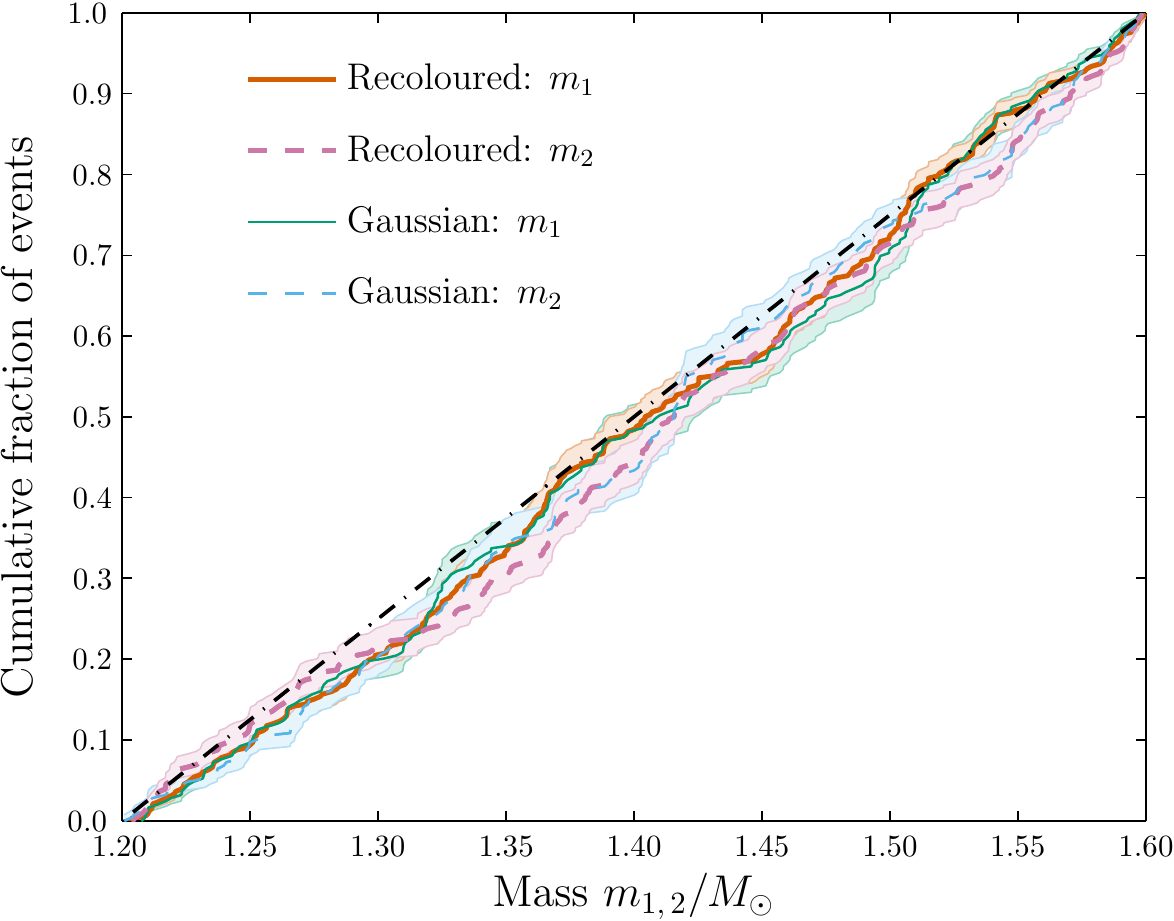}
    \caption{Cumulative fractions of detected events with component masses smaller than the abscissa value. The mass distribution for the first neutron star $m_1$ is denoted by the solid line, and the distribution for the second neutron star $m_2$ is denoted by the dashed line. Results with recoloured noise are denoted by the thicker red--purple lines, and results from the subset of $250$ events analysed with \textsc{LALInference} with Gaussian noise are denoted by the thinner blue--green lines \citep{Singer:2014qca}. The $68\%$ confidence intervals are denoted by the shaded areas. The expected distribution for component masses drawn uniformly from $m_\mathrm{min} = 1.2 M_\odot$ to $m_\mathrm{max} = 1.6 M_\odot$ is indicated by the black dot--dashed line.} 
    \label{fig:masses}
\end{figure}

The distribution of recovered (injected) component masses is shown in figure \ref{fig:masses}.
The detected events show a slight over-representation of higher-mass objects, which is the effect of selecting systems with larger chirp masses. The deviation from the injection distribution is small (a KS test with the predicted distribution gives $p$-values of $0.213$ and $0.182$ for Gaussian noise, and $0.276$ and $0.022$ for the recoloured noise), but noticeably more significant than for the spins.

\begin{figure}
  \centering
   \includegraphics[width=0.9\columnwidth]{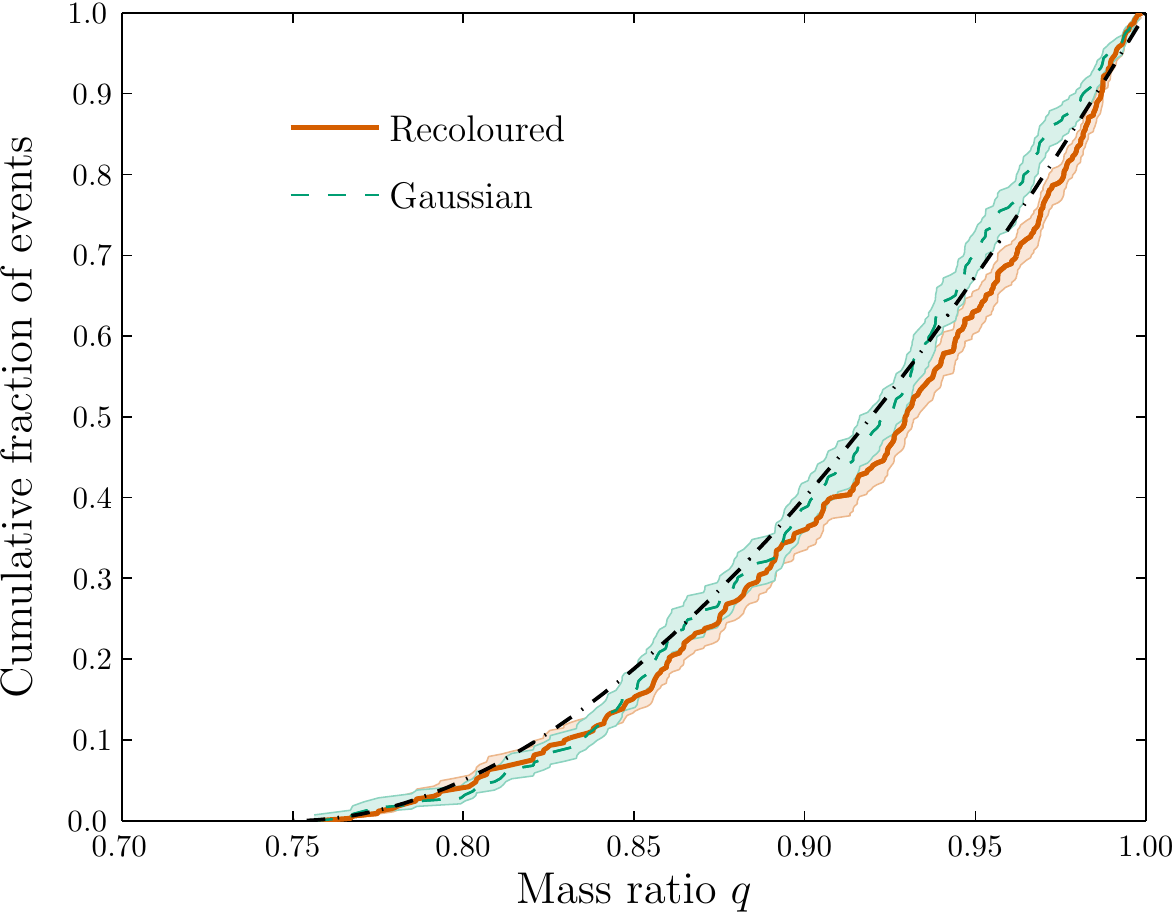}
    \caption{Cumulative fractions of detected events with asymmetric mass ratios smaller than the abscissa value. Results using recoloured noise are denoted by the solid red line, and results from the subset of $250$ events with Gaussian noise analysed with \textsc{LALInference} are denoted by the dashed green line \citep{Singer:2014qca}. The $68\%$ confidence intervals are denoted by the shaded areas. The injection distribution $C_q(q)$ is indicated by the black dot--dashed line.} 
    \label{fig:q-ratio}
\end{figure}
\begin{figure}
  \centering
   \includegraphics[width=0.9\columnwidth]{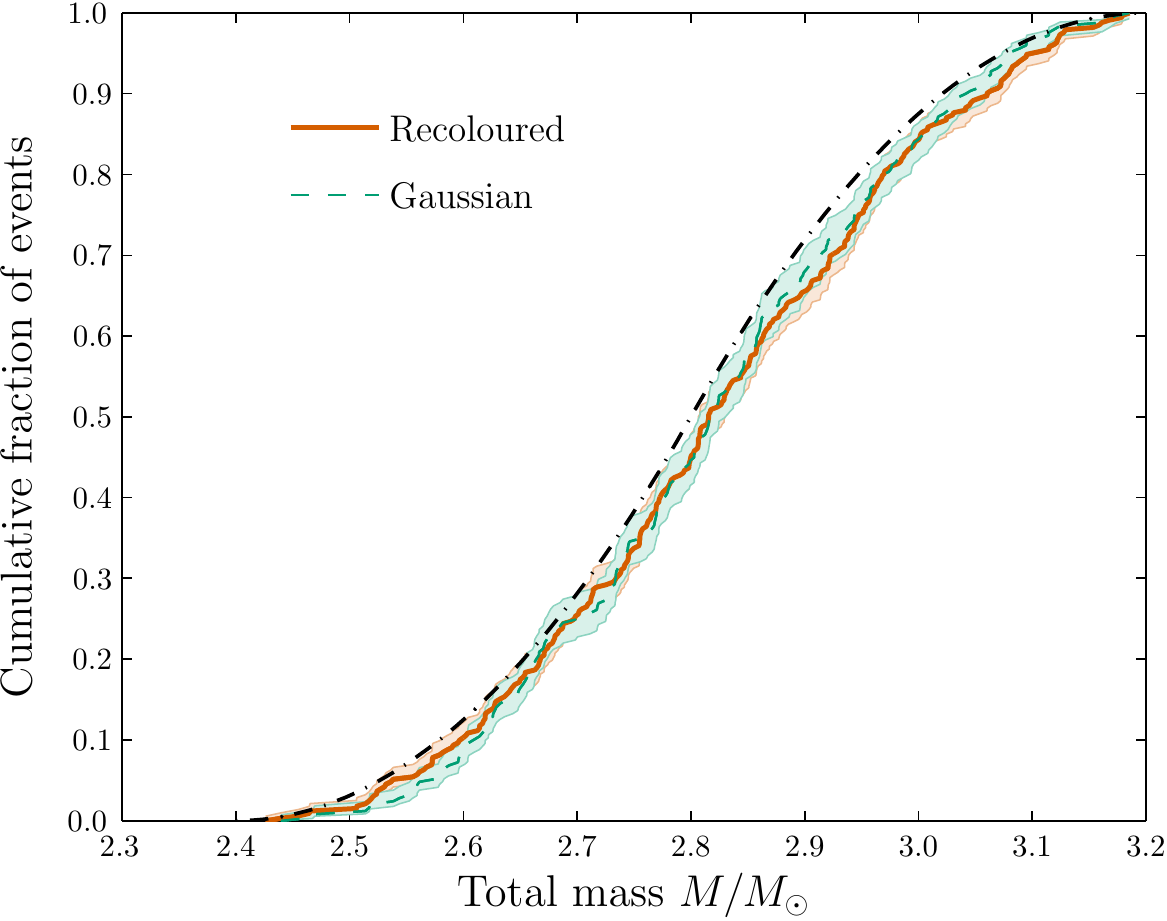}
    \caption{Cumulative fractions of detected events with total masses smaller than the abscissa value. Results using recoloured noise are denoted by the solid red line, and results from the subset of $250$ events with Gaussian noise analysed with \textsc{LALInference} are denoted by the dashed green line \citep{Singer:2014qca}. The $68\%$ confidence intervals are denoted by the shaded areas. The injection distribution $C_M(M)$ is indicated by the black dot--dashed line.} 
    \label{fig:total-mass}
\end{figure}

The asymmetric mass ratio is
\begin{equation}
q = \frac{\min\{m_1,\,m_2\}}{\max\{m_1,\,m_2\}}.
\end{equation}
For uniformly distributed $m_1$ and $m_2$ between $m_\mathrm{min}$ and $m_\mathrm{max}$, the probability density function for $q$ is
\begin{equation}
P_q(q) = \begin{cases}
\displaystyle \frac{1}{(m_\mathrm{max} - m_\mathrm{min})^2} \left(m_\mathrm{max}^2 - \frac{m_\mathrm{min}^2}{q^2}\right) & \displaystyle \frac{m_\mathrm{min}}{m_\mathrm{max}} \leq q \leq 1 \\
0 & \quad\mathrm{Otherwise}
\end{cases}\:.
\end{equation}
\begin{widetext}
\noindent{}Integrating this gives a cumulative distribution function
\begin{equation}
C_q(q) = \begin{cases}
0 & \displaystyle \hphantom{1\frac{m_\mathrm{min}}{m_\mathrm{max}} \leq\ } q \leq \frac{m_\mathrm{min}}{m_\mathrm{max}} \\
\displaystyle \frac{1}{(m_\mathrm{max} - m_\mathrm{min})^2}\left(m_\mathrm{max}^2 q - 2 m_\mathrm{min} m_\mathrm{max} + \frac{m_\mathrm{min}^2}{q}\right) & \displaystyle \hphantom{1}\frac{m_\mathrm{min}}{m_\mathrm{max}} \leq q \leq 1 \\
1 & \displaystyle \hphantom{\frac{m_\mathrm{min}}{m_\mathrm{max}}}1 \leq q
\end{cases}\:.
\end{equation}
Figure \ref{fig:q-ratio} shows the recovered distribution of mass ratios as well as the injection distribution given by $C_q(q)$.
There is a small difference between the injection and recovered distributions (a KS test with the injection distribution returns $p$-values of $0.536$ and $0.050$ for the Gaussian and recoloured noise respectively).

The probability density function for the total system mass, $M = m_1 + m_2$, is 
\begin{equation}
P_M(M) = \begin{cases}
\displaystyle \frac{1}{(m_\mathrm{max} - m_\mathrm{min})^2} \left(M - 2 m_\mathrm{min}\right) & \hphantom{m_\mathrm{max} +\ } 2 m_\mathrm{min} \leq M \leq m_\mathrm{min} + m_\mathrm{max} \\
\displaystyle \frac{1}{(m_\mathrm{max} - m_\mathrm{min})^2} \left(2 m_\mathrm{max} - M\right) & \hphantom{2}m_\mathrm{min} + m_\mathrm{max} \leq M \leq 2 m_\mathrm{max}\\
0 & \qquad\qquad\quad\mathrm{Otherwise}
\end{cases}\:.
\end{equation}
Consequently, its cumulative distribution function is
\begin{equation}
C_M(M) = \begin{cases}
0 & \displaystyle \hphantom{m_\mathrm{max} + 2 m_\mathrm{min} \leq\ } M \leq 2 m_\mathrm{min} \\
\displaystyle \frac{1}{(m_\mathrm{max} - m_\mathrm{min})^2}\left(\frac{M^2}{2} - 2 m_\mathrm{min} M + 2 m_\mathrm{min}^2\right) & \hphantom{m_\mathrm{max} +\ }2 m_\mathrm{min} \leq M \leq m_\mathrm{min} + m_\mathrm{max} \\
\displaystyle \frac{1}{(m_\mathrm{max} - m_\mathrm{min})^2}\left(2 m_\mathrm{max} M - \frac{M^2}{2} + m_\mathrm{min}^2 - 2 m_\mathrm{min} m_\mathrm{max} - m_\mathrm{max}^2\right) & \hphantom{2}m_\mathrm{min} + m_\mathrm{max} \leq M \leq 2 m_\mathrm{max} \\
1 & \hphantom{m_\mathrm{min} +\ }2 m_\mathrm{max} \leq M
\end{cases}\:.
\end{equation}
\end{widetext}
Figure \ref{fig:total-mass} shows the recovered distribution of \edit{total masses} as well as the injection distribution given by $C_M(M)$.
The distributions are similar to those seen for the chirp mass in figure \ref{fig:M-chirp}. This is not surprising, as there is a clear link between the two quantities. We are considering a narrow mass range; individual component masses can be described as $m_{1,\,2} = m_\mathrm{min}(1 + \varepsilon_{1,\,2})$, where $\varepsilon_{1,\,2} \leq (m_\mathrm{max} - m_\mathrm{min})/m_\mathrm{min} \ll 1$. The total mass is $m_\mathrm{min}(2 + \varepsilon_1 + \varepsilon_2)$; to first order in $\varepsilon_{1,\,2}$, the chirp mass can be described as $2^{-6/5} m_\mathrm{min} (2 + \varepsilon_1 + \varepsilon_2)$. Hence, the total mass is approximately proportional to the chirp mass across the range of interest. We preferentially select signals with larger total masses as these produce louder signals, although the difference between the injection and recovered distributions is not too large (a KS test with the injection distribution yields $p$-values of $0.338$ and $0.050$ for the Gaussian and recoloured noise respectively).

All the mass distributions show a difference between the injection and detected populations. This is as expected. The difference is small, such that for the numbers of events considered in this study, it is only marginally significant. The difference need not always be negligible, it would become more important when considering a larger population of events, or a set of events with a broader chirp-mass distribution. 

\section{Computational time}\label{ap:cost}

To perform rapid sky localization, we require that our analysis pipelines are expeditious. Following a detection, \textsc{bayestar} promptly returns a sky localization, and later \textsc{LALInference} returns estimates of the sky position plus further parameters. Here, we present estimates for the computational time required \edit{to run \textsc{bayestar} and \textsc{LALInference}}.

All results are specific to a two-detector network. The \textsc{LALInference} results are \edit{for} a (non-spinning) TaylorF2 analysis: this is the least expensive waveform family and provides medium-latency results. Computational times can be significantly longer using other waveforms. Efforts are being made to optimise and speed up the methods of \textsc{LALInference} \citep[e.g.,][]{Canizares:2013ywa,Canizares:2014fya,Farr:2013tia,Purrer:2014fza}.

\begin{figure}
  \centering
   \subfigure[\label{fig:time-a}]{\includegraphics[width=0.9\columnwidth]{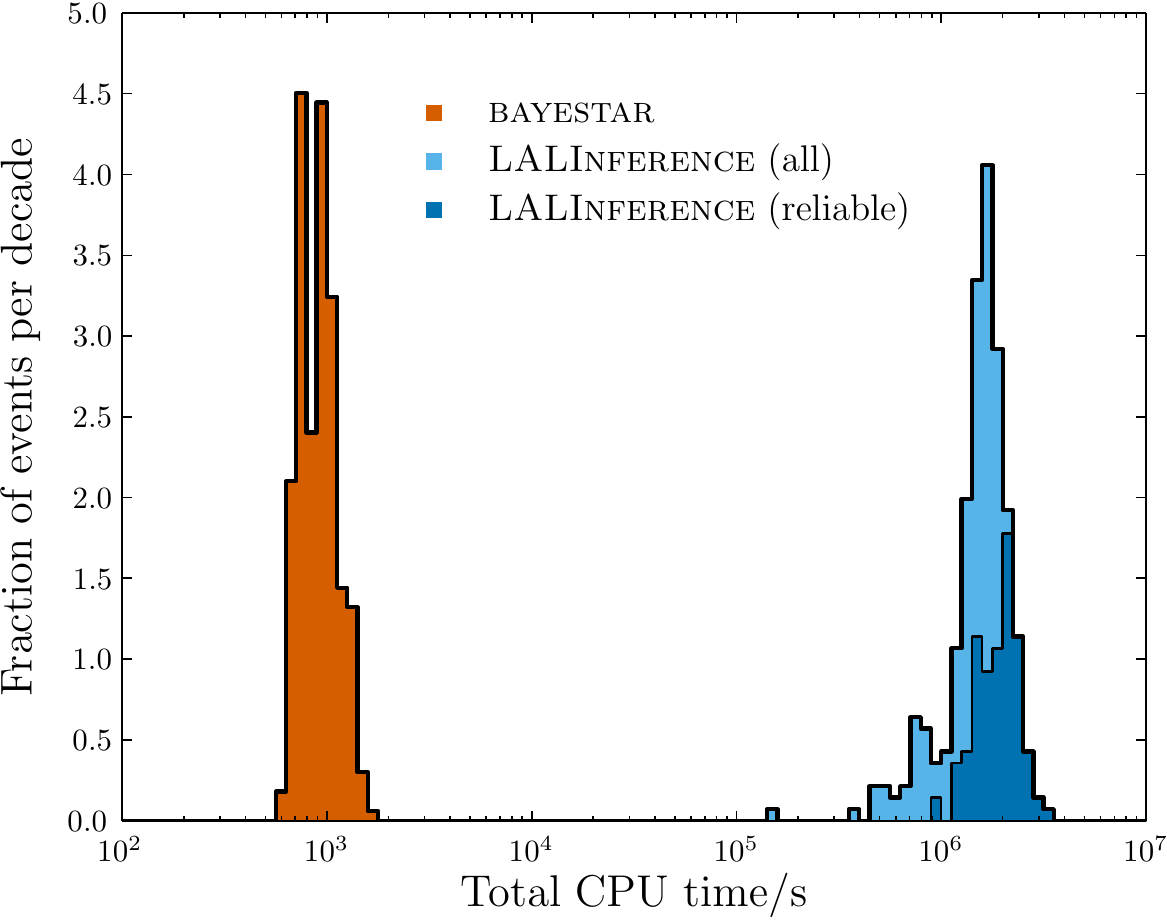}} \\
   \subfigure[\label{fig:bayes-time}]{\includegraphics[width=0.9\columnwidth]{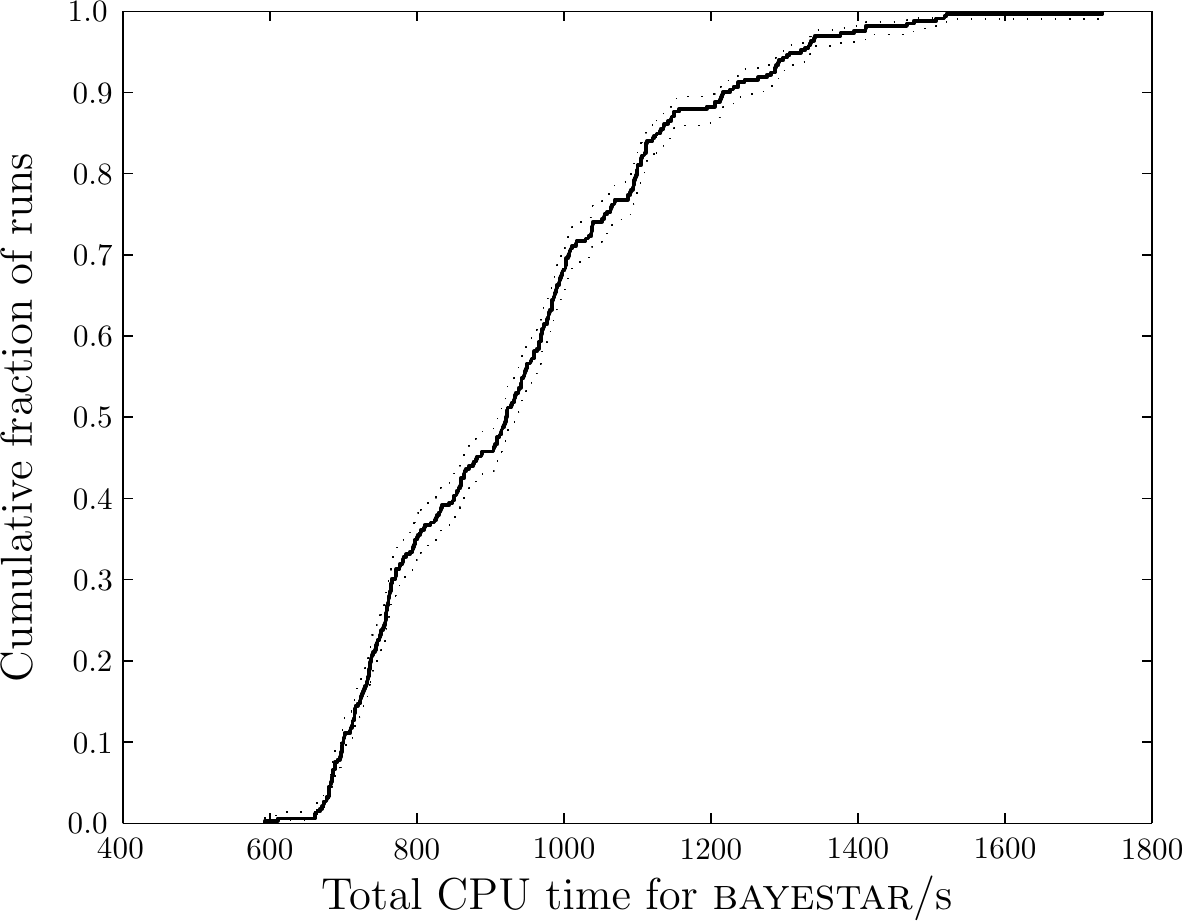}} \\
   \subfigure[\label{fig:time-b}]{\includegraphics[width=0.9\columnwidth]{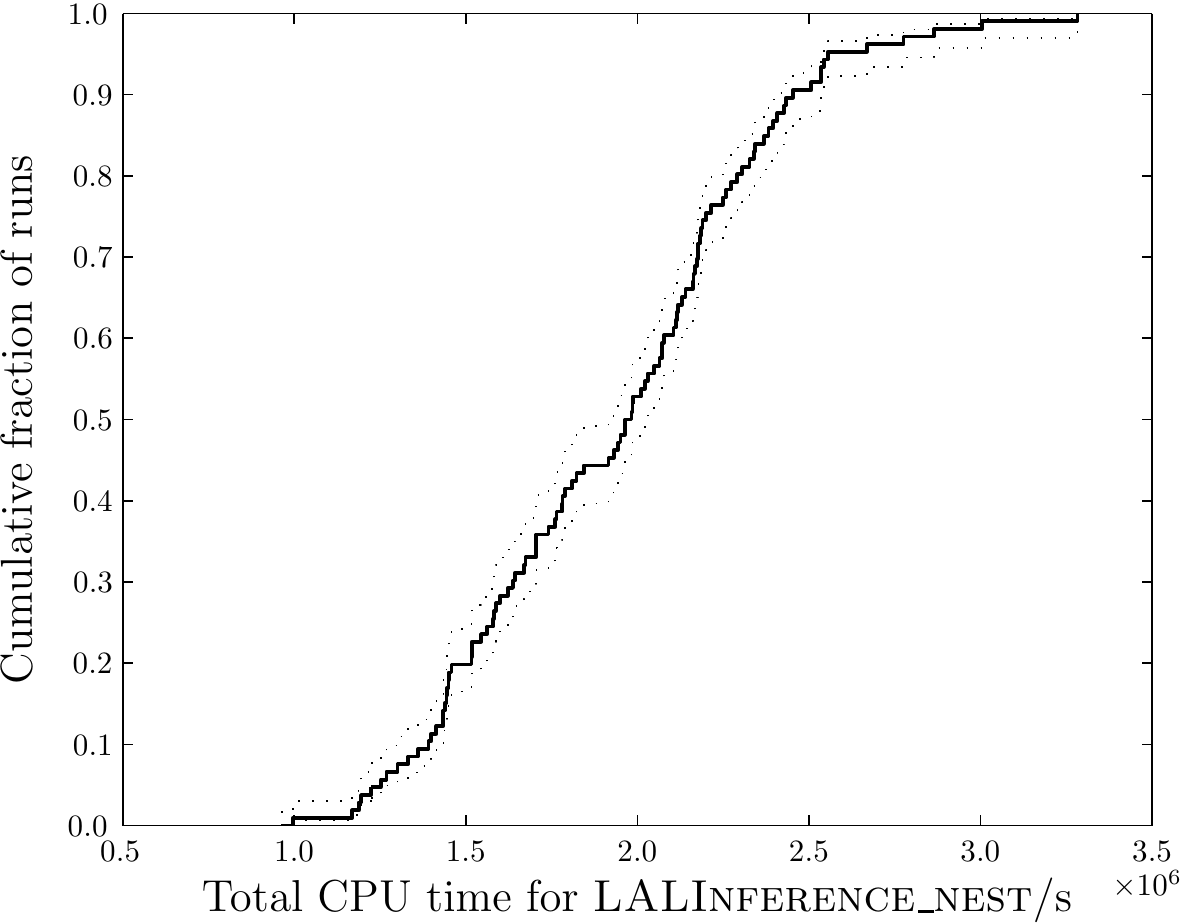}} 
    \caption{Computation time for a run measured in CPU seconds. (a) Distribution of run times. The left (red) distribution is for \textsc{bayestar} and the right (blue) distribution is for \textsc{LALInference\_nest}. \textsc{LALInference\_nest} times which are reliably estimated are shown in dark blue, while the full set of times including potentially inaccurately estimated times are shown in light blue. \edit{(b) Cumulative fractions of \textsc{bayestar} runs with computational times smaller than the abscissa value. (c) Cumulative fractions of \textsc{LALInference\_nest} runs with total CPU times smaller than the abscissa value, only reliable times are used here.} The $68\%$ confidence interval is enclosed by the dotted lines, this accounts for sampling errors and is estimated from a beta distribution \citep{Cameron:2010bh}. Each plot has a different scale.} 
    \label{fig:time}
\end{figure}

The \textsc{LALInference} PE is slower than the rapid sky localization. Distributions of estimated CPU times for the runs are shown in figure \ref{fig:time}.
The \textsc{LALInference\_nest} times are calculated from log files. This is not entirely reliable as times may not be recorded for a variety of reasons. In this case, the reported time is a lower bound on the true value. In figure \ref{fig:time-a} we show the distribution of run times for both the set of all estimated times and the subset excluding those we suspect are inaccurate due to a reported error message. The distributions are consistent with our expectation that the inaccurate times are lower bounds. In figure \ref{fig:time-b} we show the cumulative distribution of run times using only the more reliable set of estimates. The median (accurately estimated) total CPU time for \textsc{LALInference\_nest} is $1.96 \times 10^6~\mathrm{s} = 545~\mathrm{hr}$ \citep[cf.][]{Veitch:2014wba} and the median total CPU time for \textsc{bayestar} is $921~\mathrm{s} = 15.4~\mathrm{min}$. Hence, on average \textsc{LALInference\_nest} takes $\sim2000$ times as much CPU time as \edit{\textsc{bayestar}}.

The actual latency of a technique is given by the wall time, not the CPU time. Five CPU processes were used per \textsc{LALInference\_nest} run, hence the computational wall time can be estimated as a fifth of the total CPU time. This gives a median approximate wall time of $3.92 \times 10^5~\mathrm{s} = 109~\mathrm{hr}$. Some processes take longer to finish than others, so this is not an exact means of estimating the time taken for a run to finish. These calculations also neglect time spent idle rather than running, which influences the physical wall time required for a job to complete. In online mode, \textsc{bayestar} is generally deployed in a $32$--$64$-way parallel configuration. This gives a median wall time of $28.8~\mathrm{s}$ ($14.4~\mathrm{s}$) for a $32$-way ($64$-way) configuration. \textsc{bayestar} provides sky-localization $\sim10^4$ times quicker than \textsc{LALInference}, furthermore, none of our \textsc{bayestar} runs would have taken longer than a minute to complete \citep[chapter 4]{Singer:2014}.

\begin{figure}
  \centering
   \subfigure[\label{fig:speed-a}]{\includegraphics[width=0.89\columnwidth]{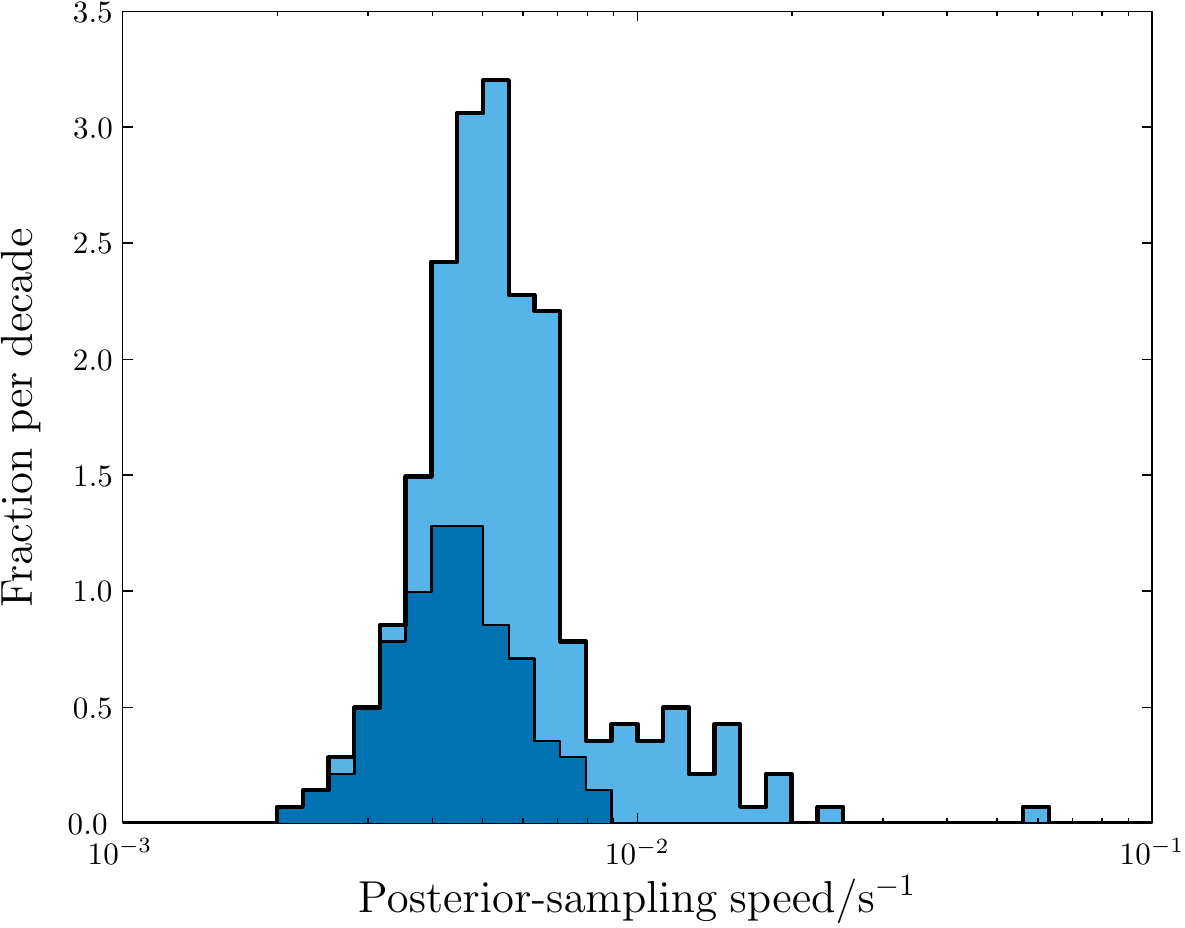}} \\
   \subfigure[\label{fig:speed-b}]{\includegraphics[width=0.9\columnwidth]{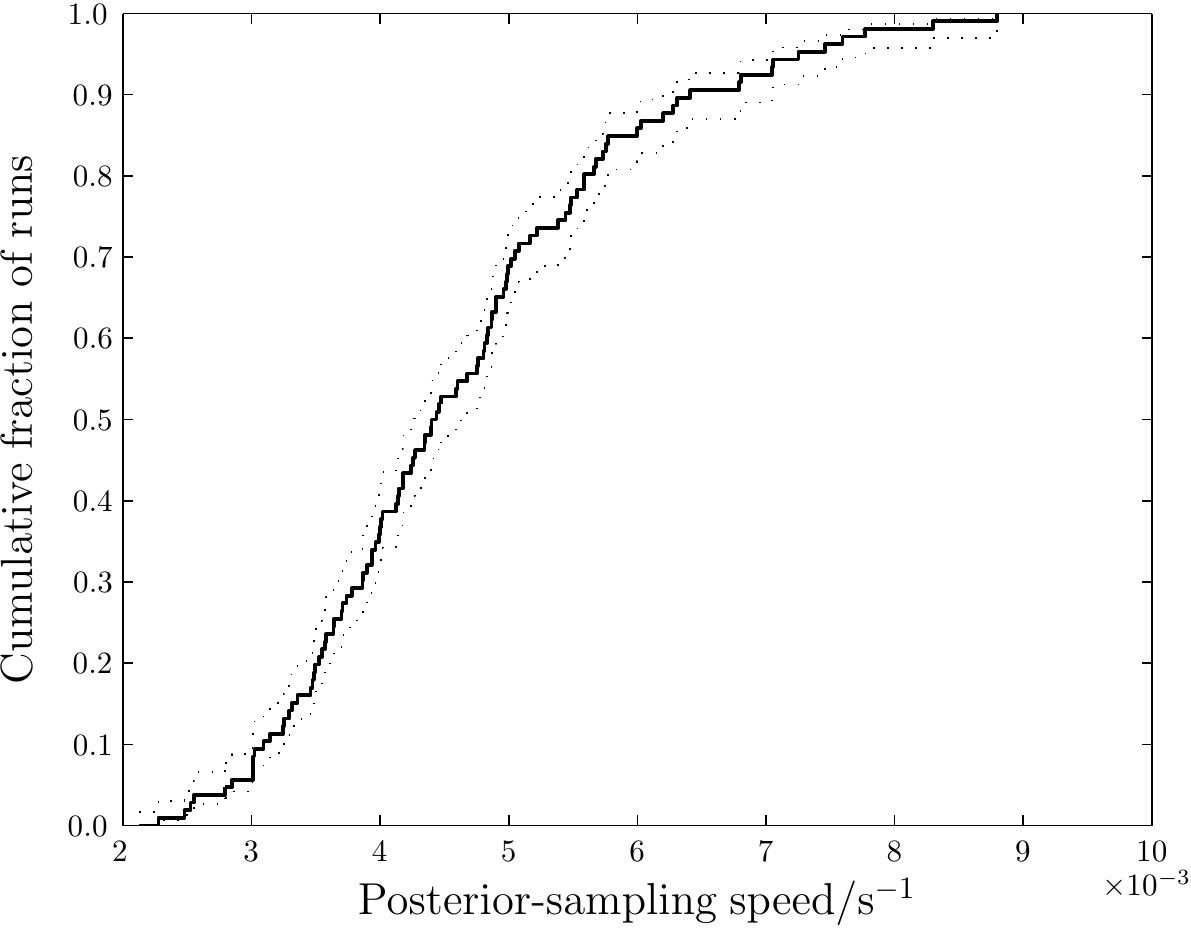}}
    \caption{Computation speed of \textsc{LALInference\_nest} runs measured in independent posterior samples per CPU second. (a) Distribution of sampling speeds. Speeds based on reliably estimated CPU times are shown in dark blue, while the full set of speeds, including those using potentially inaccurately estimated times, are shown in light blue. (b) Cumulative fractions of runs with computational speeds smaller than the abscissa value, only reliable values are used here. The $68\%$ confidence interval is enclosed by the dotted lines. All quantities are calculated based upon total CPU times, not wall times.} 
    \label{fig:speed}
\end{figure}

The length of the \textsc{LALInference} run depends upon the desired number of posterior samples. We may characterise the computational speed by the average rate at which independent samples are drawn from the posterior: the total number of (independent, as determined by \textsc{LALInference}) posterior samples divided by the total CPU time. The distribution of sampling speeds is shown in figure \ref{fig:speed}.
We use speeds calculated using both reliably estimated times and those we suspect might be lower bounds (giving upper bounds for sampling speed) in figure \ref{fig:speed-a}, but only the more reliable values in figure \ref{fig:speed-b}. The median (accurately estimated) \textsc{LALInference\_nest} sampling speed is $4.40 \times 10^{-3}~\mathrm{s^{-1}} = 15.8~\mathrm{hr^{-1}}$ corresponding to one independent posterior sample every $227~\mathrm{s} = 6.31 \times 10^{-2}~\mathrm{hr}$ of CPU time \citep[cf.][]{Sidery:2013zua,Veitch:2014wba}.

In contract, \textsc{bayestar} computes the likelihood $24576$ times. Its computation speed is thus simply inversely proportional to the total CPU time. The median \textsc{bayestar} computational speed is $26.7~\mathrm{s^{-1}}$ corresponding to one likelihood evaluation every $37.5~\mathrm{ms}$ of CPU time. The difference between the \textsc{LALInference} and \textsc{bayestar} computational speeds reflects the difference in the complexities of their likelihood functions.

The medium-latency PE runs, using the current code, finish in a few days. This is much longer than is required for \textsc{bayestar} to produce sky-localization estimates. However, \textsc{LALInference} also provides posterior probability distributions for the other parameters as well as more accurate sky localization than \textsc{bayestar} for three-detector networks \citep{Singer:2014qca}.

\section{Supplementary data}\label{ap:online-data}

Data produced for this study are available electronically, as shown in the following tables. In the print edition, only two example entries are included in these tables. Further details are explained in the appendix of \citet{Singer:2014qca}. Theses tables, along with sky maps are available online at \url{http://www.ligo.org/scientists/first2years/}. Table \ref{tab:data-sim} gives the injected (true) parameters of the $333$ simulated signals used for this study. Table \ref{tab:data-coinc} gives the detection parameters (the SNRs, FAR and masses returned by the detection pipeline), and the sky areas calculated by \textsc{bayestar} and \textsc{LALInference}. Table \ref{tab:data-pe} gives quantities related to PE for the chirp mass and distance. The second event listed in these tables is the one used for figure \ref{fig:sky-map}. Table \ref{tab:data-pe-gaussian} is the counterpart of table \ref{tab:data-pe}, but for the $250$ events using Gaussian noise. The events shown in the print edition are the same examples used by \citet{Singer:2014qca}.
\begin{deluxetable*}{cc ccccccc cc ccc ccc}
\tabletypesize{\scriptsize}
\tablecaption{\label{tab:data-sim} Simulated BNS signals of detected events for 2015 scenario using recoloured noise \citep[cf.][table 2]{Singer:2014qca}. Given are the event ID and simulation ID which specify the signal; the modified Julian date (MJD) of arrival at geocentre of the signal from last stable orbit; the sky position in terms of the right ascension $\alpha$ and declination $\delta$ (J2000); the binary's orbital-inclination angle $\iota$; the polarization angle $\psi$ \citep[appendix B]{Anderson:2000yy}; the orbital phase at coalescence $\phi_\mathrm{c}$; the source distance $D$; the component masses $m_1$ and $m_2$, and the $x$-, $y$- and $z$- components of the spins $a_1$ and $a_2$.}
\tablehead{
    \colhead{Event} &
    \colhead{Sim} &
    \multirow{2}{*}{MJD$/\mathrm{d}$} &
    \multirow{2}{*}{$\displaystyle \frac{\alpha}{\mathrm{deg}}$} &
    \multirow{2}{*}{$\displaystyle \frac{\delta}{\mathrm{deg}}$} &
    \multirow{2}{*}{$\displaystyle \frac{\iota}{\mathrm{deg}}$} &
    \multirow{2}{*}{$\displaystyle \frac{\psi}{\mathrm{deg}}$} &
    \multirow{2}{*}{$\displaystyle \frac{\phi_\mathrm{c}}{\mathrm{deg}}$} &
    \multirow{2}{*}{$\displaystyle \frac{D}{\mathrm{Mpc}}$} &
    \multirow{2}{*}{$\displaystyle \frac{m_1}{M_\odot}$} &
    \multirow{2}{*}{$\displaystyle \frac{m_2}{M_\odot}$} &
    \multirow{2}{*}{$a_1^x$} &
    \multirow{2}{*}{$a_1^y$} &
    \multirow{2}{*}{$a_1^z$} &
    \multirow{2}{*}{$a_2^x$} &
    \multirow{2}{*}{$a_2^y$} &
    \multirow{2}{*}{$a_2^z$} \\
    \colhead{ID\tablenotemark{a}} &
    \colhead{ID\tablenotemark{b}} &
    & & & & & & & & & & & & & & 
}
\startdata
 $4532$ & $ \hphantom{0}899$ & $55430.10310$ & $ \hphantom{0}99.9$ & $ -30.8$ & $ 26$ & $349$ & $ 118$ & $ 84$ & $1.25$ & $1.36$ & $-0.04$ & $-0.01$ & $-0.01$ & $\hphantom{-}0.01$ & $\hphantom{-}0.00$ & $-0.00$ \\ 
 $4572$ & $ 1243$ & $55430.52510$ & $227.5$ & $ -51.7$ & $ 48$ & $ \hphantom{0}27$ & $ 266$ & $ 61$ & $1.25$ & $1.33$ & $-0.01$ & $-0.00$ & $-0.04$ & $-0.01$ & $-0.01$ & $-0.00$ \\
\multicolumn{1}{c}{\vdots} & \multicolumn{1}{c}{\vdots} & \multicolumn{1}{c}{\vdots} & \multicolumn{1}{c}{\vdots} & \multicolumn{1}{c}{\vdots} & \multicolumn{1}{c}{\vdots} & \multicolumn{1}{c}{\vdots} & \multicolumn{1}{c}{\vdots} & \multicolumn{1}{c}{\vdots} & \multicolumn{1}{c}{\vdots} & \multicolumn{1}{c}{\vdots} & \multicolumn{1}{c}{\vdots} & \multicolumn{1}{c}{\vdots} & \multicolumn{1}{c}{\vdots} & \multicolumn{1}{c}{\vdots} & \multicolumn{1}{c}{\vdots} & \multicolumn{1}{c}{\vdots}
\enddata
\tablenotetext{a}{This identifier for detection candidates is the same value as the \texttt{coinc\_event\_id} column in the \textsc{GSTLAL} output database and the \texttt{OBJECT} cards in sky map FITS headers, with the \texttt{coinc\_event:coinc\_event\_id:} prefix stripped.}
\tablenotetext{b}{This identifier for simulateds signal is the same value as the \texttt{simulation\_id} column in the \textsc{GSTLAL} output database, with the \texttt{sim\_inspiral:simulation\_id:} prefix stripped.}
\tablecomments{Table \ref{tab:data-sim} is published in its entirety in the electronic edition of the Astrophysical Journal. A portion is shown here for guidance regarding its form and content.}
\end{deluxetable*}
\begin{deluxetable*}{cc c ccc cc ccc ccc c}
\tabletypesize{\scriptsize}
\tablecaption{\label{tab:data-coinc} Detections and sky-localization areas for 2015 scenario using recoloured noise \citep[cf.][table 3]{Singer:2014qca}. Given are the event and simulation IDs; the detector network;\tablenotemark{a} the SNR for the network $\varrho$ and for the Hanford $\varrho_\mathrm{H}$ and Livingston $\varrho_\mathrm{L}$ detectors;\tablenotemark{b} the maximum-likelihood estimates of component masses masses $m_1^\mathrm{ML}$ and $m_2^\mathrm{ML}$ as reported by \textsc{GSTLAL}; the sky areas returned by \textsc{bayestar} and \textsc{LALInference}, and the FAR corresponding to the detection.}
\tablehead{
    \colhead{} & \colhead{} & \colhead{} & \colhead{} & \colhead{} & \colhead{} & \multirow{3}{*}{$\displaystyle \frac{m_1^\mathrm{ML}}{M_\odot}$} & \multirow{3}{*}{$\displaystyle \frac{m_2^\mathrm{ML}}{M_\odot}$} &  
    \multicolumn{3}{c}{\textsc{bayestar}} &
    \multicolumn{3}{c}{\textsc{LALInference}} & \\ 
    \colhead{Event ID} &
    \colhead{Sim ID} &
    \colhead{Network} &
    \colhead{$\varrho$} &
    \colhead{$\varrho_\mathrm{H}$} &
    \colhead{$\varrho_\mathrm{L}$} &
    &
    &
    \multirow{2}{*}{$\displaystyle \frac{\mathrm{CR_{0.5}}}{\mathrm{deg^2}}$} &
    \multirow{2}{*}{$\displaystyle \frac{\mathrm{CR_{0.9}}}{\mathrm{deg^2}}$} &
    \multirow{2}{*}{$\displaystyle \frac{\mathrm{A_\ast}}{\mathrm{deg^2}}$} &
    \multirow{2}{*}{$\displaystyle \frac{\mathrm{CR_{0.5}}}{\mathrm{deg^2}}$} &
    \multirow{2}{*}{$\displaystyle \frac{\mathrm{CR_{0.9}}}{\mathrm{deg^2}}$} &
    \multirow{2}{*}{$\displaystyle \frac{\mathrm{A_\ast}}{\mathrm{deg^2}}$} &
    \colhead{FAR/$\mathrm{s^{-1}}$} \\
    & & & & & & & & & & & & & &
}
\startdata
 $4532$ & $ \hphantom{0}899$ & HL & $13.9$ & $10.1$ & $ 9.5$ & $1.60$ & $1.08$ & $ 181.76$ & $ 753.06$ & $ 186.22$ & $ 168.57$ & $ 788.15$ & $ 153.09$ & $2.14 \times 10^{-14}$ \\
 $4572$ & $ 1243$ & HL & $13.2$ & $10.0$ & $ 8.7$ & $1.73$ & $0.98$ & $ 227.91$ & $ 828.23$ & $ \hphantom{0}44.55$ & $ 203.63$ & $ 920.10$ & $ \hphantom{0}33.27$ & $1.27 \times 10^{-13}$ \\
\multicolumn{1}{c}{\vdots} & \multicolumn{1}{c}{\vdots} & \multicolumn{1}{c}{\vdots} & \multicolumn{1}{c}{\vdots} & \multicolumn{1}{c}{\vdots} & \multicolumn{1}{c}{\vdots} & \multicolumn{1}{c}{\vdots} & \multicolumn{1}{c}{\vdots} & \multicolumn{1}{c}{\vdots} & \multicolumn{1}{c}{\vdots} & \multicolumn{1}{c}{\vdots} & \multicolumn{1}{c}{\vdots} & \multicolumn{1}{c}{\vdots} & \multicolumn{1}{c}{\vdots} & \multicolumn{1}{c}{\vdots}
\enddata
\tablenotetext{a}{All detections are for a two-detector Hanford--Livingston (HL) network.}
\tablenotetext{b}{The network SNR is calculated by adding individual detectors in quadrature so $\varrho^2 = \varrho_\mathrm{H}^2 + \varrho_\mathrm{L}^2$.}
\tablecomments{Table \ref{tab:data-coinc} is published in its entirety in the electronic edition of the Astrophysical Journal. A portion is shown here for guidance regarding its form and content.}
\end{deluxetable*}
\begin{deluxetable*}{cc cc ccc ccc}
\tabletypesize{\scriptsize}
\tablecaption{\label{tab:data-pe} Parameter-estimation accuracies for 2015 scenario using recoloured noise. Given are the event and simulation IDs; the injected (true) chirp mass $\mathcal{M}_\star$ and distance $D_\star$; the posterior mean chirp mass $\bar{\mathcal{M}}_\mathrm{c}$; the chirp-mass credible intervals $\mathrm{CI}^{\mathcal{M}_\mathrm{c}}_{0.5}$ and $\mathrm{CI}^{\mathcal{M}_\mathrm{c}}_{0.9}$; the posterior mean distance $\bar{D}$, and the distance credible intervals $\mathrm{CI}^{D}_{0.5}$ and $\mathrm{CI}^{D}_{0.9}$. All parameter estimates are calculated by \textsc{LALInference}.}
\tablehead{  
    \colhead{} &
    \colhead{} &
    \multirow{3}{*}{$\displaystyle \frac{\mathcal{M}_\star}{M_\odot}$} &
    \multirow{3}{*}{$\displaystyle \frac{D_\star}{\mathrm{Mpc}}$} &
    \multirow{3}{*}{$\displaystyle \frac{\bar{\mathcal{M}}_\mathrm{c}}{M_\odot}$} &
    \multirow{3}{*}{$\displaystyle \frac{\mathrm{CI}^{\mathcal{M}_\mathrm{c}}_{0.5}}{M_\odot}$} &
    \multirow{3}{*}{$\displaystyle \frac{\mathrm{CI}^{\mathcal{M}_\mathrm{c}}_{0.9}}{M_\odot}$} &
    \multirow{3}{*}{$\displaystyle \frac{\bar{D}}{\mathrm{Mpc}}$} &
    \multirow{3}{*}{$\displaystyle \frac{\mathrm{CI}^{D}_{0.5}}{\mathrm{Mpc}}$} &
    \multirow{3}{*}{$\displaystyle \frac{\mathrm{CI}^{D}_{0.9}}{\mathrm{Mpc}}$} \\
    \colhead{Event ID} &
    \colhead{Sim ID} &
    & & & & & & & \\
    \colhead{} &
    \colhead{} &
    & & & & & & &
}
\startdata
 $4532$ & $\hphantom{0}899$ & $1.136613$ & $ 84.2$ & $1.136689$ & $0.000355$ & $0.000795$ & $ 64.6$ & $ 25.0$ & $ 53.3$ \\
 $4572$ & $1243$ & $1.123169$ & $ 60.7$ & $1.123286$ & $0.000410$ & $0.000901$ & $ 67.5$ & $ 26.7$ & $ 57.7$ \\
\multicolumn{1}{c}{\vdots} & \multicolumn{1}{c}{\vdots} & \multicolumn{1}{c}{\vdots} & \multicolumn{1}{c}{\vdots} & \multicolumn{1}{c}{\vdots} & \multicolumn{1}{c}{\vdots} & \multicolumn{1}{c}{\vdots} & \multicolumn{1}{c}{\vdots} & \multicolumn{1}{c}{\vdots} & \multicolumn{1}{c}{\vdots}
\enddata
\tablecomments{Table \ref{tab:data-pe} is published in its entirety in the electronic edition of the Astrophysical Journal. A portion is shown here for guidance regarding its form and content.}
\end{deluxetable*}
\begin{deluxetable*}{cc cc ccc ccc}
\tabletypesize{\scriptsize}
\tablecaption{\label{tab:data-pe-gaussian} Parameter-estimation accuracies for 2015 scenario using Gaussian noise. The columns are the same as in table \ref{tab:data-pe}.}
\tablehead{  
    \colhead{} &
    \colhead{} &
    \multirow{3}{*}{$\displaystyle \frac{\mathcal{M}_\star}{M_\odot}$} &
    \multirow{3}{*}{$\displaystyle \frac{D_\star}{\mathrm{Mpc}}$} &
    \multirow{3}{*}{$\displaystyle \frac{\bar{\mathcal{M}}_\mathrm{c}}{M_\odot}$} &
    \multirow{3}{*}{$\displaystyle \frac{\mathrm{CI}^{\mathcal{M}_\mathrm{c}}_{0.5}}{M_\odot}$} &
    \multirow{3}{*}{$\displaystyle \frac{\mathrm{CI}^{\mathcal{M}_\mathrm{c}}_{0.9}}{M_\odot}$} &
    \multirow{3}{*}{$\displaystyle \frac{\bar{D}}{\mathrm{Mpc}}$} &
    \multirow{3}{*}{$\displaystyle \frac{\mathrm{CI}^{D}_{0.5}}{\mathrm{Mpc}}$} &
    \multirow{3}{*}{$\displaystyle \frac{\mathrm{CI}^{D}_{0.9}}{\mathrm{Mpc}}$} \\
    \colhead{Event ID} &
    \colhead{Sim ID} &
    & & & & & & & \\
    \colhead{} &
    \colhead{} &
    & & & & & & &
}
\startdata
 $18951$ & $10807$ & $1.264368$ & $ 74.8$ & $1.264410$ & $0.000457$ & $0.001017$ & $ 70.4$ & $ 23.1$ & $ 50.6$ \\
 $20342$ & $21002$ & $1.223944$ & $ 75.0$ & $1.223740$ & $0.000444$ & $0.001034$ & $ 71.2$ & $ 26.0$ & $ 57.3$ \\
\multicolumn{1}{c}{\vdots} & \multicolumn{1}{c}{\vdots} & \multicolumn{1}{c}{\vdots} & \multicolumn{1}{c}{\vdots} & \multicolumn{1}{c}{\vdots} & \multicolumn{1}{c}{\vdots} & \multicolumn{1}{c}{\vdots} & \multicolumn{1}{c}{\vdots} & \multicolumn{1}{c}{\vdots} & \multicolumn{1}{c}{\vdots}
\enddata
\tablecomments{Table \ref{tab:data-pe-gaussian} is published in its entirety in the electronic edition of the Astrophysical Journal. A portion is shown here for guidance regarding its form and content.}
\end{deluxetable*}

\bibliographystyle{apj}
\bibliography{updateObs} 

\end{document}